\documentclass[a4paper, traditabstract, longauth]{aa} 

\def\setsymbol#1#2{\expandafter\def\csname #1\endcsname{#2}}
\def\getsymbol#1{\csname #1\endcsname}

\def\allearlypapers{\nocite{planck2011-1.1, planck2011-1.3, planck2011-1.4, planck2011-1.5, planck2011-1.6, planck2011-1.7, planck2011-1.10, planck2011-1.10sup, planck2011-5.1a, planck2011-5.1b, planck2011-5.2a, planck2011-5.2b, planck2011-5.2c, planck2011-6.1, planck2011-6.2, planck2011-6.3a, planck2011-6.4a, planck2011-6.4b, planck2011-6.6, planck2011-7.0, planck2011-7.2, planck2011-7.3, planck2011-7.7a, planck2011-7.7b, planck2011-7.12, planck2011-7.13}}

\newbox\tablebox    \newdimen\tablewidth
\def\leaderfil{\leaders\hbox to 5pt{\hss.\hss}\hfil}
\def\tablenote#1 #2\par{\begingroup \parindent=0.8em
    \abovedisplayshortskip=0pt\belowdisplayshortskip=0pt
    \noindent
    $$\hss\vbox{\hsize\tablewidth \hangindent=\parindent \hangafter=1 \noindent
    \hbox to \parindent{\sup{\rm #1}\hss}\strut#2\strut\par}\hss$$
    \endgroup}

\def\L2{\ifmmode L_2\else $L_2$\fi}

\def\DeltaT{\ifmmode \Delta T\else $\Delta T$\fi}
\def\deltat{\ifmmode \Delta t\else $\Delta t$\fi}
\def\fknee{\ifmmode f_{\rm knee}\else $f_{\rm knee}$\fi}
\def\Fmax{\ifmmode F_{\rm max}\else $F_{\rm max}$\fi}
\def\solar{\ifmmode{\rm M}_{\mathord\odot}\else${\rm M}_{\mathord\odot}$\fi}

\def\inv{\ifmmode^{-1}\else$^{-1}$\fi}
\def\mo{\ifmmode^{-1}\else$^{-1}$\fi}
\def\sup#1{\ifmmode ^{\rm #1}\else $^{\rm #1}$\fi}
\def\expo#1{\ifmmode \times 10^{#1}\else $\times 10^{#1}$\fi}
\def\,{\thinspace}
\def\lsim{\mathrel{\raise .4ex\hbox{\rlap{$<$}\lower 1.2ex\hbox{$\sim$}}}}
\def\gsim{\mathrel{\raise .4ex\hbox{\rlap{$>$}\lower 1.2ex\hbox{$\sim$}}}}

\def\simprop{\mathrel{\raise .4ex\hbox{\rlap{$\propto$}\lower 1.2ex\hbox{$\sim$}}}}
\def\deg{\ifmmode^\circ\else$^\circ$\fi}
\def\pdeg{\ifmmode $\setbox0=\hbox{$^{\circ}$}\rlap{\hskip.11\wd0 .}$^{\circ}
          \else \setbox0=\hbox{$^{\circ}$}\rlap{\hskip.11\wd0 .}$^{\circ}$\fi}
\def\arcs{\ifmmode {^{\scriptstyle\prime\prime}}
          \else $^{\scriptstyle\prime\prime}$\fi}
\def\arcm{\ifmmode {^{\scriptstyle\prime}}
          \else $^{\scriptstyle\prime}$\fi}
\newdimen\sa  \newdimen\sb
\def\parcs{\sa=.07em \sb=.03em
     \ifmmode \hbox{\rlap{.}}^{\scriptstyle\prime\kern -\sb\prime}\hbox{\kern -\sa}
     \else \rlap{.}$^{\scriptstyle\prime\kern -\sb\prime}$\kern -\sa\fi}
\def\parcm{\sa=.08em \sb=.03em
     \ifmmode \hbox{\rlap{.}\kern\sa}^{\scriptstyle\prime}\hbox{\kern-\sb}
     \else \rlap{.}\kern\sa$^{\scriptstyle\prime}$\kern-\sb\fi}
\def\ra[#1 #2 #3.#4]{#1\sup{h}#2\sup{m}#3\sup{s}\llap.#4}
\def\dec[#1 #2 #3.#4]{#1\deg#2\arcm#3\arcs\llap.#4}
\def\deco[#1 #2 #3]{#1\deg#2\arcm#3\arcs}
\def\rra[#1 #2]{#1\sup{h}#2\sup{m}}

\def\dots{\relax\ifmmode \ldots\else $\ldots$\fi}
\def\WHzsr{\ifmmode $W\,Hz\mo\,sr\mo$\else W\,Hz\mo\,sr\mo\fi}
\def\mHz{\ifmmode $\,mHz$\else \,mHz\fi}
\def\GHz{\ifmmode $\,GHz$\else \,GHz\fi}
\def\mKs{\ifmmode $\,mK\,s$^{1/2}\else \,mK\,s$^{1/2}$\fi}
\def\muKs{\ifmmode \,\mu$K\,s$^{1/2}\else \,$\mu$K\,s$^{1/2}$\fi}
\def\muKRJs{\ifmmode \,\mu$K$_{\rm RJ}$\,s$^{1/2}\else \,$\mu$K$_{\rm RJ}$\,s$^{1/2}$\fi}
\def\muKHz{\ifmmode \,\mu$K\,Hz$^{-1/2}\else \,$\mu$K\,Hz$^{-1/2}$\fi}
\def\MJysr{\ifmmode \,$MJy\,sr\mo$\else \,MJy\,sr\mo\fi}
\def\MJysrmK{\ifmmode \,$MJy\,sr\mo$\,mK$_{\rm CMB}\mo\else \,MJy\,sr\mo\,mK$_{\rm CMB}\mo$\fi}
\def\microns{\ifmmode \,\mu$m$\else \,$\mu$m\fi}

\def\muK{\ifmmode \,\mu$K$\else \,$\mu$\hbox{K}\fi}
\def\microK{\ifmmode \,\mu$K$\else \,$\mu$\hbox{K}\fi}
\def\muW{\ifmmode \,\mu$W$\else \,$\mu$\hbox{W}\fi}
\def\kms{\ifmmode $\,km\,s$^{-1}\else \,km\,s$^{-1}$\fi}
\def\kmsMpc{\ifmmode $\,\kms\,Mpc\mo$\else \,\kms\,Mpc\mo\fi}
\setsymbol{LFI:center:frequency:70GHz:units}{70.3\,GHz}
\setsymbol{LFI:center:frequency:44GHz:units}{44.1\,GHz}
\setsymbol{LFI:center:frequency:30GHz:units}{28.5\,GHz}

\setsymbol{LFI:center:frequency:70GHz}{70.3}
\setsymbol{LFI:center:frequency:44GHz}{44.1}
\setsymbol{LFI:center:frequency:30GHz}{28.5}

\setsymbol{LFI:center:frequency:LFI18:Rad:M:units}{71.7\GHz}
\setsymbol{LFI:center:frequency:LFI19:Rad:M:units}{67.5\GHz}
\setsymbol{LFI:center:frequency:LFI20:Rad:M:units}{69.2\GHz}
\setsymbol{LFI:center:frequency:LFI21:Rad:M:units}{70.4\GHz}
\setsymbol{LFI:center:frequency:LFI22:Rad:M:units}{71.5\GHz}
\setsymbol{LFI:center:frequency:LFI23:Rad:M:units}{70.8\GHz}
\setsymbol{LFI:center:frequency:LFI24:Rad:M:units}{44.4\GHz}
\setsymbol{LFI:center:frequency:LFI25:Rad:M:units}{44.0\GHz}
\setsymbol{LFI:center:frequency:LFI26:Rad:M:units}{43.9\GHz}
\setsymbol{LFI:center:frequency:LFI27:Rad:M:units}{28.3\GHz}
\setsymbol{LFI:center:frequency:LFI28:Rad:M:units}{28.8\GHz}
\setsymbol{LFI:center:frequency:LFI18:Rad:S:units}{70.1\GHz}
\setsymbol{LFI:center:frequency:LFI19:Rad:S:units}{69.6\GHz}
\setsymbol{LFI:center:frequency:LFI20:Rad:S:units}{69.5\GHz}
\setsymbol{LFI:center:frequency:LFI21:Rad:S:units}{69.5\GHz}
\setsymbol{LFI:center:frequency:LFI22:Rad:S:units}{72.8\GHz}
\setsymbol{LFI:center:frequency:LFI23:Rad:S:units}{71.3\GHz}
\setsymbol{LFI:center:frequency:LFI24:Rad:S:units}{44.1\GHz}
\setsymbol{LFI:center:frequency:LFI25:Rad:S:units}{44.1\GHz}
\setsymbol{LFI:center:frequency:LFI26:Rad:S:units}{44.1\GHz}
\setsymbol{LFI:center:frequency:LFI27:Rad:S:units}{28.5\GHz}
\setsymbol{LFI:center:frequency:LFI28:Rad:S:units}{28.2\GHz}

\setsymbol{LFI:center:frequency:LFI18:Rad:M}{71.7}
\setsymbol{LFI:center:frequency:LFI19:Rad:M}{67.5}
\setsymbol{LFI:center:frequency:LFI20:Rad:M}{69.2}
\setsymbol{LFI:center:frequency:LFI21:Rad:M}{70.4}
\setsymbol{LFI:center:frequency:LFI22:Rad:M}{71.5}
\setsymbol{LFI:center:frequency:LFI23:Rad:M}{70.8}
\setsymbol{LFI:center:frequency:LFI24:Rad:M}{44.4}
\setsymbol{LFI:center:frequency:LFI25:Rad:M}{44.0}
\setsymbol{LFI:center:frequency:LFI26:Rad:M}{43.9}
\setsymbol{LFI:center:frequency:LFI27:Rad:M}{28.3}
\setsymbol{LFI:center:frequency:LFI28:Rad:M}{28.8}
\setsymbol{LFI:center:frequency:LFI18:Rad:S}{70.1}
\setsymbol{LFI:center:frequency:LFI19:Rad:S}{69.6}
\setsymbol{LFI:center:frequency:LFI20:Rad:S}{69.5}
\setsymbol{LFI:center:frequency:LFI21:Rad:S}{69.5}
\setsymbol{LFI:center:frequency:LFI22:Rad:S}{72.8}
\setsymbol{LFI:center:frequency:LFI23:Rad:S}{71.3}
\setsymbol{LFI:center:frequency:LFI24:Rad:S}{44.1}
\setsymbol{LFI:center:frequency:LFI25:Rad:S}{44.1}
\setsymbol{LFI:center:frequency:LFI26:Rad:S}{44.1}
\setsymbol{LFI:center:frequency:LFI27:Rad:S}{28.5}
\setsymbol{LFI:center:frequency:LFI28:Rad:S}{28.2}

\setsymbol{LFI:white:noise:sensitivity:70GHz:units}{134.7\muKs}
\setsymbol{LFI:white:noise:sensitivity:44GHz:units}{164.7\muKs}
\setsymbol{LFI:white:noise:sensitivity:30GHz:units}{143.4\muKs}

\setsymbol{LFI:white:noise:sensitivity:70GHz}{134.7}
\setsymbol{LFI:white:noise:sensitivity:44GHz}{164.7}
\setsymbol{LFI:white:noise:sensitivity:30GHz}{143.4}

\setsymbol{LFI:white:noise:sensitivity:LFI18:Rad:M:units}{512.0\muKs}
\setsymbol{LFI:white:noise:sensitivity:LFI19:Rad:M:units}{581.4\muKs}
\setsymbol{LFI:white:noise:sensitivity:LFI20:Rad:M:units}{590.8\muKs}
\setsymbol{LFI:white:noise:sensitivity:LFI21:Rad:M:units}{455.2\muKs}
\setsymbol{LFI:white:noise:sensitivity:LFI22:Rad:M:units}{492.0\muKs}
\setsymbol{LFI:white:noise:sensitivity:LFI23:Rad:M:units}{507.7\muKs}
\setsymbol{LFI:white:noise:sensitivity:LFI24:Rad:M:units}{462.2\muKs}
\setsymbol{LFI:white:noise:sensitivity:LFI25:Rad:M:units}{413.6\muKs}
\setsymbol{LFI:white:noise:sensitivity:LFI26:Rad:M:units}{478.6\muKs}
\setsymbol{LFI:white:noise:sensitivity:LFI27:Rad:M:units}{277.7\muKs}
\setsymbol{LFI:white:noise:sensitivity:LFI28:Rad:M:units}{312.3\muKs}
\setsymbol{LFI:white:noise:sensitivity:LFI18:Rad:S:units}{465.7\muKs}
\setsymbol{LFI:white:noise:sensitivity:LFI19:Rad:S:units}{555.6\muKs}
\setsymbol{LFI:white:noise:sensitivity:LFI20:Rad:S:units}{623.2\muKs}
\setsymbol{LFI:white:noise:sensitivity:LFI21:Rad:S:units}{564.1\muKs}
\setsymbol{LFI:white:noise:sensitivity:LFI22:Rad:S:units}{534.4\muKs}
\setsymbol{LFI:white:noise:sensitivity:LFI23:Rad:S:units}{542.4\muKs}
\setsymbol{LFI:white:noise:sensitivity:LFI24:Rad:S:units}{399.2\muKs}
\setsymbol{LFI:white:noise:sensitivity:LFI25:Rad:S:units}{392.6\muKs}
\setsymbol{LFI:white:noise:sensitivity:LFI26:Rad:S:units}{418.6\muKs}
\setsymbol{LFI:white:noise:sensitivity:LFI27:Rad:S:units}{302.9\muKs}
\setsymbol{LFI:white:noise:sensitivity:LFI28:Rad:S:units}{285.3\muKs}

\setsymbol{LFI:white:noise:sensitivity:LFI18:Rad:M}{512.0}
\setsymbol{LFI:white:noise:sensitivity:LFI19:Rad:M}{581.4}
\setsymbol{LFI:white:noise:sensitivity:LFI20:Rad:M}{590.8}
\setsymbol{LFI:white:noise:sensitivity:LFI21:Rad:M}{455.2}
\setsymbol{LFI:white:noise:sensitivity:LFI22:Rad:M}{492.0}
\setsymbol{LFI:white:noise:sensitivity:LFI23:Rad:M}{507.7}
\setsymbol{LFI:white:noise:sensitivity:LFI24:Rad:M}{462.2}
\setsymbol{LFI:white:noise:sensitivity:LFI25:Rad:M}{413.6}
\setsymbol{LFI:white:noise:sensitivity:LFI26:Rad:M}{478.6}
\setsymbol{LFI:white:noise:sensitivity:LFI27:Rad:M}{277.7}
\setsymbol{LFI:white:noise:sensitivity:LFI28:Rad:M}{312.3}
\setsymbol{LFI:white:noise:sensitivity:LFI18:Rad:S}{465.7}
\setsymbol{LFI:white:noise:sensitivity:LFI19:Rad:S}{555.6}
\setsymbol{LFI:white:noise:sensitivity:LFI20:Rad:S}{623.2}
\setsymbol{LFI:white:noise:sensitivity:LFI21:Rad:S}{564.1}
\setsymbol{LFI:white:noise:sensitivity:LFI22:Rad:S}{534.4}
\setsymbol{LFI:white:noise:sensitivity:LFI23:Rad:S}{542.4}
\setsymbol{LFI:white:noise:sensitivity:LFI24:Rad:S}{399.2}
\setsymbol{LFI:white:noise:sensitivity:LFI25:Rad:S}{392.6}
\setsymbol{LFI:white:noise:sensitivity:LFI26:Rad:S}{418.6}
\setsymbol{LFI:white:noise:sensitivity:LFI27:Rad:S}{302.9}
\setsymbol{LFI:white:noise:sensitivity:LFI28:Rad:S}{285.3}

\setsymbol{LFI:knee:frequency:70GHz:units}{29.5\mHz}
\setsymbol{LFI:knee:frequency:44GHz:units}{56.2\mHz}
\setsymbol{LFI:knee:frequency:30GHz:units}{113.7\mHz}

\setsymbol{LFI:knee:frequency:70GHz}{29.5}
\setsymbol{LFI:knee:frequency:44GHz}{56.2}
\setsymbol{LFI:knee:frequency:30GHz}{113.7}

\setsymbol{LFI:knee:frequency:LFI18:Rad:M:units}{16.3\mHz}
\setsymbol{LFI:knee:frequency:LFI19:Rad:M:units}{15.1\mHz}
\setsymbol{LFI:knee:frequency:LFI20:Rad:M:units}{18.7\mHz}
\setsymbol{LFI:knee:frequency:LFI21:Rad:M:units}{37.2\mHz}
\setsymbol{LFI:knee:frequency:LFI22:Rad:M:units}{12.7\mHz}
\setsymbol{LFI:knee:frequency:LFI23:Rad:M:units}{34.6\mHz}
\setsymbol{LFI:knee:frequency:LFI24:Rad:M:units}{46.2\mHz}
\setsymbol{LFI:knee:frequency:LFI25:Rad:M:units}{24.9\mHz}
\setsymbol{LFI:knee:frequency:LFI26:Rad:M:units}{67.6\mHz}
\setsymbol{LFI:knee:frequency:LFI27:Rad:M:units}{187.4\mHz}
\setsymbol{LFI:knee:frequency:LFI28:Rad:M:units}{122.2\mHz}
\setsymbol{LFI:knee:frequency:LFI18:Rad:S:units}{17.7\mHz}
\setsymbol{LFI:knee:frequency:LFI19:Rad:S:units}{22.0\mHz}
\setsymbol{LFI:knee:frequency:LFI20:Rad:S:units}{8.7\mHz}
\setsymbol{LFI:knee:frequency:LFI21:Rad:S:units}{25.9\mHz}
\setsymbol{LFI:knee:frequency:LFI22:Rad:S:units}{15.8\mHz}
\setsymbol{LFI:knee:frequency:LFI23:Rad:S:units}{129.8\mHz}
\setsymbol{LFI:knee:frequency:LFI24:Rad:S:units}{100.9\mHz}
\setsymbol{LFI:knee:frequency:LFI25:Rad:S:units}{38.9\mHz}
\setsymbol{LFI:knee:frequency:LFI26:Rad:S:units}{58.9\mHz}
\setsymbol{LFI:knee:frequency:LFI27:Rad:S:units}{104.4\mHz}
\setsymbol{LFI:knee:frequency:LFI28:Rad:S:units}{40.7\mHz}

\setsymbol{LFI:knee:frequency:LFI18:Rad:M}{16.3}
\setsymbol{LFI:knee:frequency:LFI19:Rad:M}{15.1}
\setsymbol{LFI:knee:frequency:LFI20:Rad:M}{18.7}
\setsymbol{LFI:knee:frequency:LFI21:Rad:M}{37.2}
\setsymbol{LFI:knee:frequency:LFI22:Rad:M}{12.7}
\setsymbol{LFI:knee:frequency:LFI23:Rad:M}{34.6}
\setsymbol{LFI:knee:frequency:LFI24:Rad:M}{46.2}
\setsymbol{LFI:knee:frequency:LFI25:Rad:M}{24.9}
\setsymbol{LFI:knee:frequency:LFI26:Rad:M}{67.6}
\setsymbol{LFI:knee:frequency:LFI27:Rad:M}{187.4}
\setsymbol{LFI:knee:frequency:LFI28:Rad:M}{122.2}
\setsymbol{LFI:knee:frequency:LFI18:Rad:S}{17.7}
\setsymbol{LFI:knee:frequency:LFI19:Rad:S}{22.0}
\setsymbol{LFI:knee:frequency:LFI20:Rad:S}{8.7}
\setsymbol{LFI:knee:frequency:LFI21:Rad:S}{25.9}
\setsymbol{LFI:knee:frequency:LFI22:Rad:S}{15.8}
\setsymbol{LFI:knee:frequency:LFI23:Rad:S}{129.8}
\setsymbol{LFI:knee:frequency:LFI24:Rad:S}{100.9}
\setsymbol{LFI:knee:frequency:LFI25:Rad:S}{38.9}
\setsymbol{LFI:knee:frequency:LFI26:Rad:S}{58.9}
\setsymbol{LFI:knee:frequency:LFI27:Rad:S}{104.4}
\setsymbol{LFI:knee:frequency:LFI28:Rad:S}{40.7}

\setsymbol{LFI:slope:70GHz:units}{$-1.03$\mHz}
\setsymbol{LFI:slope:44GHz:units}{$-0.89$\mHz}
\setsymbol{LFI:slope:30GHz:units}{$-0.87$\mHz}

\setsymbol{LFI:slope:70GHz}{$-1.03$}
\setsymbol{LFI:slope:44GHz}{$-0.89$}
\setsymbol{LFI:slope:30GHz}{$-0.87$}

\setsymbol{LFI:slope:LFI18:Rad:M:units}{$-1.04$\mHz}
\setsymbol{LFI:slope:LFI19:Rad:M:units}{$-1.09$\mHz}
\setsymbol{LFI:slope:LFI20:Rad:M:units}{$-0.69$\mHz}
\setsymbol{LFI:slope:LFI21:Rad:M:units}{$-1.56$\mHz}
\setsymbol{LFI:slope:LFI22:Rad:M:units}{$-1.01$\mHz}
\setsymbol{LFI:slope:LFI23:Rad:M:units}{$-0.96$\mHz}
\setsymbol{LFI:slope:LFI24:Rad:M:units}{$-0.83$\mHz}
\setsymbol{LFI:slope:LFI25:Rad:M:units}{$-0.91$\mHz}
\setsymbol{LFI:slope:LFI26:Rad:M:units}{$-0.95$\mHz}
\setsymbol{LFI:slope:LFI27:Rad:M:units}{$-0.87$\mHz}
\setsymbol{LFI:slope:LFI28:Rad:M:units}{$-0.88$\mHz}
\setsymbol{LFI:slope:LFI18:Rad:S:units}{$-1.15$\mHz}
\setsymbol{LFI:slope:LFI19:Rad:S:units}{$-1.00$\mHz}
\setsymbol{LFI:slope:LFI20:Rad:S:units}{$-0.95$\mHz}
\setsymbol{LFI:slope:LFI21:Rad:S:units}{$-0.92$\mHz}
\setsymbol{LFI:slope:LFI22:Rad:S:units}{$-1.01$\mHz}
\setsymbol{LFI:slope:LFI23:Rad:S:units}{$-0.95$\mHz}
\setsymbol{LFI:slope:LFI24:Rad:S:units}{$-0.73$\mHz}
\setsymbol{LFI:slope:LFI25:Rad:S:units}{$-1.16$\mHz}
\setsymbol{LFI:slope:LFI26:Rad:S:units}{$-0.79$\mHz}
\setsymbol{LFI:slope:LFI27:Rad:S:units}{$-0.82$\mHz}
\setsymbol{LFI:slope:LFI28:Rad:S:units}{$-0.91$\mHz}

\setsymbol{LFI:slope:LFI18:Rad:M}{$-1.04$}
\setsymbol{LFI:slope:LFI19:Rad:M}{$-1.09$}
\setsymbol{LFI:slope:LFI20:Rad:M}{$-0.69$}
\setsymbol{LFI:slope:LFI21:Rad:M}{$-1.56$}
\setsymbol{LFI:slope:LFI22:Rad:M}{$-1.01$}
\setsymbol{LFI:slope:LFI23:Rad:M}{$-0.96$}
\setsymbol{LFI:slope:LFI24:Rad:M}{$-0.83$}
\setsymbol{LFI:slope:LFI25:Rad:M}{$-0.91$}
\setsymbol{LFI:slope:LFI26:Rad:M}{$-0.95$}
\setsymbol{LFI:slope:LFI27:Rad:M}{$-0.87$}
\setsymbol{LFI:slope:LFI28:Rad:M}{$-0.88$}
\setsymbol{LFI:slope:LFI18:Rad:S}{$-1.15$}
\setsymbol{LFI:slope:LFI19:Rad:S}{$-1.00$}
\setsymbol{LFI:slope:LFI20:Rad:S}{$-0.95$}
\setsymbol{LFI:slope:LFI21:Rad:S}{$-0.92$}
\setsymbol{LFI:slope:LFI22:Rad:S}{$-1.01$}
\setsymbol{LFI:slope:LFI23:Rad:S}{$-0.95$}
\setsymbol{LFI:slope:LFI24:Rad:S}{$-0.73$}
\setsymbol{LFI:slope:LFI25:Rad:S}{$-1.16$}
\setsymbol{LFI:slope:LFI26:Rad:S}{$-0.79$}
\setsymbol{LFI:slope:LFI27:Rad:S}{$-0.82$}
\setsymbol{LFI:slope:LFI28:Rad:S}{$-0.91$}

\setsymbol{LFI:FWHM:70GHz:units}{13\parcm01}
\setsymbol{LFI:FWHM:44GHz:units}{27\parcm92}
\setsymbol{LFI:FWHM:30GHz:units}{32\parcm65}

\setsymbol{LFI:FWHM:70GHz}{13.01}
\setsymbol{LFI:FWHM:44GHz}{27.92}
\setsymbol{LFI:FWHM:30GHz}{32.65}

\setsymbol{LFI:FWHM:LFI18:units}{13\parcm39}
\setsymbol{LFI:FWHM:LFI19:units}{13\parcm01}
\setsymbol{LFI:FWHM:LFI20:units}{12\parcm75}
\setsymbol{LFI:FWHM:LFI21:units}{12\parcm74}
\setsymbol{LFI:FWHM:LFI22:units}{12\parcm87}
\setsymbol{LFI:FWHM:LFI23:units}{13\parcm27}
\setsymbol{LFI:FWHM:LFI24:units}{22\parcm98}
\setsymbol{LFI:FWHM:LFI25:units}{30\parcm46}
\setsymbol{LFI:FWHM:LFI26:units}{30\parcm31}
\setsymbol{LFI:FWHM:LFI27:units}{32\parcm65}
\setsymbol{LFI:FWHM:LFI28:units}{32\parcm66}

\setsymbol{LFI:FWHM:LFI18}{13.39}
\setsymbol{LFI:FWHM:LFI19}{13.01}
\setsymbol{LFI:FWHM:LFI20}{12.75}
\setsymbol{LFI:FWHM:LFI21}{12.74}
\setsymbol{LFI:FWHM:LFI22}{12.87}
\setsymbol{LFI:FWHM:LFI23}{13.27}
\setsymbol{LFI:FWHM:LFI24}{22.98}
\setsymbol{LFI:FWHM:LFI25}{30.46}
\setsymbol{LFI:FWHM:LFI26}{30.31}
\setsymbol{LFI:FWHM:LFI27}{32.65}
\setsymbol{LFI:FWHM:LFI28}{32.66}

\setsymbol{LFI:FWHM:uncertainty:LFI18:units}{0.170\arcm}
\setsymbol{LFI:FWHM:uncertainty:LFI19:units}{0.174\arcm}
\setsymbol{LFI:FWHM:uncertainty:LFI20:units}{0.170\arcm}
\setsymbol{LFI:FWHM:uncertainty:LFI21:units}{0.156\arcm}
\setsymbol{LFI:FWHM:uncertainty:LFI22:units}{0.164\arcm}
\setsymbol{LFI:FWHM:uncertainty:LFI23:units}{0.171\arcm}
\setsymbol{LFI:FWHM:uncertainty:LFI24:units}{0.652\arcm}
\setsymbol{LFI:FWHM:uncertainty:LFI25:units}{1.075\arcm}
\setsymbol{LFI:FWHM:uncertainty:LFI26:units}{1.131\arcm}
\setsymbol{LFI:FWHM:uncertainty:LFI27:units}{1.266\arcm}
\setsymbol{LFI:FWHM:uncertainty:LFI28:units}{1.287\arcm}

\setsymbol{LFI:FWHM:uncertainty:LFI18}{0.170}
\setsymbol{LFI:FWHM:uncertainty:LFI19}{0.174}
\setsymbol{LFI:FWHM:uncertainty:LFI20}{0.170}
\setsymbol{LFI:FWHM:uncertainty:LFI21}{0.156}
\setsymbol{LFI:FWHM:uncertainty:LFI22}{0.164}
\setsymbol{LFI:FWHM:uncertainty:LFI23}{0.171}
\setsymbol{LFI:FWHM:uncertainty:LFI24}{0.652}
\setsymbol{LFI:FWHM:uncertainty:LFI25}{1.075}
\setsymbol{LFI:FWHM:uncertainty:LFI26}{1.131}
\setsymbol{LFI:FWHM:uncertainty:LFI27}{1.266}
\setsymbol{LFI:FWHM:uncertainty:LFI28}{1.287}

\setsymbol{HFI:center:frequency:100GHz:units}{100\,GHz}
\setsymbol{HFI:center:frequency:143GHz:units}{143\,GHz}
\setsymbol{HFI:center:frequency:217GHz:units}{217\,GHz}
\setsymbol{HFI:center:frequency:353GHz:units}{353\,GHz}
\setsymbol{HFI:center:frequency:545GHz:units}{545\,GHz}
\setsymbol{HFI:center:frequency:857GHz:units}{857\,GHz}

\setsymbol{HFI:center:frequency:100GHz}{100}
\setsymbol{HFI:center:frequency:143GHz}{143}
\setsymbol{HFI:center:frequency:217GHz}{217}
\setsymbol{HFI:center:frequency:353GHz}{353}
\setsymbol{HFI:center:frequency:545GHz}{545}
\setsymbol{HFI:center:frequency:857GHz}{857}

\setsymbol{HFI:Ndetectors:100GHz}{8}
\setsymbol{HFI:Ndetectors:143GHz}{11}
\setsymbol{HFI:Ndetectors:217GHz}{12}
\setsymbol{HFI:Ndetectors:353GHz}{12}
\setsymbol{HFI:Ndetectors:545GHz}{3}
\setsymbol{HFI:Ndetectors:857GHz}{4}

\setsymbol{HFI:FWHM:Maps:100GHz:units}{9\parcm88}
\setsymbol{HFI:FWHM:Maps:143GHz:units}{7\parcm18}
\setsymbol{HFI:FWHM:Maps:217GHz:units}{4\parcm87}
\setsymbol{HFI:FWHM:Maps:353GHz:units}{4\parcm65}
\setsymbol{HFI:FWHM:Maps:545GHz:units}{4\parcm72}
\setsymbol{HFI:FWHM:Maps:857GHz:units}{4\parcm39}
\setsymbol{HFI:FWHM:Maps:100GHz}{9.88}
\setsymbol{HFI:FWHM:Maps:143GHz}{7.18}
\setsymbol{HFI:FWHM:Maps:217GHz}{4.87}
\setsymbol{HFI:FWHM:Maps:353GHz}{4.65}
\setsymbol{HFI:FWHM:Maps:545GHz}{4.72}
\setsymbol{HFI:FWHM:Maps:857GHz}{4.39}

\setsymbol{HFI:beam:ellipticity:Maps:100GHz}{1.15}
\setsymbol{HFI:beam:ellipticity:Maps:143GHz}{1.01}
\setsymbol{HFI:beam:ellipticity:Maps:217GHz}{1.06}
\setsymbol{HFI:beam:ellipticity:Maps:353GHz}{1.05}
\setsymbol{HFI:beam:ellipticity:Maps:545GHz}{1.14}
\setsymbol{HFI:beam:ellipticity:Maps:857GHz}{1.19}

\setsymbol{HFI:FWHM:Mars:100GHz:units}{9\parcm37}
\setsymbol{HFI:FWHM:Mars:143GHz:units}{7\parcm04}
\setsymbol{HFI:FWHM:Mars:217GHz:units}{4\parcm68}
\setsymbol{HFI:FWHM:Mars:353GHz:units}{4\parcm43}
\setsymbol{HFI:FWHM:Mars:545GHz:units}{3\parcm80}
\setsymbol{HFI:FWHM:Mars:857GHz:units}{3\parcm67}

\setsymbol{HFI:FWHM:Mars:100GHz}{9.37}
\setsymbol{HFI:FWHM:Mars:143GHz}{7.04}
\setsymbol{HFI:FWHM:Mars:217GHz}{4.68}
\setsymbol{HFI:FWHM:Mars:353GHz}{4.43}
\setsymbol{HFI:FWHM:Mars:545GHz}{3.80}
\setsymbol{HFI:FWHM:Mars:857GHz}{3.67}

\setsymbol{HFI:beam:ellipticity:Mars:100GHz}{1.18}
\setsymbol{HFI:beam:ellipticity:Mars:143GHz}{1.03}
\setsymbol{HFI:beam:ellipticity:Mars:217GHz}{1.14}
\setsymbol{HFI:beam:ellipticity:Mars:353GHz}{1.09}
\setsymbol{HFI:beam:ellipticity:Mars:545GHz}{1.25}
\setsymbol{HFI:beam:ellipticity:Mars:857GHz}{1.03}

\setsymbol{HFI:CMB:relative:calibration:100GHz}{$\lsim 1\%$}
\setsymbol{HFI:CMB:relative:calibration:143GHz}{$\lsim 1\%$}
\setsymbol{HFI:CMB:relative:calibration:217GHz}{$\lsim 1\%$}
\setsymbol{HFI:CMB:relative:calibration:353GHz}{$\lsim 1\%$}
\setsymbol{HFI:CMB:relative:calibration:545GHz}{}
\setsymbol{HFI:CMB:relative:calibration:857GHz}{}

\setsymbol{HFI:CMB:absolute:calibration:100GHz}{$\lsim 2\%$}
\setsymbol{HFI:CMB:absolute:calibration:143GHz}{$\lsim 2\%$}
\setsymbol{HFI:CMB:absolute:calibration:217GHz}{$\lsim 2\%$}
\setsymbol{HFI:CMB:absolute:calibration:353GHz}{$\lsim 2\%$}
\setsymbol{HFI:CMB:absolute:calibration:545GHz}{}
\setsymbol{HFI:CMB:absolute:calibration:857GHz}{}

\setsymbol{HFI:FIRAS:gain:calibration:accuracy:statistical:100GHz}{}
\setsymbol{HFI:FIRAS:gain:calibration:accuracy:statistical:143GHz}{}
\setsymbol{HFI:FIRAS:gain:calibration:accuracy:statistical:217GHz}{}
\setsymbol{HFI:FIRAS:gain:calibration:accuracy:statistical:353GHz}{2.5\%}
\setsymbol{HFI:FIRAS:gain:calibration:accuracy:statistical:545GHz}{1\%}
\setsymbol{HFI:FIRAS:gain:calibration:accuracy:statistical:857GHz}{0.5\%}

\setsymbol{HFI:FIRAS:gain:calibration:accuracy:systematic:100GHz}{}
\setsymbol{HFI:FIRAS:gain:calibration:accuracy:systematic:143GHz}{}
\setsymbol{HFI:FIRAS:gain:calibration:accuracy:systematic:217GHz}{}
\setsymbol{HFI:FIRAS:gain:calibration:accuracy:systematic:353GHz}{}
\setsymbol{HFI:FIRAS:gain:calibration:accuracy:systematic:545GHz}{7\%}
\setsymbol{HFI:FIRAS:gain:calibration:accuracy:systematic:857GHz}{7\%}

\setsymbol{HFI:FIRAS:zero:point:accuracy:100GHz:units}{0.8\MJysr}
\setsymbol{HFI:FIRAS:zero:point:accuracy:143GHz:units}{}
\setsymbol{HFI:FIRAS:zero:point:accuracy:217GHz:units}{}
\setsymbol{HFI:FIRAS:zero:point:accuracy:353GHz:units}{1.4\MJysr}
\setsymbol{HFI:FIRAS:zero:point:accuracy:545GHz:units}{2.2\MJysr}
\setsymbol{HFI:FIRAS:zero:point:accuracy:857GHz:units}{1.7\MJysr}

\setsymbol{HFI:FIRAS:zero:point:accuracy:100GHz}{0.8}
\setsymbol{HFI:FIRAS:zero:point:accuracy:143GHz}{}
\setsymbol{HFI:FIRAS:zero:point:accuracy:217GHz}{}
\setsymbol{HFI:FIRAS:zero:point:accuracy:353GHz}{1.4}
\setsymbol{HFI:FIRAS:zero:point:accuracy:545GHz}{2.2}
\setsymbol{HFI:FIRAS:zero:point:accuracy:857GHz}{1.7}

\setsymbol{HFI:unit:conversion:100GHz:units}{0.2415\MJysrmK}
\setsymbol{HFI:unit:conversion:143GHz:units}{0.3694\MJysrmK}
\setsymbol{HFI:unit:conversion:217GHz:units}{0.4811\MJysrmK}
\setsymbol{HFI:unit:conversion:353GHz:units}{0.2883\MJysrmK}
\setsymbol{HFI:unit:conversion:545GHz:units}{0.05826\MJysrmK}
\setsymbol{HFI:unit:conversion:857GHz:units}{0.002238\MJysrmK}

\setsymbol{HFI:unit:conversion:100GHz}{0.2415}
\setsymbol{HFI:unit:conversion:143GHz}{0.3694}
\setsymbol{HFI:unit:conversion:217GHz}{0.4811}
\setsymbol{HFI:unit:conversion:353GHz}{0.2883}
\setsymbol{HFI:unit:conversion:545GHz}{0.05826}
\setsymbol{HFI:unit:conversion:857GHz}{0.002238}

\setsymbol{HFI:colour:correction:alpha=-2:V1.01:100GHz}{0.9893}
\setsymbol{HFI:colour:correction:alpha=-2:V1.01:143GHz}{0.9759}
\setsymbol{HFI:colour:correction:alpha=-2:V1.01:217GHz}{1.0007}
\setsymbol{HFI:colour:correction:alpha=-2:V1.01:353GHz}{1.0028}
\setsymbol{HFI:colour:correction:alpha=-2:V1.01:545GHz}{1.0019}
\setsymbol{HFI:colour:correction:alpha=-2:V1.01:857GHz}{0.9889}

\setsymbol{HFI:colour:correction:alpha=0:V1.01:100GHz}{1.0008}
\setsymbol{HFI:colour:correction:alpha=0:V1.01:143GHz}{1.0148}
\setsymbol{HFI:colour:correction:alpha=0:V1.01:217GHz}{0.9909}
\setsymbol{HFI:colour:correction:alpha=0:V1.01:353GHz}{0.9888}
\setsymbol{HFI:colour:correction:alpha=0:V1.01:545GHz}{0.9878}
\setsymbol{HFI:colour:correction:alpha=0:V1.01:857GHz}{1.0014}

\usepackage{graphicx}
\usepackage{txfonts}
\usepackage[breaklinks, colorlinks, citecolor=blue]{hyperref}
\usepackage{url}
\usepackage{natbib}
\usepackage{color}
\usepackage{amssymb}
\usepackage{booktabs}
\usepackage{subfig}
\usepackage[table]{xcolor}
\definecolor{tableShade}{HTML}{F1F5FA}   
\definecolor{tableShade2}{HTML}{ECF3FE}  
\definecolor{light-grey}{gray}{0.90}

\newfont{\gwpfont}{cmssq8 scaled 1000}
\newcommand{\rexcess}{{\gwpfont REXCESS}}

\def\msol{{M$_{\odot}$}}

\def\xmm{{\it XMM-Newton}}
\def\planck{{\it Planck}}

\def\M500{M_{500}}
\def\R500{R_{500}}
\def\Mgv{M_{\rm g,500}}

\def\YX {Y_{\rm X, 500}}
\def\TX {T_{\rm X}}

\def\Mv {M_{\rm 500}}
\def \Rv {R_{500}}
\def\keV {\rm keV}
\def\Yv {Y_{500}}
\def\LX {L_{\rm X, 500}}

\def\msol {{\rm M_{\odot}}}

\def\lesssim{\mathrel{\hbox{\rlap{\hbox{\lower4pt\hbox{$\sim$}}}\hbox{$<$}}}}
\def\gtrsim{\mathrel{\hbox{\rlap{\hbox{\lower4pt\hbox{$\sim$}}}\hbox{$>$}}}}

\newcommand{\propsim}{\lower 3pt \hbox{$\, \buildrel {\textstyle
       \propto}\over {\textstyle \sim}\,$}}

\begin{document}
\author{\small
Planck Collaboration:
P.~A.~R.~Ade\inst{73}
\and
N.~Aghanim\inst{48}
\and
M.~Arnaud\inst{59}
\and
M.~Ashdown\inst{57, 4}
\and
J.~Aumont\inst{48}
\and
C.~Baccigalupi\inst{71}
\and
A.~Balbi\inst{31}
\and
A.~J.~Banday\inst{78, 7, 64}
\and
R.~B.~Barreiro\inst{54}
\and
M.~Bartelmann\inst{77, 64}
% ...
}
\institute{\small
Aalto University Mets\"{a}hovi Radio Observatory, Mets\"{a}hovintie 114, FIN-02540 Kylm\"{a}l\"{a}, Finland\\
\and
Agenzia Spaziale Italiana Science Data Center, c/o ESRIN, via Galileo Galilei, Frascati, Italy\\
\and
Astroparticule et Cosmologie, CNRS (UMR7164), Universit\'{e} Denis Diderot Paris 7, B\^{a}timent Condorcet, 10 rue A. Domon et L\'{e}onie Duquet, Paris, France\\
\and
Astrophysics Group, Cavendish Laboratory, University of Cambridge, J J Thomson Avenue, Cambridge CB3 0HE, U.K.\\
\and
Atacama Large Millimeter/submillimeter Array, ALMA Santiago Central Offices, Alonso de Cordova 3107, Vitacura, Casilla 763 0355, Santiago, Chile\\
\and
CITA, University of Toronto, 60 St. George St., Toronto, ON M5S 3H8, Canada\\
\and
CNRS, IRAP, 9 Av. colonel Roche, BP 44346, F-31028 Toulouse cedex 4, France\\
\and
California Institute of Technology, Pasadena, California, U.S.A.\\
\and
Centre of Mathematics for Applications, University of Oslo, Blindern, Oslo, Norway\\
\and
Centro de Astrof\'{\i}sica, Universidade do Porto, Rua das Estrelas, 4150-762 Porto, Portugal\\
\and
DAMTP, University of Cambridge, Centre for Mathematical Sciences, Wilberforce Road, Cambridge CB3 0WA, U.K.\\
\and
DSM/Irfu/SPP, CEA-Saclay, F-91191 Gif-sur-Yvette Cedex, France\\
\and
DTU Space, National Space Institute, Juliane Mariesvej 30, Copenhagen, Denmark\\
\and
Departamento de F\'{\i}sica, Universidad de Oviedo, Avda. Calvo Sotelo s/n, Oviedo, Spain\\
\and
Department of Astronomy and Astrophysics, University of Toronto, 50 Saint George Street, Toronto, Ontario, Canada\\
\and
Department of Physics \& Astronomy, University of British Columbia, 6224 Agricultural Road, Vancouver, British Columbia, Canada\\
\and
Department of Physics and Astronomy, University of Southern California, Los Angeles, California, U.S.A.\\
\and
Department of Physics and Astronomy, University of Sussex, Brighton BN1 9QH, U.K.\\
\and
Department of Physics, Gustaf H\"{a}llstr\"{o}min katu 2a, University of Helsinki, Helsinki, Finland\\
\and
Department of Physics, Princeton University, Princeton, New Jersey, U.S.A.\\
\and
Department of Physics, Purdue University, 525 Northwestern Avenue, West Lafayette, Indiana, U.S.A.\\
\and
Department of Physics, University of California, Berkeley, California, U.S.A.\\
\and
Department of Physics, University of California, One Shields Avenue, Davis, California, U.S.A.\\
\and
Department of Physics, University of California, Santa Barbara, California, U.S.A.\\
\and
Department of Physics, University of Illinois at Urbana-Champaign, 1110 West Green Street, Urbana, Illinois, U.S.A.\\
\and
Dipartimento di Fisica G. Galilei, Universit\`{a} degli Studi di Padova, via Marzolo 8, 35131 Padova, Italy\\
\and
Dipartimento di Fisica, Universit\`{a} La Sapienza, P. le A. Moro 2, Roma, Italy\\
\and
Dipartimento di Fisica, Universit\`{a} degli Studi di Milano, Via Celoria, 16, Milano, Italy\\
\and
Dipartimento di Fisica, Universit\`{a} degli Studi di Trieste, via A. Valerio 2, Trieste, Italy\\
\and
Dipartimento di Fisica, Universit\`{a} di Ferrara, Via Saragat 1, 44122 Ferrara, Italy\\
\and
Dipartimento di Fisica, Universit\`{a} di Roma Tor Vergata, Via della Ricerca Scientifica, 1, Roma, Italy\\
\and
Discovery Center, Niels Bohr Institute, Blegdamsvej 17, Copenhagen, Denmark\\
\and
Dpto. Astrof\'{i}sica, Universidad de La Laguna (ULL), E-38206 La Laguna, Tenerife, Spain\\
\and
European Southern Observatory, ESO Vitacura, Alonso de Cordova 3107, Vitacura, Casilla 19001, Santiago, Chile\\
\and
European Space Agency, ESAC, Planck Science Office, Camino bajo del Castillo, s/n, Urbanizaci\'{o}n Villafranca del Castillo, Villanueva de la Ca\~{n}ada, Madrid, Spain\\
\and
European Space Agency, ESTEC, Keplerlaan 1, 2201 AZ Noordwijk, The Netherlands\\
\and
Helsinki Institute of Physics, Gustaf H\"{a}llstr\"{o}min katu 2, University of Helsinki, Helsinki, Finland\\
\and
INAF - Osservatorio Astronomico di Padova, Vicolo dell'Osservatorio 5, Padova, Italy\\
\and
INAF - Osservatorio Astronomico di Roma, via di Frascati 33, Monte Porzio Catone, Italy\\
\and
INAF - Osservatorio Astronomico di Trieste, Via G.B. Tiepolo 11, Trieste, Italy\\
\and
INAF/IASF Bologna, Via Gobetti 101, Bologna, Italy\\
\and
INAF/IASF Milano, Via E. Bassini 15, Milano, Italy\\
\and
INRIA, Laboratoire de Recherche en Informatique, Universit\'{e} Paris-Sud 11, B\^{a}timent 490, 91405 Orsay Cedex, France\\
\and
IPAG: Institut de Plan\'{e}tologie et d'Astrophysique de Grenoble, Universit\'{e} Joseph Fourier, Grenoble 1 / CNRS-INSU, UMR 5274, Grenoble, F-38041, France\\
\and
Imperial College London, Astrophysics group, Blackett Laboratory, Prince Consort Road, London, SW7 2AZ, U.K.\\
\and
Infrared Processing and Analysis Center, California Institute of Technology, Pasadena, CA 91125, U.S.A.\\
\and
Institut N\'{e}el, CNRS, Universit\'{e} Joseph Fourier Grenoble I, 25 rue des Martyrs, Grenoble, France\\
\and
Institut d'Astrophysique Spatiale, CNRS (UMR8617) Universit\'{e} Paris-Sud 11, B\^{a}timent 121, Orsay, France\\
\and
Institut d'Astrophysique de Paris, CNRS UMR7095, Universit\'{e} Pierre \& Marie Curie, 98 bis boulevard Arago, Paris, France\\
\and
Institute of Astronomy and Astrophysics, Academia Sinica, Taipei, Taiwan\\
\and
Institute of Astronomy, University of Cambridge, Madingley Road, Cambridge CB3 0HA, U.K.\\
\and
Institute of Theoretical Astrophysics, University of Oslo, Blindern, Oslo, Norway\\
\and
Instituto de Astrof\'{\i}sica de Canarias, C/V\'{\i}a L\'{a}ctea s/n, La Laguna, Tenerife, Spain\\
\and
Instituto de F\'{\i}sica de Cantabria (CSIC-Universidad de Cantabria), Avda. de los Castros s/n, Santander, Spain\\
\and
Jet Propulsion Laboratory, California Institute of Technology, 4800 Oak Grove Drive, Pasadena, California, U.S.A.\\
\and
Jodrell Bank Centre for Astrophysics, Alan Turing Building, School of Physics and Astronomy, The University of Manchester, Oxford Road, Manchester, M13 9PL, U.K.\\
\and
Kavli Institute for Cosmology Cambridge, Madingley Road, Cambridge, CB3 0HA, U.K.\\
\and
LERMA, CNRS, Observatoire de Paris, 61 Avenue de l'Observatoire, Paris, France\\
\and
Laboratoire AIM, IRFU/Service d'Astrophysique - CEA/DSM - CNRS - Universit\'{e} Paris Diderot, B\^{a}t. 709, CEA-Saclay, F-91191 Gif-sur-Yvette Cedex, France\\
\and
Laboratoire Traitement et Communication de l'Information, CNRS (UMR 5141) and T\'{e}l\'{e}com ParisTech, 46 rue Barrault F-75634 Paris Cedex 13, France\\
\and
Laboratoire de Physique Subatomique et de Cosmologie, CNRS/IN2P3, Universit\'{e} Joseph Fourier Grenoble I, Institut National Polytechnique de Grenoble, 53 rue des Martyrs, 38026 Grenoble cedex, France\\
\and
Laboratoire de l'Acc\'{e}l\'{e}rateur Lin\'{e}aire, Universit\'{e} Paris-Sud 11, CNRS/IN2P3, Orsay, France\\
\and
Lawrence Berkeley National Laboratory, Berkeley, California, U.S.A.\\
\and
Max-Planck-Institut f\"{u}r Astrophysik, Karl-Schwarzschild-Str. 1, 85741 Garching, Germany\\
\and
Max-Planck-Institut f\"{u}r Extraterrestrische Physik, Giessenbachstra{\ss}e, 85748 Garching, Germany\\
\and
MilliLab, VTT Technical Research Centre of Finland, Tietotie 3, Espoo, Finland\\
\and
National University of Ireland, Department of Experimental Physics, Maynooth, Co. Kildare, Ireland\\
\and
Niels Bohr Institute, Blegdamsvej 17, Copenhagen, Denmark\\
\and
Observational Cosmology, Mail Stop 367-17, California Institute of Technology, Pasadena, CA, 91125, U.S.A.\\
\and
Optical Science Laboratory, University College London, Gower Street, London, U.K.\\
\and
SISSA, Astrophysics Sector, via Bonomea 265, 34136, Trieste, Italy\\
\and
SUPA, Institute for Astronomy, University of Edinburgh, Royal Observatory, Blackford Hill, Edinburgh EH9 3HJ, U.K.\\
\and
School of Physics and Astronomy, Cardiff University, Queens Buildings, The Parade, Cardiff, CF24 3AA, U.K.\\
\and
Space Research Institute (IKI), Russian Academy of Sciences, Profsoyuznaya Str, 84/32, Moscow, 117997, Russia\\
\and
Space Sciences Laboratory, University of California, Berkeley, California, U.S.A.\\
\and
Stanford University, Dept of Physics, Varian Physics Bldg, 382 Via Pueblo Mall, Stanford, California, U.S.A.\\
\and
Universit\"{a}t Heidelberg, Institut f\"{u}r Theoretische Astrophysik, Albert-\"{U}berle-Str. 2, 69120, Heidelberg, Germany\\
\and
Universit\'{e} de Toulouse, UPS-OMP, IRAP, F-31028 Toulouse cedex 4, France\\
% ...
}

\title{\textit{Planck} Early Results XI: Calibration of the local galaxy cluster Sunyaev-Zeldovich scaling relations}

\authorrunning{Planck Collaboration}
\titlerunning{Local galaxy cluster SZ scaling relations}

\date{Received 7 January 2011; accepted 5 May 2011}

\abstract{
We present precise Sunyaev-Zeldovich (SZ) effect measurements in the direction of 62 nearby galaxy clusters ($z <0.5$) detected at high signal-to-noise in the first \planck\ all-sky data set.  The sample spans approximately a decade in total mass, $2 \times 10^{14}\,{\rm M_{\odot}} < M_{500} < 2\times10^{15}\,{\rm M_{\odot}}$, where $M_{500}$ is the mass corresponding to a total density contrast of 500.  Combining these high quality \planck\ measurements with deep \xmm\ X-ray data, we investigate the relations between $D_{\rm A}^2\, Y_{500}$, the integrated Compton parameter due to the SZ effect, and the X-ray-derived gas mass $M_{\rm g,500}$, temperature $\TX$, luminosity $\LX$, SZ signal analogue $Y_{\rm X,500} = M_{\rm g,500} \times T_{\rm X}$, and total mass $\M500$.  After correction for the effect of selection bias on the scaling relations, we find results that are in excellent agreement with both X-ray predictions and recently-published ground-based data derived from smaller samples. The present data yield an exceptionally robust, high-quality local reference, and illustrate \planck's unique capabilities for all-sky statistical studies of galaxy clusters.}
   \keywords{Cosmology: observations,   Galaxies: cluster: general, Galaxies: clusters: intracluster medium, Cosmic background radiation, X-rays: galaxies: clusters, Planck satellite}

   \maketitle
%

\section{Introduction}
The X-ray emitting gas in galaxy clusters induces inverse Compton scattering of Cosmic Microwave Background (CMB) photons, shifting their frequency distribution towards higher energies.  First discussed in 1972 by \citeauthor{sun72}, the scattering produces a characteristic distortion of the CMB spectrum in the direction of a cluster known as the thermal Sunyaev-Zeldovich (SZ) effect.  It is directly proportional to the Compton parameter $y$, a measure of the thermal electron pressure of the intracluster medium (ICM) gas along the line of sight: $y = (\sigma_{\rm T}/m_{\rm e} c^2) \int{P\, dl}$.  Here $P \propto n_{\rm e}\,T$ is the ICM thermal electron pressure, where $n_e$ is the density and $T$ is the temperature, $\sigma_{\rm T}$ is the Thomson cross section, $m_{\rm e}$ is the electron rest mass, and $c$ is the speed of light.  
The SZ signal integrated over the cluster extent is proportional to the integrated Compton parameter, such that $D_{\rm A}^2\, Y_{\rm SZ} =  (\sigma_{\rm T}/m_{\rm e} c^2) \int{P\, dV}$, where $D_{\rm A}$ is the angular distance to the source.  

Clusters are currently thought to form via the hierarchical gravitational collapse of dark matter haloes, so that their number as a function of mass and redshift is a sensitive indicator of the underlying cosmology.   The ICM is formed when gas falls into the dark matter gravitational potential and is heated to X-ray emitting temperatures by shocks and compression.   The scale-free nature of this process implies that simple power law relationships exist between the total halo mass and various other physical properties \citep[e.g.,][]{ber85,kai86} such as X-ray temperature $T$ or luminosity $L$ \citep[e.g.,][]{voi05,arn05,arn07,pra09,vik09}.  As the total mass is not directly observable, such mass proxies are needed to leverage the statistical power of various large-scale surveys for cosmological applications.  Since the gas pressure is directly related to the depth of the gravitational potential, the quantity $D_{\rm A}^2\,Y_{\rm SZ}$ is expected to scale particularly closely with the total mass, a claim supported by recent numerical simulations \citep[e.g.,][]{whi02,das04,mot05,nag06,wik08,agh09}.  SZ surveys for galaxy clusters thus have great potential to produce competitive cosmological constraints.

In a few short years, SZ observations have progressed from the first spatially resolved observations of individual objects \citep{poi99,poi01,kom99,kom01}, to the first discoveries of new objects \citep{sta10}, to large-scale survey projects for cosmology such as the {\it Atacama Cosmology Telescope} \citep[][{\it ACT}]{kos03} and the {\it South Pole Telescope} \citep[][{\it SPT}]{car11}.  Indeed, first cosmological results from these surveys have started appearing \citep{van10,seh11}.   Attention is now focussing on the shape and normalisation of the pressure profile \citep[e.g.,][]{nag07,arn10,kom11}, calibration of the relationship between $D_{\rm A}^2\, Y_{\rm SZ}$ and the total mass for cosmological applications \citep[e.g.,][]{bon08,mar09,arn10,mel11}, comparison of the measured SZ signal to X-ray predictions \citep[][]{lie06,bie07,afs07,kom11,mel11}, and the relationship between $D_{\rm A}^2\, Y_{\rm SZ}$ and its X-ray analogue $\YX$ \citep[e.g.,][]{and10}. First introduced by \citet{kra06}, the latter quantity is defined as the product of $M_{\rm g}$, the gas mass, and $\TX$, the spectroscopic temperature excluding the core regions.   As the link between $\YX$ and $D_{\rm A}^2\, Y_{\rm SZ}$ depends on the relationship between the gas mass weighted and X-ray spectroscopic temperatures, it is a sensitive probe of cluster astrophysics.

In the following, we use a subsample of \planck\footnote{\planck\ (\url{http://www.esa.int/Planck}) is a project of the European Space Agency (ESA) with instruments provided by two scientific consortia funded by ESA member 
states (in particular the lead countries France and Italy), with contributions from NASA (USA) and telescope reflectors provided by a collaboration between ESA and a scientific consortium led and funded by Denmark.} Early Release Compact Source Catalogue SZ (ESZ) clusters, consisting of high signal-to-noise ratio \planck\ detections with deep \xmm\ archive observations, to investigate the local ($z \lesssim 0.5$) SZ scaling relations.  Given its all-sky coverage and high sensitivity, \planck\ is uniquely suited to this task, allowing high signal-to-noise ratio detection of many hot, massive systems that do not appear in other SZ surveys due simply to their limited area; correspondingly, the large field of view and collecting power of \xmm\ make it the ideal instrument to observe these objects in X-rays out to a significant fraction of the virial radius.  Here we investigate the relationship between SZ quantities and X-ray quantities, making full use of the exceptional quality of both data sets. Two complementary companion papers (\citealt{planck2011-5.2a} and \citealt{planck2011-5.2c}) harness the statistical power of the \planck\ survey by analysing the SZ flux--X-ray luminosity and SZ flux--optical richness relations, respectively, using a bin-averaging approach. Two further companion papers present the parent catalogue \citep{planck2011-5.1a} and \xmm\ validation observations of newly-discovered clusters \citep{planck2011-5.1b}.  

In this paper we adopt a $\Lambda$CDM cosmology with $H_0=70$~km~s$^{-1}$~Mpc$^{-1}$, $\Omega_{\rm M}=0.3$ and $\Omega_\Lambda=0.7$.  The factor $E(z) = \sqrt{\Omega_{\rm M} (1+z)^3+\Omega_\Lambda}$ is the ratio of the Hubble constant at redshift $z$ to its present day value.  The variables $M_{500}$ and $R_{500}$ are the total mass and radius corresponding to a total density contrast $500 \, \rho_{\rm c}(z)$, where $\rho_{\rm c}(z)$ is the critical 
density of the Universe at the cluster redshift; thus $M_{500} = (4\pi/3)\,500\,\rho_{\rm c}(z)\,R_{500}^3$.  The quantity $\YX$ is defined as the product of  $\Mgv$, the gas mass within $\Rv$, and $\TX$, the spectroscopic temperature measured in the $[0.15$--$0.75]~\Rv$ aperture.  The SZ signal is denoted $Y_{500}$ throughout.  This quantity is defined by $D_{\rm A}^2\, Y_{500}  \equiv (\sigma_{\rm T}/m_{\rm e} c^2)\int P dV$.  Here $D_{\rm A}$ is the angular distance to the system, $\sigma_{\rm T}$ is the Thomson cross-section, $c$ the speed of light, $m_{\rm e}$ the electron rest mass, $P\propto n_{\rm e} T$ is the pressure (the product of the electron number density and temperature), and the integration is performed over a sphere of radius  $R_{500}$.  The quantity 
$D_{\rm A}^2\,Y_{500}$ is the spherically integrated Compton parameter, and $Y_{500}$ is proportional to the flux of the SZ signal within $R_{500}$. 

\section{The ESZ catalogue and the \planck-\xmm\ archive subsample}

\subsection{\planck\ and the ESZ Catalogue}

\planck\ \citep{tauber2010a,planck2011-1.1} is the third generation space mission to measure the anisotropy of the cosmic microwave background (CMB).  It observes the sky in nine frequency bands covering 30--857\,GHz with high sensitivity and angular resolution from 31\arcmin\ to 5\arcmin.  The Low Frequency Instrument (LFI; \citealt{Mandolesi2010,Bersanelli2010,planck2011-1.4} covers the 30, 44, and 70\,GHz bands with amplifiers cooled to 20\,\hbox{K}.  The High Frequency Instrument (HFI; \citealt{Lamarre2010, planck2011-1.5}) covers the 100, 143, 217, 353, 545, and 857\,GHz bands with bolometers cooled to 0.1\,\hbox{K}.   Polarization is measured in all but the highest two bands \citep{Leahy2010, Rosset2010}.  A combination of radiative cooling and three mechanical coolers produces the temperatures needed for the detectors and optics \citep{planck2011-1.3}.  Two Data Processing Centres (DPCs) check and calibrate the data and make maps of the sky \citep{planck2011-1.7, planck2011-1.6}.  \planck's sensitivity, angular resolution, and frequency coverage make it a powerful instrument for galactic and extragalactic astrophysics as well as cosmology.  Early astrophysics results are given in Planck Collaboration (2011e-u) \allearlypapers.

The basic data set used in the present paper is the \planck\ Early Release Compact Source Catalogue SZ (ESZ) sample, described in detail in \citet{planck2011-5.1a}.  The sample is derived from the highest signal-to-noise ratio detections (S/N $ > 6$) in a blind multi-frequency search in the all-sky maps from observations obtained in the first ten months of the \planck\ mission.

\begin{figure}[]
\begin{centering}
\includegraphics[scale=1.,angle=0,keepaspectratio,width=0.95\columnwidth]{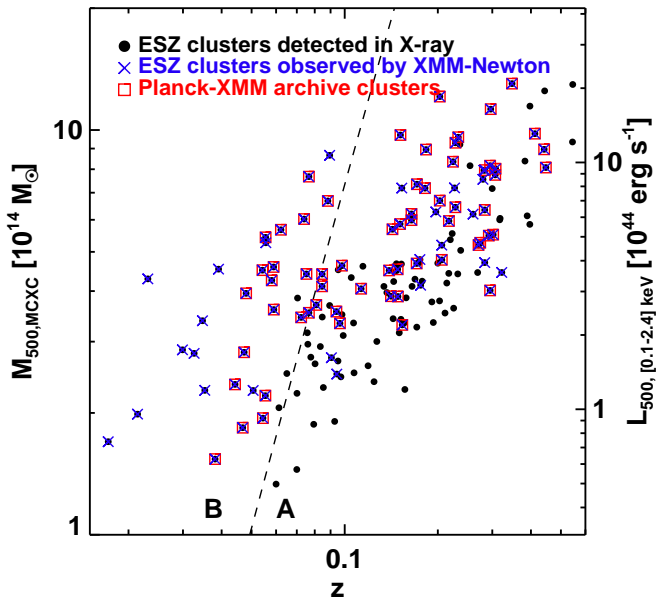}
\end{centering}
\caption{{\footnotesize The 158 \planck\ ESZ clusters already observed in X-rays.  Masses are estimated from their X-ray luminosity as described in \citet{pif10}.  The 62 clusters analysed and discussed in this paper are indicated by the red squares.  The dashed line represents the locus at which $R_{500} \sim 12\arcm$.}}\label{fig:Mz}
\end{figure}


\subsection{The \planck-\xmm\ archive subsample}

Cross-correlation of the ESZ subsample with the Meta Catalogue of X-ray Clusters (MCXC; \citealt{pif10}) produced 158 matches with known X-ray clusters.  As shown in Fig.~\ref{fig:Mz}, these objects lie at a redshift $z \lesssim 0.5$ and cover approximately a decade in mass\footnote{Estimated from the X-ray luminosity-mass relation of \citet{pra09}, as detailed in \citet{pif10}.}. A search for these clusters in the \xmm\ Science Archive\footnote{\url{http://xmm.esac.esa.int/xsa/}} produced a combined \planck-\xmm\ archive sample of 88 objects as of July 2010, indicated by blue crosses in Fig.~\ref{fig:Mz}. 

As detailed below in Sect.~\ref{sec:xproc}, we used different X-ray data processing techniques depending on cluster angular extent.  More specifically, if the source extent lies well within the \xmm\ field of view then the X-ray background can be characterised using a source-free region of the observation, while clusters with a larger angular extent require simultaneous source and background modelling.  We label these classes of clusters as A and B, respectively.  The dashed line in Fig.~\ref{fig:Mz} illustrates the radius at which $R_{500} \lesssim 12\arcmin$, corresponding to the maximum angular extent within which the X-ray background can be characterised in a single \xmm\ field of view.  Using this criterion, we divide the \planck-\xmm\ archive sample into 58 A clusters and 30 B clusters.  

Not all of the clusters in the full \planck-\xmm\ archive sample are used in the present paper.  Some observations in the A cluster list were excluded because soft proton solar flare contamination had rendered the observations unusable, or because the object had not yet been observed at the time of the archive search, or because the target was a clear multiple system unsuited to a spherically-symmetric analysis.  For the B clusters, in addition to the high-luminosity systems already published in \citet{bou08}, we prioritised those where the \xmm\ field of view was expected to cover the largest possible fraction of $R_{500}$, corresponding to objects with the lowest estimated mass in Fig.~\ref{fig:Mz}.  The final sample of 62 systems consists of 44 A objects and 18 B objects.  While the sample is neither representative nor complete, it represents the largest, highest-quality SZ-X-ray data set currently-available.


\begin{figure*}[]
\begin{centering}
\includegraphics[scale=1.,angle=0,keepaspectratio,width=0.975\columnwidth]{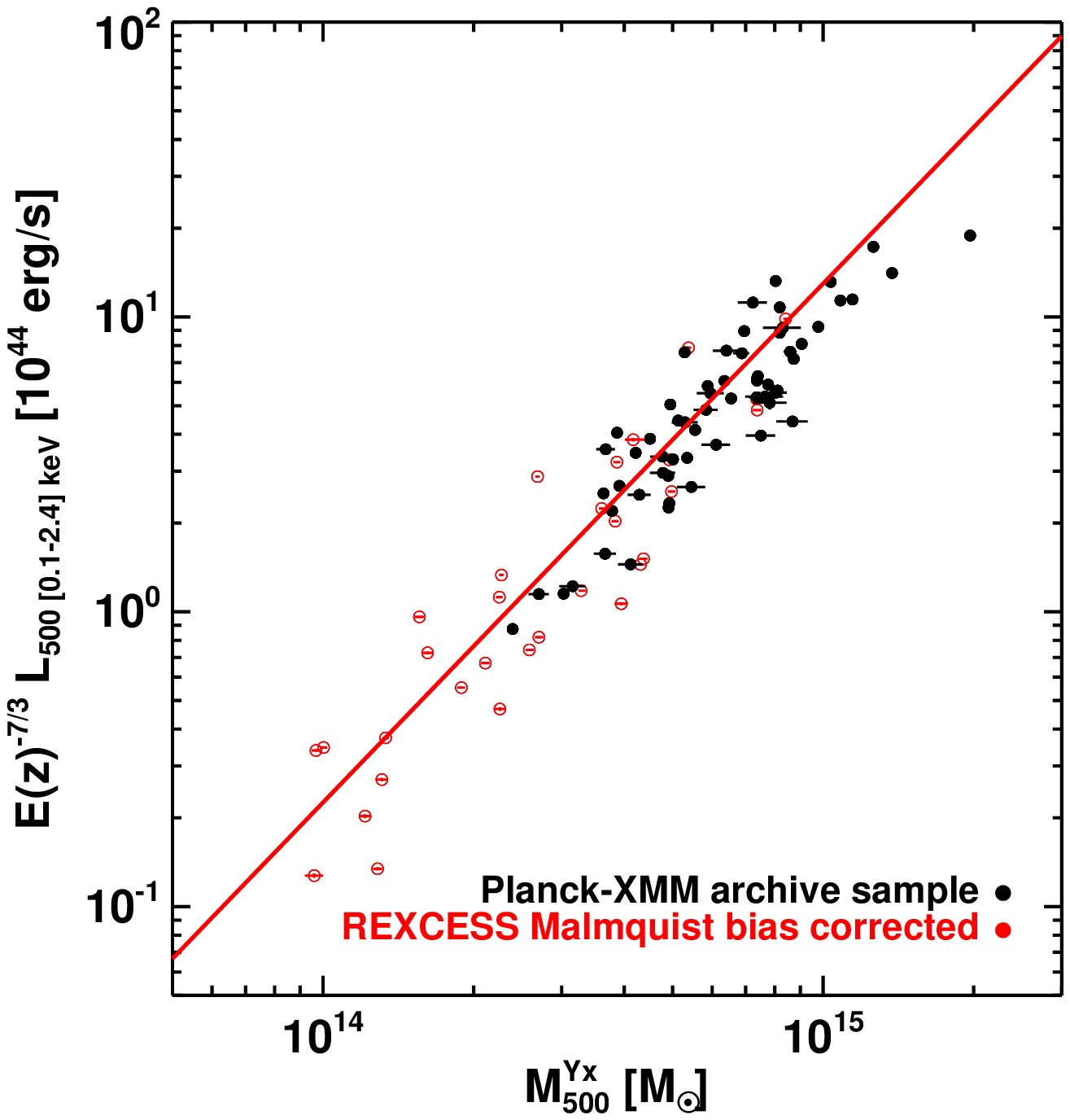}
\hfill
\includegraphics[angle=0,keepaspectratio,width=0.975\columnwidth]{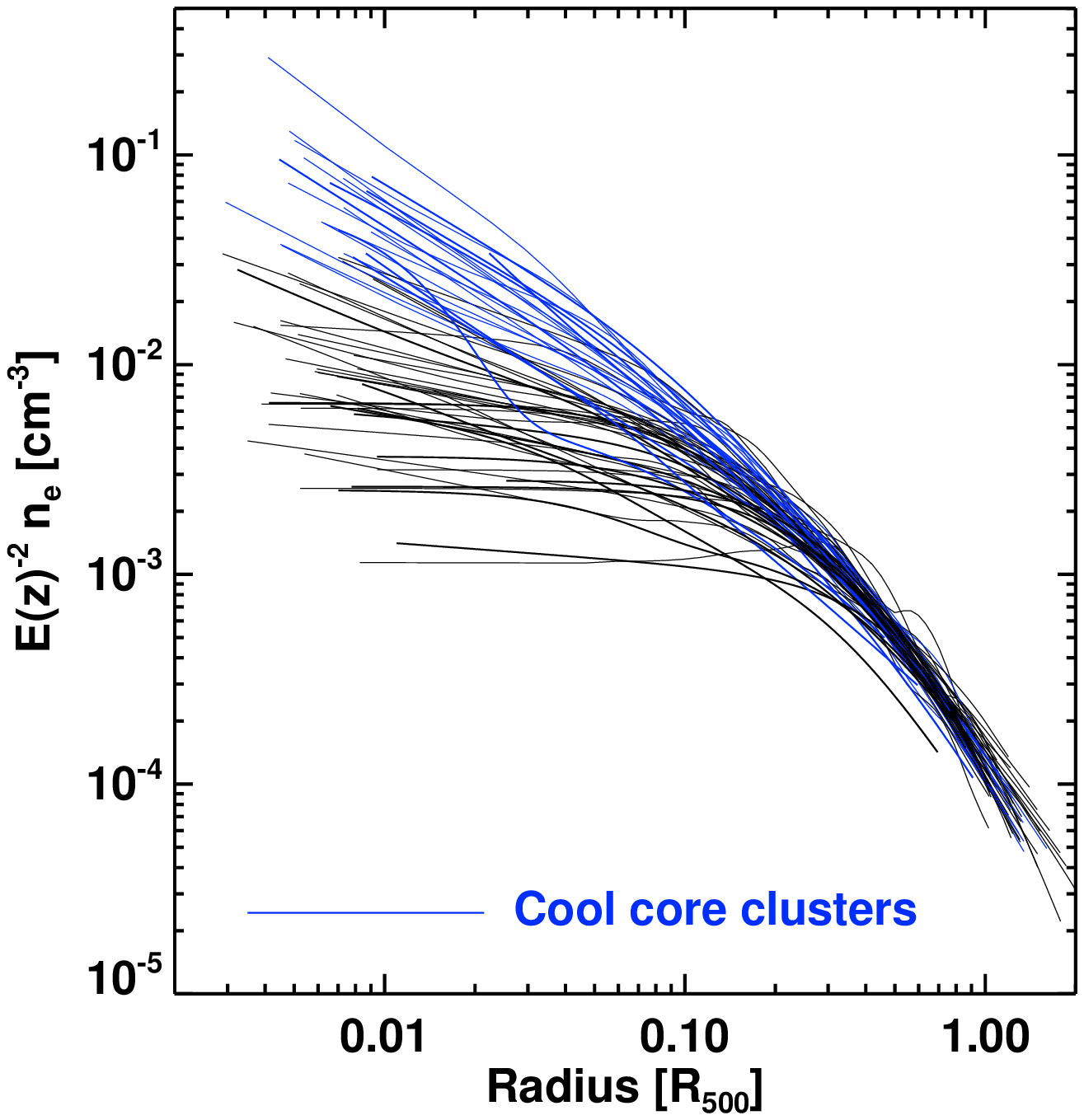}
\end{centering}
\caption{{\footnotesize {\it Left:} $L_{\rm X,500}-M_{500}$ relation of the \planck-\xmm\ archive sample compared to \rexcess, a representative X-ray cluster sample.  Luminosity is estimated interior to $R_{500}$ in the [0.1--2.4] keV band, and mass from the $M_{500}-Y_{\rm X,500}$ relation of \citet{arn10}.  The solid red line is the fit to the \rexcess\ sample only \citep{pra09}.  {\it Right:} Scaled density profiles of the 62 systems in the \planck-\xmm\ archive sample.  Profiles have been corrected for projection and PSF effects as described in the text.  Systems classified as cool cores are indicated by blue lines.}}\label{fig:neprofs}
\end{figure*}

\section{X-ray cluster properties\label{sec:xproc}}

\subsection{X-ray data processing: A clusters\label{sec:xprocp0}}

For the A clusters, we use the latest version (v10.0) of the \xmm-SAS, ensuring that the most recent calibration corrections are applied to the X-ray photons.  Event lists are processed, cleaned for periods of high background, {\sc pattern}-selected and corrected for vignetting as detailed in \citet{pra07}.

The quiescent \xmm\ background is dominated by events due to charged particles.  We subtract this component by using a background file built from stacked observations obtained with the filter wheel in the CLOSED position, recast to the source position and renormalised using the count rate in the high energy band free of cluster emission\footnote{We excluded a 5\arcmin\ region around the cluster centre to avoid contamination from residual cluster emission.}.  The remaining components are due to the cosmic X-ray background produced by unresolved sources and a diffuse soft X-ray contribution attributable to the Galaxy.  The A sample is selected so that $R_{500} \lesssim 12\arcmin$, allowing us to model these remaining components using emission from an annular region external to the cluster emission as detailed in \citet[][]{cro08} and \citet{pra10}.   

Point sources were identified from the small scales of wavelet-decomposed images in the [0.3--2] and [2--5] keV bands.  These sources were excluded, with the exclusion radius matched to the variation of the PSF size across the detector.  We also masked any  well-defined substructures that produced prominent secondary maxima and were visible in the larger scales of the wavelet decomposition process.  

Surface brightness profiles were extracted from the data in $3\farcs3$ bins centred on the X-ray peak.  Finally, a non-parametric regularisation method was used to derive the deprojected, PSF-corrected density profiles, $n_{\rm e}(r)$, as described in \citet{cro08}.


\subsection{X-ray data processing: B clusters\label{sec:xprocp1}}

For each object in the B cluster sample, a merged energy-position photon cube was built from the various observations of a given target.  The cube was built from soft proton-cleaned events from each camera, generated with v10.0 of the \xmm-SAS, to which an effective exposure and a background noise array were added.  The exposure array was computed from the effective exposure time, with corrections for spatially variable mirror effective areas, filter transmissions, CCD pixel area, chip gaps and bad pixels, as appropriate.  The background noise array was modelled as the sum of components accounting for the Galactic foreground and cosmic X-ray background, plus charged particle-induced and out-of-time events.  Full details of the method are given in \citet{bou08}.  

The Galactic foreground is a critical model component in the case of the B clusters.  These objects often extend over the full \xmm\ field of view, so that the cluster emission cannot be spatially separated from the foreground components.  We thus constrained the foreground components using a joint fit of cluster emissivity and temperature in an external annulus corresponding to $\sim R_{500}$.  Despite the degeneracy of this estimate with the cluster emissivity itself, in all cases the temperature obtained in this annulus was found to be lower than the average cluster temperature, as is commonly observed in clusters allowing full cluster-foreground spatial separation \citep[e.g.,][]{pra07,lec08}.

The ICM density profiles in the B cluster sample were then derived using the analytic distributions of ICM density and temperature introduced by \citet{vik06}.  These parametric distributions were projected along the line of sight, convolved with the \xmm\ PSF, and fitted to the observed projected cluster brightness and temperature profiles.  The resulting density profiles, $n_{\rm e}(r)$, were used to derive X-ray quantities for each cluster as described below.


\begin{table*}[!t]
\caption{{\footnotesize X-ray and SZ properties. }\label{tab:xray}} 
\newcolumntype{L}{>{\columncolor{white}[0pt][\tabcolsep]}l}
\newcolumntype{R}{>{\columncolor{white}[\tabcolsep][0pt]}l}
\rowcolors{3}{light-grey}{white}
\resizebox{\textwidth}{!} {
\begin{tabular}{@{}LrrrrrrrrrrrR@{}}
\toprule
\toprule
\multicolumn{1}{c}{Name} & \multicolumn{1}{c}{RA} & 
\multicolumn{1}{c}{Dec} & \multicolumn{1}{c}{$z$} &
\multicolumn{1}{c}{$R_{500}$} & \multicolumn{1}{c}{$\TX$} &
\multicolumn{1}{c}{$\Mgv$} & \multicolumn{1}{c}{$Y_{X,500}$} &
\multicolumn{1}{c}{$D_{\rm A}^2\, Y_{500}$} & \multicolumn{1}{c}{$M_{500}$} &
\multicolumn{1}{c}{$L_{\rm X,500}$} & \multicolumn{1}{c}{CC} \\

\noalign{\smallskip}
\multicolumn{1}{c}{} & \multicolumn{1}{c}{[deg]} & 
\multicolumn{1}{c}{[deg]} & \multicolumn{1}{c}{} &
\multicolumn{1}{c}{[kpc]} & \multicolumn{1}{c}{[keV]} &
\multicolumn{1}{c}{[$10^{14}$ M$_{\odot}$]} & \multicolumn{1}{c}{[$10^{14}$ M$_{\odot}$ keV]} &
\multicolumn{1}{c}{[$10^{-4}$ Mpc$^2$]} & \multicolumn{1}{c}{$[10^{14}$ M$_{\odot}$]} &
\multicolumn{1}{c}{[10$^{44}$ erg s$^{-1}$]} & \multicolumn{1}{c}{} \\


\midrule

RXC\,J0014.3-3022 & 3.58 & -30.38 & 0.307 & 1358 & 7.72$\pm$ 0.25 & 1.65$\pm$ 0.01 & 12.73$\pm$ 0.51 & 1.74$\pm$ 0.21 & 9.78$\pm$ 0.21 & 13.35$\pm$ 0.09 & \ldots \\
A85 & 10.44 & -9.37 & 0.052 & 1206 & 5.78$\pm$ 0.22 & 0.66$\pm$ 0.01 & 3.84$\pm$ 0.19 & 0.47$\pm$ 0.05 & 5.30$\pm$ 0.31 & 4.65$\pm$ 0.02 & \checkmark \\
RXC\,J0043.4-2037 & 10.84 & -20.61 & 0.292 & 1152 & 5.82$\pm$ 0.20 & 0.88$\pm$ 0.01 & 5.10$\pm$ 0.20 & 1.40$\pm$ 0.17 & 5.88$\pm$ 0.14 & 8.26$\pm$ 0.08 & \ldots \\
A119 & 14.02 & -1.30 & 0.044 & 1114 & 5.40$\pm$ 0.23 & 0.45$\pm$ 0.01 & 2.45$\pm$ 0.14 & 0.27$\pm$ 0.03 & 4.12$\pm$ 0.23 & 1.52$\pm$ 0.01 & \ldots \\
RXC\,J0232.2-4420 & 38.06 & -44.37 & 0.284 & 1223 & 6.41$\pm$ 0.20 & 1.07$\pm$ 0.01 & 6.86$\pm$ 0.26 & 0.86$\pm$ 0.13 & 6.95$\pm$ 0.15 & 12.53$\pm$ 0.09 & \checkmark \\
A401 & 44.73 & 13.56 & 0.075 & 1355 & 7.26$\pm$ 0.44 & 1.02$\pm$ 0.04 & 7.43$\pm$ 0.58 & 0.83$\pm$ 0.08 & 7.65$\pm$ 0.67 & 5.82$\pm$ 0.04 & \ldots \\
RXC\,J0303.8-7752 & 46.00 & -77.88 & 0.274 & 1251 & 7.88$\pm$ 0.36 & 0.96$\pm$ 0.02 & 7.58$\pm$ 0.45 & 1.09$\pm$ 0.13 & 7.37$\pm$ 0.25 & 7.39$\pm$ 0.07 & \ldots \\
A3112 & 49.51 & -44.26 & 0.070 & 1062 & 5.02$\pm$ 0.15 & 0.40$\pm$ 0.01 & 2.03$\pm$ 0.07 & 0.18$\pm$ 0.03 & 3.67$\pm$ 0.16 & 3.84$\pm$ 0.02 & \checkmark \\
A3158 & 55.72 & -53.60 & 0.060 & 1124 & 5.00$\pm$ 0.18 & 0.53$\pm$ 0.01 & 2.66$\pm$ 0.12 & 0.35$\pm$ 0.03 & 4.29$\pm$ 0.23 & 2.66$\pm$ 0.01 & \ldots \\
A478 & 63.35 & 10.45 & 0.088 & 1326 & 6.43$\pm$ 0.19 & 1.06$\pm$ 0.03 & 6.81$\pm$ 0.26 & 0.92$\pm$ 0.08 & 7.23$\pm$ 0.48 & 12.33$\pm$ 0.05 & \checkmark \\
A3266 & 67.83 & -61.42 & 0.059 & 1354 & 7.46$\pm$ 0.22 & 0.96$\pm$ 0.02 & 7.17$\pm$ 0.30 & 0.90$\pm$ 0.07 & 7.51$\pm$ 0.51 & 4.22$\pm$ 0.01 & \ldots \\
A520 & 73.55 & 2.96 & 0.203 & 1325 & 7.74$\pm$ 0.22 & 1.13$\pm$ 0.01 & 8.75$\pm$ 0.32 & 0.99$\pm$ 0.14 & 8.11$\pm$ 0.16 & 7.11$\pm$ 0.04 & \ldots \\
RXC\,J0516.7-5430 & 79.17 & -54.52 & 0.295 & 1266 & 7.11$\pm$ 0.67 & 1.20$\pm$ 0.06 & 8.50$\pm$ 1.06 & 1.29$\pm$ 0.10 & 7.82$\pm$ 0.60 & 7.27$\pm$ 0.38 & \ldots \\
RXC\,J0528.9-3927 & 82.22 & -39.44 & 0.284 & 1218 & 6.04$\pm$ 0.32 & 1.11$\pm$ 0.02 & 6.73$\pm$ 0.46 & 1.18$\pm$ 0.13 & 6.88$\pm$ 0.25 & 10.55$\pm$ 0.11 & \checkmark \\
RXC\,J0532.9-3701 & 83.23 & -37.02 & 0.275 & 1190 & 6.84$\pm$ 0.26 & 0.85$\pm$ 0.01 & 5.82$\pm$ 0.28 & 0.97$\pm$ 0.13 & 6.35$\pm$ 0.17 & 8.40$\pm$ 0.07 & \checkmark \\
RXC\,J0547.6-3152 & 86.89 & -31.90 & 0.148 & 1150 & 6.10$\pm$ 0.14 & 0.60$\pm$ 0.01 & 3.63$\pm$ 0.10 & 0.45$\pm$ 0.07 & 5.01$\pm$ 0.08 & 3.89$\pm$ 0.02 & \ldots \\
A3376 & 90.47 & -39.99 & 0.045 & 930 & 3.39$\pm$ 0.09 & 0.28$\pm$ 0.01 & 0.94$\pm$ 0.03 & 0.10$\pm$ 0.02 & 2.39$\pm$ 0.06 & 0.92$\pm$ 0.01 & \ldots \\
RXC\,J0605.8-3518 & 91.48 & -35.29 & 0.139 & 1059 & 4.93$\pm$ 0.11 & 0.46$\pm$ 0.01 & 2.29$\pm$ 0.07 & 0.47$\pm$ 0.06 & 3.87$\pm$ 0.06 & 4.74$\pm$ 0.02 & \checkmark \\
RXC\,J0645.4-5413 & 101.39 & -54.21 & 0.164 & 1303 & 7.26$\pm$ 0.18 & 1.01$\pm$ 0.01 & 7.33$\pm$ 0.24 & 1.09$\pm$ 0.07 & 7.40$\pm$ 0.14 & 7.59$\pm$ 0.04 & \checkmark \\
RXC\,J0658.5-5556 & 104.63 & -55.96 & 0.296 & 1527 & 11.19$\pm$ 0.25 & 2.08$\pm$ 0.02 & 23.22$\pm$ 0.64 & 2.66$\pm$ 0.14 & 13.73$\pm$ 0.21 & 20.05$\pm$ 0.10 & \ldots \\
A665 & 127.75 & 65.88 & 0.182 & 1331 & 7.64$\pm$ 0.46 & 1.12$\pm$ 0.03 & 8.55$\pm$ 0.61 & 1.09$\pm$ 0.11 & 8.04$\pm$ 0.37 & 6.81$\pm$ 0.10 & \ldots \\
A754 & 137.24 & -9.65 & 0.054 & 1423 & 8.93$\pm$ 0.24 & 1.04$\pm$ 0.03 & 9.28$\pm$ 0.39 & 0.86$\pm$ 0.05 & 8.69$\pm$ 0.63 & 4.68$\pm$ 0.02 & \ldots \\
A773 & 139.49 & 51.69 & 0.217 & 1228 & 6.78$\pm$ 0.16 & 0.89$\pm$ 0.01 & 6.01$\pm$ 0.18 & 0.86$\pm$ 0.11 & 6.55$\pm$ 0.11 & 6.80$\pm$ 0.04 & \ldots \\
A781 & 140.09 & 30.49 & 0.298 & 1114 & 5.72$\pm$ 0.10 & 0.76$\pm$ 0.01 & 4.32$\pm$ 0.10 & 0.72$\pm$ 0.14 & 5.35$\pm$ 0.07 & 4.75$\pm$ 0.03 & \ldots \\
A868 & 146.36 & -8.64 & 0.153 & 1058 & 4.63$\pm$ 0.16 & 0.51$\pm$ 0.01 & 2.34$\pm$ 0.08 & 0.41$\pm$ 0.07 & 3.91$\pm$ 0.10 & 3.18$\pm$ 0.03 & \ldots \\
A963 & 154.24 & 39.01 & 0.206 & 1123 & 5.49$\pm$ 0.11 & 0.66$\pm$ 0.01 & 3.63$\pm$ 0.09 & 0.41$\pm$ 0.09 & 4.95$\pm$ 0.07 & 6.40$\pm$ 0.03 & \checkmark \\
RXC\,J1131.9-1955 & 173.00 & -19.92 & 0.308 & 1300 & 7.75$\pm$ 0.31 & 1.30$\pm$ 0.02 & 10.11$\pm$ 0.53 & 1.30$\pm$ 0.23 & 8.59$\pm$ 0.26 & 11.01$\pm$ 0.09 & \ldots \\
A1413 & 178.81 & 23.39 & 0.143 & 1242 & 6.57$\pm$ 0.07 & 0.82$\pm$ 0.01 & 5.41$\pm$ 0.05 & 0.74$\pm$ 0.08 & 6.27$\pm$ 0.04 & 6.85$\pm$ 0.02 & \checkmark \\
RXC\,J1206.2-0848 & 181.59 & -8.81 & 0.441 & 1334 & 10.15$\pm$ 0.32 & 1.59$\pm$ 0.02 & 16.13$\pm$ 0.63 & 1.70$\pm$ 0.30 & 10.83$\pm$ 0.24 & 19.65$\pm$ 0.12 & \checkmark \\
ZwCl1215 & 184.41 & 3.65 & 0.077 & 1211 & 6.45$\pm$ 0.27 & 0.63$\pm$ 0.02 & 4.09$\pm$ 0.21 & 0.46$\pm$ 0.07 & 5.45$\pm$ 0.35 & 2.88$\pm$ 0.01 & \ldots \\
A1576 & 189.23 & 63.19 & 0.302 & 1145 & 6.32$\pm$ 0.47 & 0.80$\pm$ 0.03 & 5.05$\pm$ 0.49 & 0.79$\pm$ 0.11 & 5.83$\pm$ 0.32 & 6.94$\pm$ 0.18 & \ldots \\
A3528S & 193.65 & -29.21 & 0.053 & 966 & 4.11$\pm$ 0.21 & 0.28$\pm$ 0.01 & 1.16$\pm$ 0.07 & 0.22$\pm$ 0.03 & 2.70$\pm$ 0.13 & 1.22$\pm$ 0.01 & \checkmark \\
A1644 & 194.30 & -17.40 & 0.047 & 1070 & 4.86$\pm$ 0.20 & 0.41$\pm$ 0.01 & 1.99$\pm$ 0.11 & 0.25$\pm$ 0.04 & 3.66$\pm$ 0.19 & 1.66$\pm$ 0.05 & \checkmark \\
A3532 & 194.39 & -30.41 & 0.056 & 1015 & 4.44$\pm$ 0.30 & 0.34$\pm$ 0.01 & 1.53$\pm$ 0.12 & 0.21$\pm$ 0.04 & 3.16$\pm$ 0.19 & 1.30$\pm$ 0.01 & \ldots \\
A1650 & 194.67 & -1.76 & 0.084 & 1110 & 5.11$\pm$ 0.06 & 0.51$\pm$ 0.01 & 2.61$\pm$ 0.04 & 0.44$\pm$ 0.06 & 4.22$\pm$ 0.03 & 3.79$\pm$ 0.01 & \checkmark \\
A1651 & 194.88 & -4.20 & 0.084 & 1135 & 5.23$\pm$ 0.12 & 0.56$\pm$ 0.01 & 2.94$\pm$ 0.08 & 0.36$\pm$ 0.06 & 4.51$\pm$ 0.07 & 4.23$\pm$ 0.02 & \ldots \\
A1689 & 197.88 & -1.35 & 0.183 & 1339 & 8.17$\pm$ 0.12 & 1.08$\pm$ 0.01 & 8.84$\pm$ 0.15 & 1.37$\pm$ 0.15 & 8.19$\pm$ 0.08 & 13.29$\pm$ 0.03 & \checkmark \\
A3558 & 202.00 & -31.51 & 0.047 & 1170 & 4.78$\pm$ 0.13 & 0.67$\pm$ 0.02 & 3.21$\pm$ 0.13 & 0.42$\pm$ 0.05 & 4.77$\pm$ 0.26 & 3.54$\pm$ 0.01 & \ldots \\
A1763 & 203.80 & 41.00 & 0.223 & 1275 & 6.55$\pm$ 0.17 & 1.14$\pm$ 0.01 & 7.44$\pm$ 0.27 & 1.28$\pm$ 0.12 & 7.37$\pm$ 0.15 & 8.00$\pm$ 0.05 & \ldots \\
A1795 & 207.24 & 26.58 & 0.062 & 1254 & 6.60$\pm$ 0.21 & 0.73$\pm$ 0.02 & 4.79$\pm$ 0.18 & 0.46$\pm$ 0.04 & 5.96$\pm$ 0.37 & 5.90$\pm$ 0.02 & \checkmark \\
A1914 & 216.49 & 37.83 & 0.171 & 1345 & 8.26$\pm$ 0.19 & 1.07$\pm$ 0.01 & 8.80$\pm$ 0.26 & 1.00$\pm$ 0.09 & 8.19$\pm$ 0.13 & 10.73$\pm$ 0.05 & \ldots \\
A2034 & 227.53 & 33.49 & 0.151 & 1330 & 7.01$\pm$ 0.15 & 1.13$\pm$ 0.01 & 7.94$\pm$ 0.23 & 0.74$\pm$ 0.10 & 7.76$\pm$ 0.13 & 6.99$\pm$ 0.04 & \ldots \\
A2029 & 227.73 & 5.75 & 0.078 & 1392 & 7.70$\pm$ 0.41 & 1.12$\pm$ 0.05 & 8.63$\pm$ 0.60 & 0.81$\pm$ 0.07 & 8.30$\pm$ 0.72 & 10.00$\pm$ 0.05 & \checkmark \\
A2065 & 230.61 & 27.70 & 0.072 & 1160 & 5.36$\pm$ 0.20 & 0.60$\pm$ 0.02 & 3.24$\pm$ 0.15 & 0.39$\pm$ 0.05 & 4.78$\pm$ 0.28 & 3.20$\pm$ 0.02 & \checkmark \\
A2163 & 243.95 & -6.13 & 0.203 & 1781 & 13.40$\pm$ 0.45 & 3.17$\pm$ 0.04 & 42.51$\pm$ 1.82 & 4.55$\pm$ 0.21 & 19.68$\pm$ 0.48 & 23.86$\pm$ 0.15 & \ldots \\
A2204 & 248.18 & 5.59 & 0.152 & 1345 & 7.75$\pm$ 0.21 & 1.09$\pm$ 0.02 & 8.45$\pm$ 0.28 & 1.11$\pm$ 0.10 & 8.04$\pm$ 0.15 & 15.73$\pm$ 0.06 & \checkmark \\
A2218 & 248.99 & 66.21 & 0.171 & 1151 & 5.23$\pm$ 0.10 & 0.73$\pm$ 0.01 & 3.82$\pm$ 0.10 & 0.77$\pm$ 0.06 & 5.13$\pm$ 0.08 & 5.41$\pm$ 0.03 & \ldots \\
A2219 & 250.10 & 46.71 & 0.228 & 1473 & 9.37$\pm$ 0.22 & 1.74$\pm$ 0.02 & 16.33$\pm$ 0.47 & 2.34$\pm$ 0.14 & 11.44$\pm$ 0.20 & 14.94$\pm$ 0.10 & \ldots \\
A2256 & 256.13 & 78.63 & 0.058 & 1265 & 6.40$\pm$ 0.25 & 0.78$\pm$ 0.02 & 4.98$\pm$ 0.23 & 0.71$\pm$ 0.04 & 6.11$\pm$ 0.40 & 3.92$\pm$ 0.02 & \ldots \\
A2255 & 258.24 & 64.05 & 0.081 & 1169 & 5.79$\pm$ 0.15 & 0.59$\pm$ 0.01 & 3.42$\pm$ 0.11 & 0.52$\pm$ 0.04 & 4.91$\pm$ 0.09 & 2.47$\pm$ 0.02 & \ldots \\
RXC\,J1720.1+2638 & 260.03 & 26.61 & 0.164 & 1165 & 5.78$\pm$ 0.12 & 0.70$\pm$ 0.01 & 4.02$\pm$ 0.10 & 0.63$\pm$ 0.08 & 5.28$\pm$ 0.08 & 9.14$\pm$ 0.04 & \checkmark \\
A2261 & 260.61 & 32.14 & 0.224 & 1216 & 6.23$\pm$ 0.55 & 0.93$\pm$ 0.04 & 5.79$\pm$ 0.61 & 1.18$\pm$ 0.12 & 6.41$\pm$ 0.41 & 9.97$\pm$ 0.27 & \checkmark \\
A2390 & 328.41 & 17.69 & 0.231 & 1423 & 8.89$\pm$ 0.24 & 1.54$\pm$ 0.02 & 13.68$\pm$ 0.46 & 1.66$\pm$ 0.13 & 10.35$\pm$ 0.20 & 17.20$\pm$ 0.09 & \checkmark \\
A3827 & 330.46 & -59.95 & 0.099 & 1210 & 6.19$\pm$ 0.10 & 0.69$\pm$ 0.01 & 4.28$\pm$ 0.09 & 0.63$\pm$ 0.05 & 5.55$\pm$ 0.07 & 4.62$\pm$ 0.02 & \ldots \\
RXC\,J2217.7-3543 & 334.46 & -35.73 & 0.149 & 1034 & 4.68$\pm$ 0.10 & 0.44$\pm$ 0.01 & 2.05$\pm$ 0.05 & 0.35$\pm$ 0.06 & 3.64$\pm$ 0.06 & 2.98$\pm$ 0.01 & \ldots \\
RXC\,J2218.6-3853 & 334.68 & -38.89 & 0.141 & 1147 & 6.19$\pm$ 0.19 & 0.57$\pm$ 0.01 & 3.51$\pm$ 0.13 & 0.34$\pm$ 0.06 & 4.92$\pm$ 0.11 & 2.74$\pm$ 0.02 & \ldots \\
RXC\,J2228.6+2036 & 337.12 & 20.62 & 0.412 & 1256 & 8.16$\pm$ 0.30 & 1.33$\pm$ 0.02 & 10.86$\pm$ 0.52 & 1.34$\pm$ 0.23 & 8.73$\pm$ 0.24 & 11.96$\pm$ 0.10 & \ldots \\
RXC\,J2234.5-3744 & 338.62 & -37.75 & 0.151 & 1307 & 7.34$\pm$ 0.12 & 0.99$\pm$ 0.01 & 7.24$\pm$ 0.15 & 0.90$\pm$ 0.07 & 7.37$\pm$ 0.09 & 7.21$\pm$ 0.05 & \ldots \\
MACS\,J2243.3-0935 & 340.84 & -9.58 & 0.444 & 1256 & 7.98$\pm$ 0.12 & 1.47$\pm$ 0.01 & 11.75$\pm$ 0.22 & 1.91$\pm$ 0.24 & 9.06$\pm$ 0.10 & 14.05$\pm$ 0.05 & \ldots \\
A3911 & 341.60 & -52.72 & 0.097 & 1066 & 4.52$\pm$ 0.06 & 0.48$\pm$ 0.01 & 2.16$\pm$ 0.04 & 0.38$\pm$ 0.04 & 3.78$\pm$ 0.03 & 2.45$\pm$ 0.01 & \ldots \\
AS1063 & 342.21 & -44.53 & 0.347 & 1456 & 10.73$\pm$ 0.25 & 1.89$\pm$ 0.02 & 20.33$\pm$ 0.58 & 2.21$\pm$ 0.16 & 12.60$\pm$ 0.20 & 26.32$\pm$ 0.13 & \ldots \\
A3921 & 342.49 & -64.42 & 0.094 & 1071 & 5.01$\pm$ 0.07 & 0.44$\pm$ 0.01 & 2.20$\pm$ 0.02 & 0.36$\pm$ 0.03 & 3.82$\pm$ 0.03 & 2.57$\pm$ 0.02 & \ldots \\

\bottomrule
\end{tabular}
}

\tablefoot{The temperature $\TX$ is measured in the $[0.15-0.75]\,\Rv$ region, and the luminosity $L_{\rm X,500}$ is measured interior to $\Rv$ in the $[0.1-2.4]$ keV band. The mass $M_{500}$ is estimated from the $M_{500}-Y_{\rm X,500}$ relation given in Eqn.~\ref{eqn:Yx}. The final column indicates whether the cluster is classified as a cool core system, defined as described in Sect.~\ref{sec:denscc}.}
\end{table*}

\subsection{X-ray quantities\label{sec:xquan}}

For the current generation of high-resolution X-ray telescopes, the effective limiting radius for high-quality nearby observations of the type discussed here is $R_{500}$.  Beyond this radius, the effect of the variable background becomes dominant and the uncertainties begin to become difficult to quantify.  In addition, as shown by \citet{evr96}, $\Rv$ is also the radius within which clusters are relatively relaxed.  We estimate the X-ray quantities for each cluster self-consistently within $R_{500}$ using the $M_{500}-Y_{\rm X,500}$ relation given in \citet[][see also \citealt{pra10}]{arn10}, assuming standard evolution, viz.,

{\small
\begin{equation}
E(z)^{2/5}\M500 = 10^{14.567 \pm 0.010} \left[\frac{\YX}{2\times10^{14}\,{\msol}\,\keV}\right]^{0.561 \pm 0.018}\, \msol.\label{eqn:Yx}
\end{equation}
}

\noindent The radius $R_{500}$ was calculated iteratively as described in \citet{kra06}.  Using Eq.~\ref{eqn:Yx} and the definition of $M_{500}$ and $\YX$, an equation of the form $R^3_{500} = C[ M_{\rm g,500}\,\TX]^\alpha$ must be solved.  Starting from an initial temperature measurement, the equation is solved for $R_{500}$, with the gas mass computed from the density profiles discussed above in Sects.~\ref{sec:xprocp0} and~\ref{sec:xprocp1}.  A new temperature is then estimated within $[0.15-0.75]\, R_{500}$ and the procedure is repeated until convergence. Once converged, we measure the temperature within an aperture corresponding to $R_{500}$ and calculate the $[0.1-2.4]\,{\rm keV}$ and $[0.5-2]\,{\rm keV}$ band luminosities as described in \citet{pra09}. The resulting X-ray quantities are listed in Table~\ref{tab:xray}. 

The left-hand panel of Fig.~\ref{fig:neprofs} shows the $L_{\rm X,500}-M_{500}$ relation of the \planck-\xmm\ archive sample, where $L_{\rm X,500}$ is the X-ray luminosity estimated interior to $R_{500}$ in the [0.1--2.4] keV band\footnote{The standard $[0.1-2.4]$ keV band self-similar evolution factor is  $E(z)^{-2}$. However, the observed scaling relations are steeper than self-similar \citep[see e.g.,][]{pra09}, leading to a dependence of the evolution factor on the quantity to which it is applied. In the present work we use an evolution factor of $E(z)^{-7/3}$, appropriate for bolometric luminosities, for comparison to \rexcess\ (see also \citealt{mau07}).}, and the mass is estimated from the $M_{500}-Y_{\rm X,500}$ relation given in Equation~\ref{eqn:Yx}. The data are compared to the equivalent relation from \rexcess, a sample designed to be representative of the X-ray cluster population \citep{boe07}.  One can see that the \planck-\xmm\ archive clusters are all massive, luminous systems, as expected for objects detected in SZ at high S/N by \planck.  They follow the general trend exhibited by \rexcess\ \citep{pra09}, but extend to higher mass and luminosity.


\subsection{Scaled gas density profiles and cool core subsample}\label{sec:denscc}

The scaled gas density profiles of the full sample of 62 clusters are shown in the right-hand panel of Fig.~\ref{fig:neprofs}, where each profile has been corrected for evolution and scaled to $R_{500}$.  As has been seen in other cluster samples \citep[e.g.,][]{cro08}, there is a large amount of scatter in the central regions, extending out to $\sim 0.15\,R_{500}$, beyond which the profiles rapidly converge.

It is well-known that some clusters exhibit so-called cool cores, central regions of very dense gas where the cooling time is less than the Hubble time \citep[e.g.,][]{jon84}.  Such objects have very high X-ray luminosities and extremely low central entropies that tend to set them apart from the rest of the X-ray cluster population \citep[e.g.,][]{fab94,pra10}.  In addition, the current consensus is that these systems represent a generally more  relaxed subset of the cluster population (although see \citealt{bur08} for a dissenting view).  Following \citet{pra09}, we estimated the central gas density $n_{\rm e,0}$ using a $\beta$ model fit to the gas density profile interior to $0.05\,R_{500}$, and classified objects with $E(z)^{-2}\, n_{\rm e,0} > 4 \times 10^{-2}$ cm$^{-3}$ as cool core systems.  In total, 22/62 clusters in the present sample are classified as such.  These are plotted in blue in Fig.~\ref{fig:neprofs} and in all following plots.


\section{SZ cluster properties}

\subsection{Optimisation of the SZ flux extraction}

The basic SZ signal extraction procedure is described in full in \citet[][]{planck2011-5.1a}.  In brief, this procedure consists of applying multi-frequency matched filters (MMF, \citealt{mel06}), that incorporate prior knowledge of the signal, to the \planck\ maps.  Specifically, the ICM pressure is assumed to follow the universal profile shape derived from the \rexcess\ sample by \citet{arn10}.  The SZ flux is computed by integrating along the line-of-sight and normalising the universal pressure profile.  Each profile is truncated at $5 \times R_{500}$ , effectively giving a measure of the flux within a cylinder of aperture radius $5 \times R_{500}$, and then converted to the value in a sphere of radius $\Rv$ for direct comparison with the X-ray prediction.  This is the fundamental SZ quantity used in the present 
paper\footnote{Note that $\Yv$ is the directly observed `apparent' quantity, while $D_{\rm A}^2\, \Yv$ is the corresponding `absolute' quantity, intrinsic to the cluster.}, and we refer to it throughout as $\Yv$.  

Section~6 of \citet[][]{planck2011-5.1a} shows that the cluster flux derived from blind application of the MMF algorithm is systematically larger than X-ray expectations.  This discrepancy is a result of overestimation of the cluster size $\theta_{500}$ due to the freedom to optimise significance with position and size.  As shown in \citet[][]{planck2011-5.1a}, if the SZ signal is instead extracted from a region centred on the X-ray position with size $\theta_{500}$ estimated from the X-ray luminosity-mass relation, the SZ flux is in better agreement with X-ray expectations.  When additional constraints on the cluster size are available, the SZ flux extraction can be further optimised.  

With the present cluster sample we can make use of the higher-quality estimate of the X-ray size $\theta_{500}$, derived from $R_{500}$, measured using the $M_{500}-Y_{\rm X,500}$ relation as detailed in Sect.~\ref{sec:xquan}.  Appendix~\ref{ap:opt} details the improvement in SZ flux extraction when these higher-quality size estimates are used. For each cluster in the sample, we thus re-ran the SZ flux extraction, calculating $\Yv$ with the X-ray position and size fixed to the refined values derived from the high-quality \xmm\ observation.  
 

\subsection{Robustness tests specific to local sample}

Section~6 of \citet[][]{planck2011-5.1a} details various robustness tests relevant to all \planck\ SZ papers, including investigation of the cluster size--flux degeneracy discussed above, the impact of the assumed pressure profile used for cluster detection, beam-shape effects, colour corrections, contamination by point sources, and discussion of the overall error budget.  For the present sample we undertake two further robustness tests: the first is  related to the impact of radio source contamination; the second examines the impact of the assumed pressure profile shape on the derived $\Yv$.

\begin{figure}[]
\begin{centering}
\includegraphics[scale=1.,angle=0,keepaspectratio,width=\columnwidth]{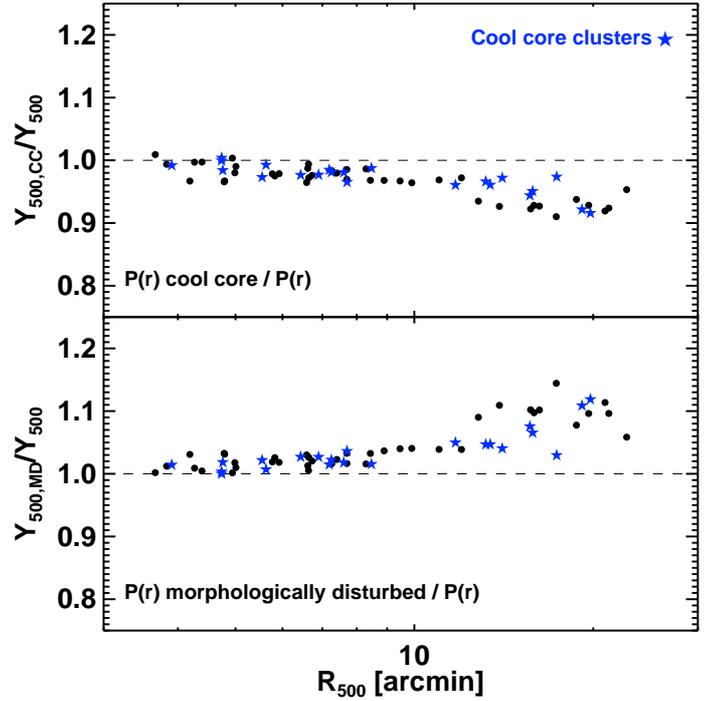}
\end{centering}
\caption{{\footnotesize Comparison of $\Yv$ from extraction using the baseline pressure profile with that from the best fitting \rexcess\ cool core and morphologically disturbed pressure profiles given in Appendix~C of \citet{arn10}.  Cool core systems are marked as blue stars, other systems as black dots.  {\it Top}: universal pressure profile vs cool-core pressure profile; {\it Bottom}: universal pressure profile vs morphologically-disturbed pressure profile.  The trend with $R_{500}$ is due to the inability of the \planck\ beam to resolve different profiles at small angular size.  The effect is small (maximum $\sim 10$ per cent) and quasi-symmetric (as expected), so no bias is introduced.}}\label{fig:extr}
\end{figure}


\subsubsection{Contamination by point sources}\label{sec:rsource}

Contamination by point sources can affect extraction of the SZ parameters, and have implications for astrophysical studies of clusters or further cosmological applications \citep{agh05,dou06}. We have thus checked the possible effect of radio galaxies on the derived $\Yv$ by combining data from SUMSS \citep[][a catalogue of radio sources at 0.85 GHz]{boc99}, NVSS \citep[][a catalogue of radio sources at 1.4 GHz]{con98}, and data from the \planck\ LFI and HFI.  Two clusters in our sample exhibit relatively bright ($S(1.4\ {\rm GHz}) \gtrsim 1$ Jy), flat spectrum radio sources within a radius of $15 \arcmin$ from the X-ray peak.  These sources are clearly seen in LFI data and could potentially affect the SZ measurement.  However, as we discuss below in Sect.~\ref{sec:screl}, inclusion or exclusion of these objects has a negligible effect on the derived scaling relations.


\subsubsection{Impact  of assumptions on pressure profile and scaling}\label{sec:pprof}

The blind SZ signal detection method used to detect and extract the ESZ clusters from the Planck survey \citep{planck2011-5.1a} implements the universal pressure profile from \citet{arn10}.  More specifically, the baseline model makes use of the generalised NFW profile fit to the 31 individual \rexcess\ cluster pressure profiles, after removal of the mass dependence by scaling according to the $M_{500}-Y_{\rm X,500}$ relation given in Eqn.~\ref{eqn:Yx}. However, \citeauthor{arn10} showed that the scatter of the individual cluster pressure profiles about the universal form increases toward the central regions, since cool core systems are more peaked, and morphologically disturbed systems are shallower, respectively, than the mean. In their Appendix~C, \citeauthor{arn10} give the best fitting GNFW model parameters for the average scaled profiles of the \rexcess\ cool core and morphologically disturbed subsamples.

As our cluster sample contains both cool core and morphologically disturbed systems, it is pertinent to investigate the effect of the baseline pressure profile assumption on the resulting $\Yv$ values.  We thus re-ran the $\Yv$ extraction process separately for each object using the cool core and morphologically disturbed cluster profiles given in Appendix~C of \citet{arn10}. The X-ray size $\theta_{500}$ is kept the same in each case so that we are investigating the impact of the pressure profile shape within a fixed aperture.  Figure ~\ref{fig:extr} shows the ratio of the $\Yv$ of the cool core and morphologically disturbed profile extractions to that of the baseline model.

Clear trends are seen in both cases: the ratio tends to increase (decrease) with $\theta_{\rm 500}$ if the morphologically disturbed (cool core) profile is used instead of the baseline universal profile.  Up to $\theta_{500}\sim 10\arcm$ the ratio differs from unity only by 2 per cent on average.  Beyond $\theta_{500}\sim 10\arcm$, the derived $\Yv$ starts to differ gradually from the baseline value.  This effect can be traced to the influence of the \planck\ angular resolution.  Since the SZ signal extraction uses all \planck-HFI channels, the effective angular resolution is that of the channel with the largest ${\rm FWHM}\ (\sim 10$~arcmin at 100~GHz).  Below this angular scale the profile shape is washed out by the convolution with the \planck\ beam, while above it, clusters are increasingly well-resolved.  The two panels of Fig.~\ref{fig:extr} show that at the largest $\theta_{500}$ the maximum excursion is $\sim 10$ per cent.  Beyond 10~arcmin, the average excursions are $\sim 6$ and $\sim7$ per cent, respectively, for cool-core and morphologically disturbed profiles.  Note that the effect is symmetric, in that for large $\theta_{500}$ a cool core profile and a morphologically disturbed profile return a value of $\Yv$ that differs from the baseline value by approximately the same amount, but the former is lower and the latter is higher.

In the following, the difference in $\Yv$ derived from extraction with the cool core and morphologically disturbed cluster profiles is added in quadrature to the uncertainty on the $\Yv$ from the baseline extraction. We expect this conservative error estimate to account for any difference in the underlying pressure profile shape from the universal baseline model. As detailed below in Sect.~\ref{sec:screl}, we have further checked the effect of the pressure profile assumption on the derived scaling relation fits, finding it to be entirely negligible.


\section{SZ scaling relations\label{sec:screl}}

We fitted the parameters governing a scaling relation between $D_{\rm A}^2\, \Yv$, the spherically-integrated SZ signal within $\Rv$, and its X-ray analogue $Y_{\rm X,500}$. We also fitted parameters governing scaling relations between $D_{\rm A}^2\, \Yv$ and various other X-ray-derived quantities including $M_{\rm g,500}$, $\TX$ and $L_{\rm X,500}$.  We further investigated the relation between $D_{\rm A}^2\, \Yv$ and the total mass, $M_{500}$, using the $M_{500}-Y_{\rm X,500}$ calibration given in Eqn.~\ref{eqn:Yx}.


\subsection{Fitting method}
\label{sec:fit}

For each set of observables $(B,A)$, we fitted a power law relation of
the form $E(z)^\gamma \, D_{\rm A}^2\, \Yv = 10^A\, [E(z)^\kappa \, X/X_0]^B$, where
$E(z)$ is the Hubble constant normalised to its present day value and
$\gamma$ and $\kappa $ were fixed to their expected self-similar scalings with
$z$. The fit was undertaken using linear regression in the log-log
plane, taking the uncertainties in both variables into account, and
the scatter was computed as described in \citet{pra09}.  In brief, assuming a relation of the form $Y= aX + b$, and a sample of $N$ data points ($Y_i,X_i$) with errors $\sigma_{Y_i}$ and $\sigma_{X_i}$, the raw scatter was estimated using the error-weighted distances to the regression line:

\begin{equation}
\sigma^2_{\rm raw} = \frac{1}{N-2} \sum_{i=1}^{N} w_i\, (Y_i -aX_i -b)^2\label{eqn:disp1}
\end{equation}

\noindent where

\begin{equation}
w_i = \frac{1/\sigma_i^2}{(1/N) \sum_{i=1}^N 1/\sigma_i^2}\ \ \ \ \ \ \ {\rm and}\ \ \ \ \ \ \sigma_i^2 = \sigma^2_{Y_i} + a^2 \sigma^2_{X_i}.\label{eqn:disp2}
\end{equation}

\noindent  The intrinsic scatter $\sigma_{\rm i}$ was computed from the quadratic difference between the raw scatter and that expected from the statistical uncertainties.  

We use the BCES regression method \citep{akr96}, which takes into account
measurement errors in both coordinates and intrinsic scatter in the
data and is widely used in astronomical regression, giving results
that may easily be compared with other data sets fitted using the same
method. We fitted all relations using orthogonal BCES regression.

\begin{table*}[!t]
\newcolumntype{L}{>{\columncolor{white}[0pt][\tabcolsep]}l}
\newcolumntype{R}{>{\columncolor{white}[\tabcolsep][0pt]}l}
\rowcolors{0}{light-grey}{white}
\caption{{\footnotesize Best fitting orthogonal BCES parameters for scaling relations.} 
\label{tab:screl}}          
\centering  
\begin{tabular}{@{}Lcccccccl}
\toprule
\toprule

\multicolumn{1}{c}{Relation} & \multicolumn{1}{c}{$A_{obs}$} &
\multicolumn{1}{c}{$B_{\rm obs}$} & \multicolumn{1}{c}{$\sigma_{\rm log,i}$} &\multicolumn{1}{c}{$A_{\rm corr}$} & \multicolumn{1}{c}{$B_{\rm corr}$}  &
\multicolumn{1}{c}{$\gamma$} & \multicolumn{1}{c}{$\kappa$} &
\multicolumn{1}{c}{$X_0$}\\

\midrule
$D_{\rm A}^2\, Y_{500}- c^\ast \YX$ & $-4.02\pm 0.01$ & $0.95\pm 0.04$ & $0.10\pm 0.01$ & $-4.02$ &  0.96 & -- & -- &  $1 \times 10^{-4}$~Mpc$^2$ \\

$D_{\rm A}^2\,Y_{500}-M_{500}$ & $-4.19\pm 0.01$ & $1.74\pm 0.08$ &
$0.10\pm 0.01$ & $-4.22$ & 1.74 & $-2/3$ & -- &  $6\times 10^{14}$~M$_\odot$ \\

$D_{\rm A}^2\,Y_{500}-\TX$ & $-4.27\pm 0.02$ & $2.82\pm 0.18$ & $0.14\pm 0.02$ &  $-4.22$ & 2.92 &  $-1$ &  --  & $6$~keV \\

$D_{\rm A}^2\,Y_{500}-\Mgv$ & $-4.05\pm 0.01$ & $1.39\pm 0.06$ & $0.08\pm 0.01$ & $-4.03$ & 1.48 & $-2/3$ & -- &  $1 \times 10^{14}$~M$_\odot$ \\

$D_{\rm A}^2\,Y_{500}-L\,[0.1-2.4]_{\rm X, 500}$ & $-4.00\pm 0.02$ & $1.02\pm 0.07$ & $0.14\pm 0.02$ &  $-3.97$ &  1.12 & $-2/3$ & $-7/3$ & $7\times 10^{44}$~erg s$^{-1}$ \\

$D_{\rm A}^2\,Y_{500}-L\,[0.5-2]_{\rm X, 500}$ & $-3.79\pm 0.03$ & $1.02\pm 0.07$ & $0.14\pm 0.02$ & $-3.75$ & 1.12 & $-2/3$ & $-7/3$ & $7\times 10^{44}$~erg s$^{-1}$ \\

\bottomrule
\end{tabular}
\tablefoot{$c^\ast =(\sigma_T/m_ec^2)/(\mu_em_p)$. \\
Relations are expressed as $E(z)^\gamma\, [D_{\rm A}^2\, \Yv] =10^A\, [E(z)^\kappa\, X/X_{0}]^B$.  The logarithmic intrinsic scatter of the relation is denoted by $\sigma_{\rm log,i}$.}
\end{table*}

\begin{figure*}[]
\begin{centering}
\includegraphics[scale=1.,angle=0,keepaspectratio,width=0.985\columnwidth]{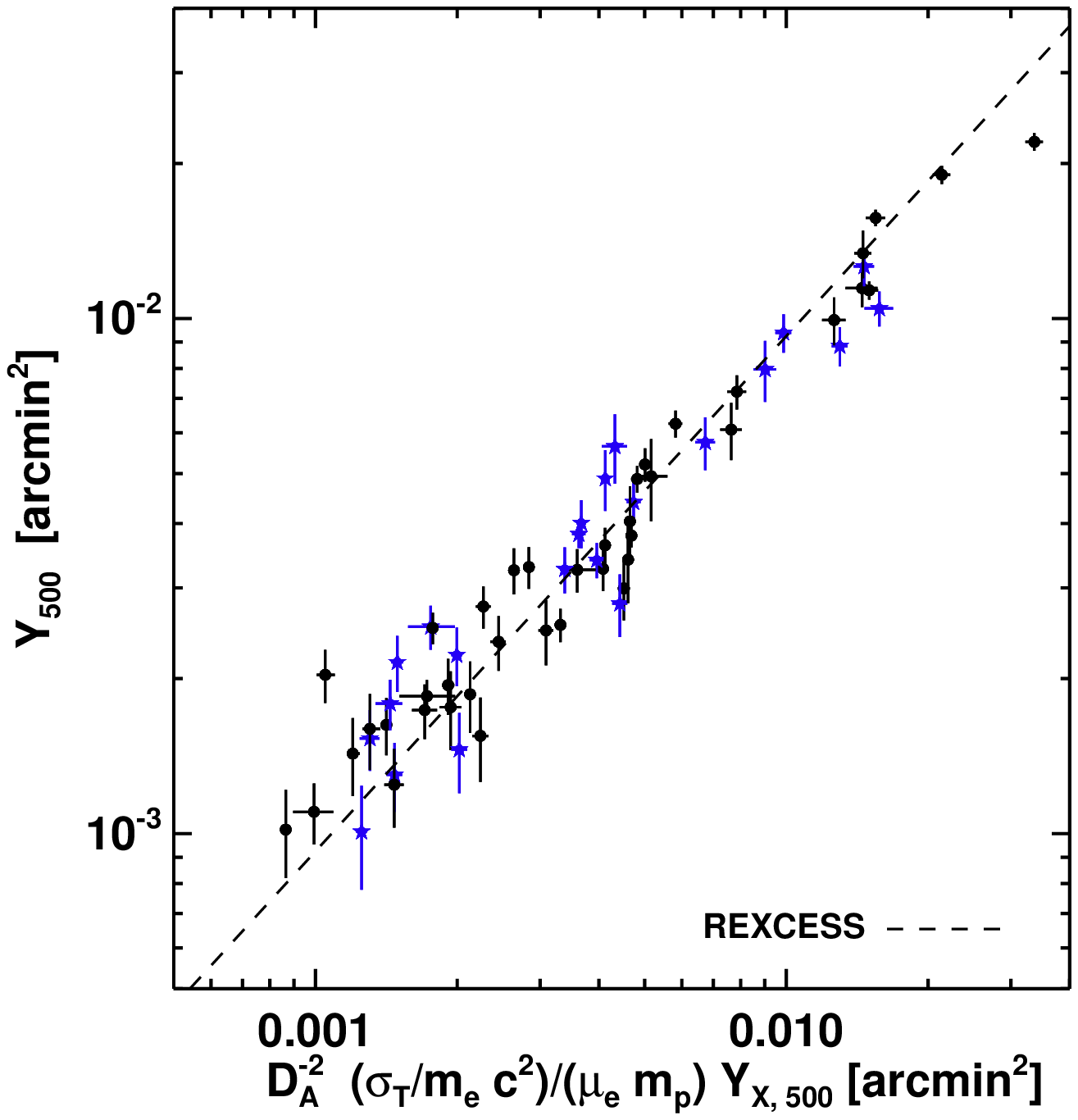}
\hfill
\includegraphics[scale=1.,angle=0,keepaspectratio,width=0.985\columnwidth]{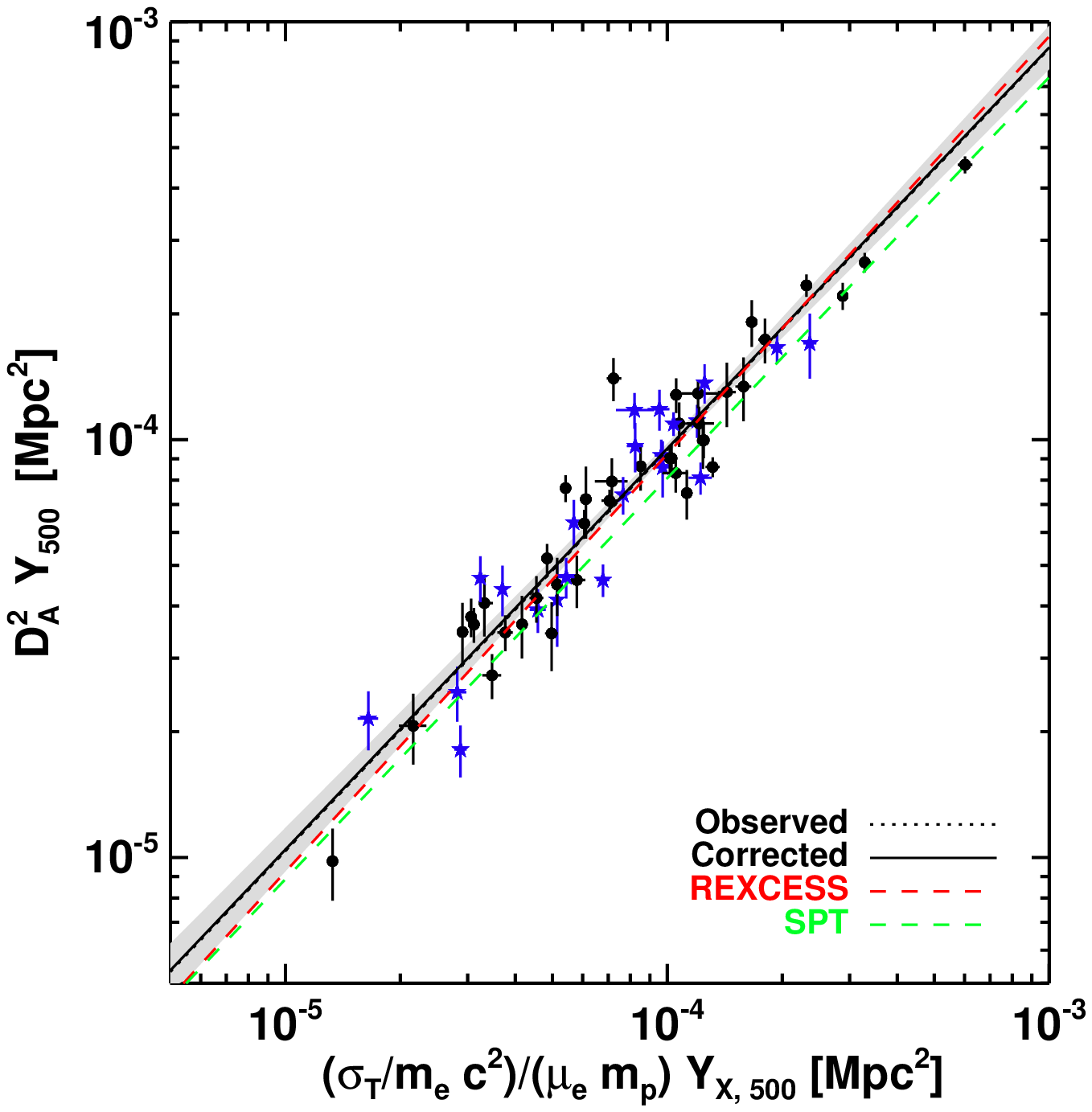}
\end{centering}
\caption{{\footnotesize SZ flux vs X-ray prediction. Blue stars indicate cool core systems. {\it Left panel:} Relation plotted in units of arcmin$^2$. The dashed line is the prediction from \rexcess\ X-ray observations \citep{arn10}. {\it Right panel:} Relation plotted in units of Mpc$^2$. The {\it SPT\/} results are taken from \citet{and10}.}}\label{fig:yszyx_arcmin}
\end{figure*}


\subsection{Effect of point sources and choice of baseline pressure profile}

We fitted the SZ scaling relations excluding the two objects with significant radio source contamination (Sect.~\ref{sec:rsource}).  For all relations the change in normalisation, slope and intrinsic scatter is negligible compared to their associated uncertainties.  We therefore consider the contamination by radio sources to have a negligible effect on scaling relation fits and proceed with the full sample of 62 clusters.  

We have also checked whether the best fitting scaling relations are affected by the choice of baseline pressure profile, as follows.  For the cool core subsample, we assigned the $\Yv$ derived from extraction using the cool core pressure profile.  Of the remaining 40 systems, for the 20 objects with the lowest central density (Fig.~\ref{fig:neprofs}), we assigned the $\Yv$ value derived from extraction using the morphologically disturbed profile. We then re-fitted all the scaling relations. The resulting best fits are in full agreement with those derived from the baseline universal profile; i.e., the difference in best fitting parameters (2 per cent maximum) is again smaller than their respective uncertainties.  


\subsection{Correction for selection bias}

It is well known that scaling-relation determinations of the sort we
are considering can be biased by selection effects of Malmquist and
Eddington type when a significant part of the sample lies near a
selection cut (for discussions in a cluster context see e.g., \citealt{vik09,pra09,man10} and \citealt{and10}). We estimated the effect of the
\planck\ SZ selection as follows. In order to impose a selection cut
on the mock catalogues, we used the {\em observed} relation between
$D_{\rm A}^2\, Y_{500}$ and S/N from the region significantly above the
selection cut and extrapolated below it, along with an estimate of
scatter again from observations, carried out in several redshift
bins. We then constructed large mock catalogues of clusters through
drawing of Poisson samples from a suitably-normalised \citet{jen01}
mass function; to each cluster we assigned a value of $D_{\rm A}^2\, Y_{500}$
by adopting scaling relations with scatter that are consistent with
the observed values.  This procedure leads to a predicted S/N
value that can be used to impose selection cuts on the mock
sample.  We applied it to the full 158 cluster sample as the
only X-ray information needed was the position for SZ signal
re-extraction.

\begin{figure*}
\begin{centering}
\label{fig:YMg}\includegraphics[width=0.975\columnwidth]{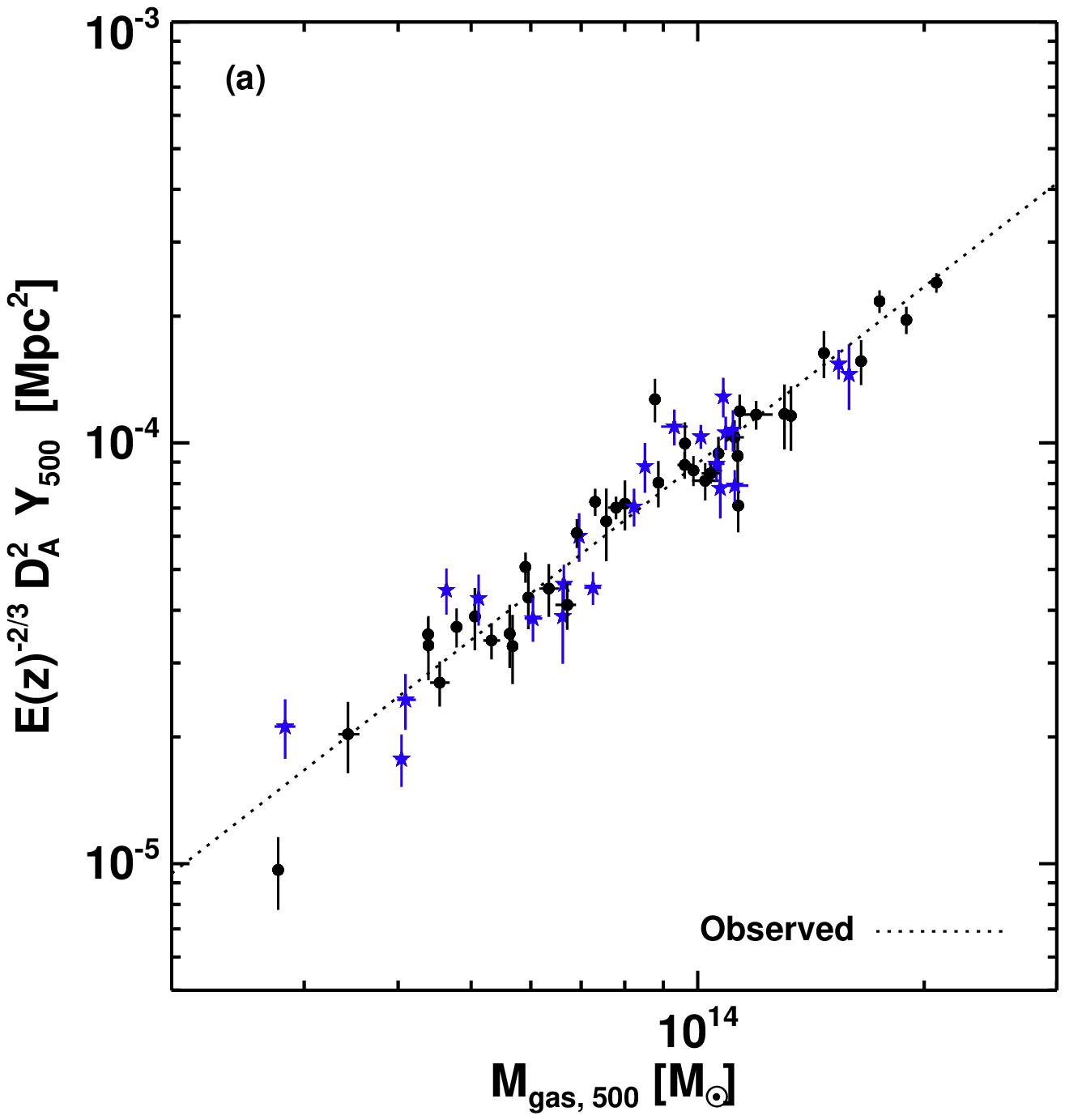}\qquad
\label{fig:YM}\includegraphics[width=0.975\columnwidth]{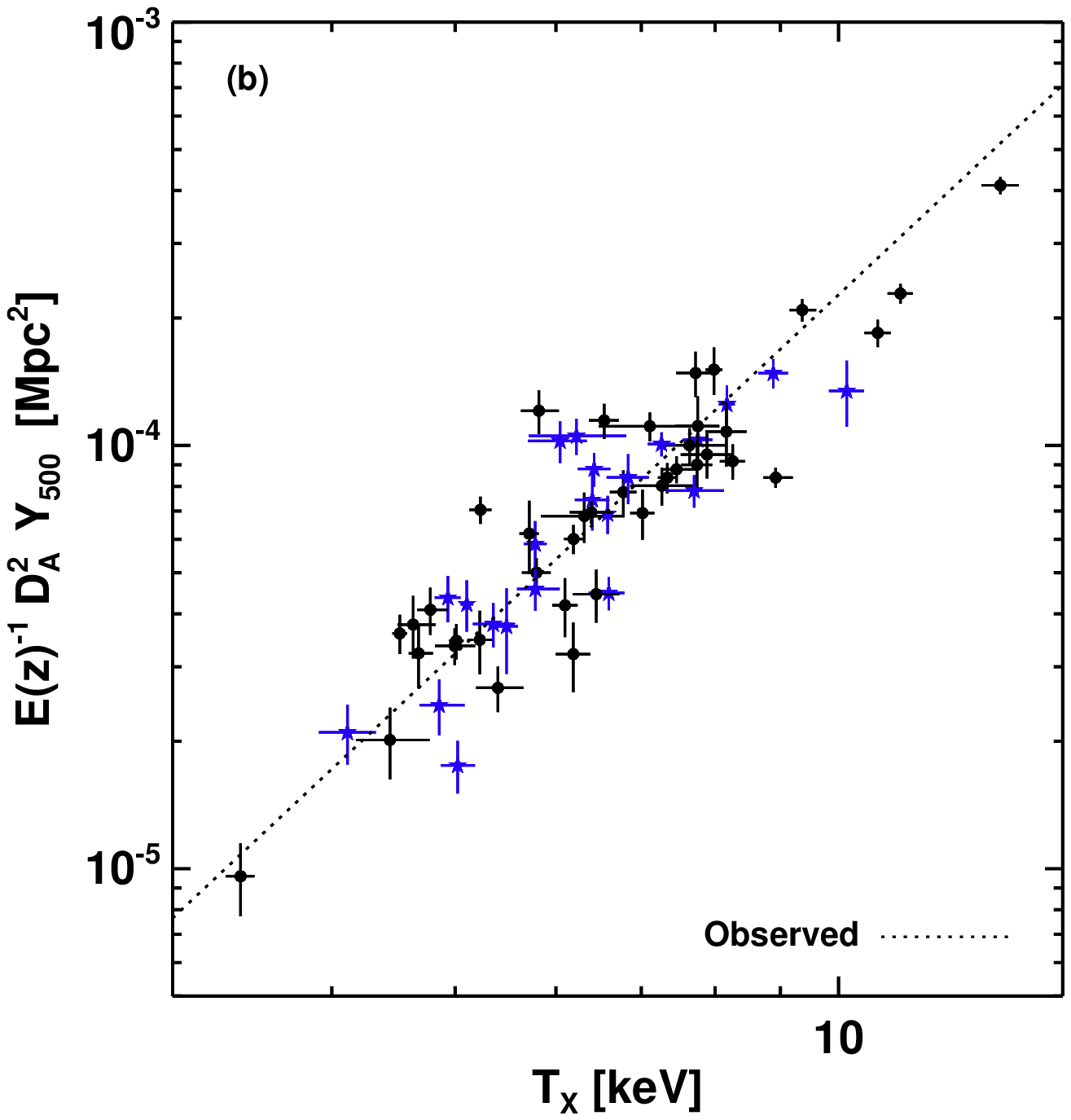}\\
\label{fig:YYx}\includegraphics[width=0.975\columnwidth]{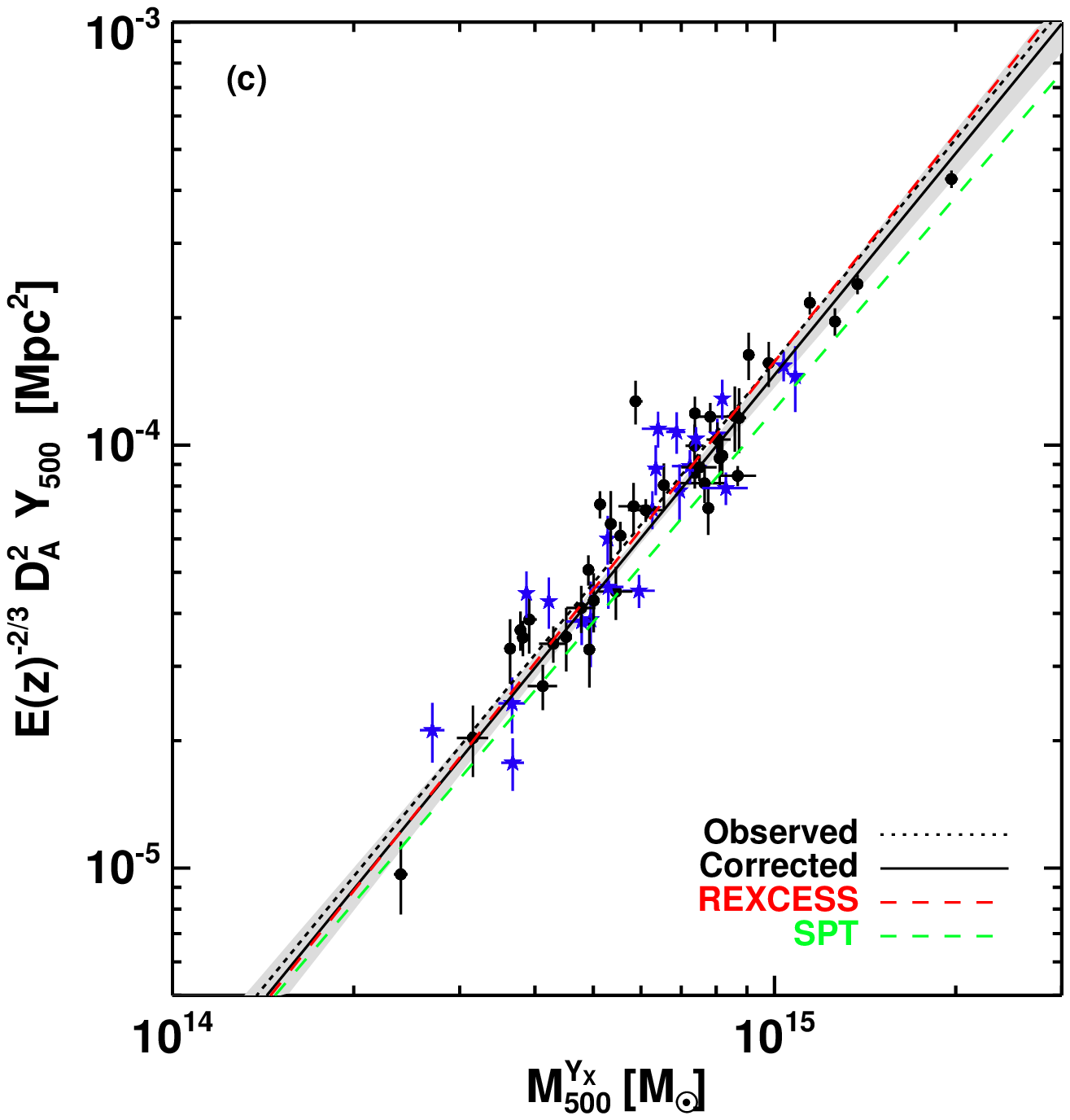}\qquad
\label{fig:LY}\includegraphics[width=0.975\columnwidth]{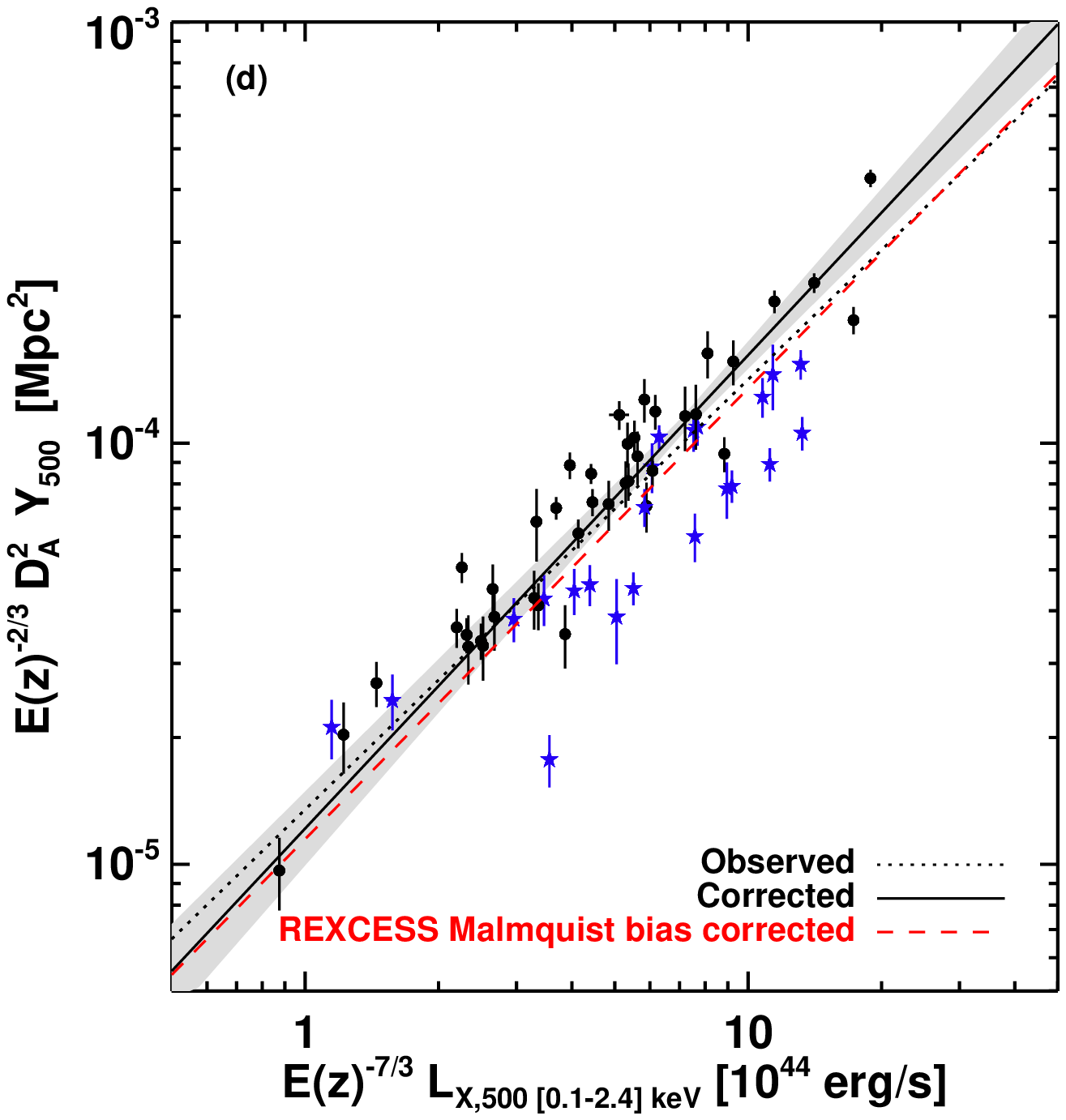}\\
\end{centering}
\caption{{\footnotesize Scaling relations for the 62 clusters in the \planck-\xmm\ archive sample; fits are given in Table~\ref{tab:screl}.  Cool core systems are plotted as blue stars, other systems as black dots.  In the upper panels, the dotted line denotes the observed scaling relation fit.  In the lower panels, the dotted line denotes the observed scaling relation fit, while the solid line shows the fit once the effects of selection bias are taken into account.  The grey shaded area indicates the 1$\sigma$ uncertainty.  The {\it SPT\/} results are taken from \citet{and10}.}}\label{fig:scaling}
\end{figure*}

The effect on scaling relations is then assessed by assigning further
physical properties to the mock catalogue.  Following the methods of
the X-ray analysis, $Y_{\rm X,500}$ is obtained directly from the mass
using Eqn.~\ref{eqn:Yx}, while $M_{\rm g,500}$ and $L_{\rm X,500}$ are
obtained from assumed input scaling relations including
scatter. Finally $\TX$ is simply obtained from $Y_{\rm X,500}/M_{\rm g,500}$
on a cluster-by-cluster basis. The input scaling relation slopes and
amplitudes are then adjusted until the mock observed samples match
those recovered from actual data in Table~\ref{tab:screl}. The input
slopes then provide an estimate of the bias-corrected slope that would
have been obtained had the bias been absent.  The original and
bias-corrected estimates are shown in Fig.~\ref{fig:yszyx_arcmin}
and~\ref{fig:scaling}, and the best-fitting parameters for each
relation are given in Table~\ref{tab:screl}.  Note that the slopes of
the $Y_{\rm X,500}$ and $\TX$ relations are derived quantities fixed by
the other scalings we have chosen.

As seen in Table~\ref{tab:screl} and in Fig.~\ref{fig:scaling}, the
importance of the Malmquist correction depends on the relation under
consideration.  In the case of $Y_{\rm X,500}$ and $M_{500}$ it is negligible, due to the very small scatter seen in these relations.  For
the other relations, however, the Malmquist corrections can be comparable
to the quoted observational uncertainties, indicating that despite the
dynamic range of the \planck\ sample, there are  biases
introduced by the selection cut.  The bias-corrected slopes in
Table~\ref{tab:screl} are thus our best current estimates of the true
underlying slopes.  One should also bear in mind that the bias
correction itself carries uncertainty, which we have not been able to
estimate, and this increases the uncertainty on the underlying slope.

Note in particular that the bias correction leaves the $\Yv$-$\YX$
relation completely consistent with the expected slope of unity, while
the relation to $M_{500}$ remains consistent with a slope $5/3$.


%

\section{Discussion}
\label{sec:disc}

\subsection{SZ flux vs X-ray prediction}

Figure~\ref{fig:yszyx_arcmin} shows the fundamental relation probed by the present study, that between the measured quantities $Y_{\rm X,500}$ and $\Yv$.  We recall that the link between these two quantities is sensitive to the structure in temperature and density.  Note that X-ray information is used to determine the radius of integration for the SZ signal (i.e., $\Rv$) and its overall shape (i.e., the underlying  universal pressure profile).  However, as we have shown above in Sect.~\ref{sec:pprof}, the amplitude of the SZ signal is relatively insensitive to the assumed pressure profile shape, so that the use of X-ray priors reduces to a choice of integration aperture.  Thus we regard the X-ray and SZ quantities as quasi-independent.

In the left-hand panel of Fig.~\ref{fig:yszyx_arcmin} the relation is plotted in units of arcmin$^2$, and shows the excellent agreement between the observed $\Yv$, $Y_{\rm X,500}$ and the X-ray prediction from \rexcess\ (dashed line).  Indeed, fitting the relation with the slope fixed to unity yields a normalisation $\Yv / Y_{\rm X,500} = 0.95\pm0.03$, perfectly consistent with the value $0.924 \pm 0.004$ found for the \rexcess\ sample \citep{arn10}, and less than unity as expected for radially-decreasing temperature profiles \citep[e.g.,][]{pra07}.  Furthermore, the relation is quite tight (see below), and  there is no indication that cool core systems differ systematically from the other systems.

The right-hand panel of Fig.~\ref{fig:yszyx_arcmin} shows the relation between $Y_{\rm X,500}$ and the spherically-integrated Compton parameter $D_{\rm A}^2\, \Yv$.  Note in particular that the slope of the bias-corrected relation is completely consistent with unity, and that the intrinsic scatter (calculated as described above in Eqns.~\ref{eqn:disp1} and~\ref{eqn:disp2}) is exceptionally small, $\sigma_{\rm log,i} = 0.09\pm0.01$.  A comparison with recent results obtained by {\it SPT\/} shows a slight difference in normalisation, although it is not significant given the larger uncertainties in the latter measurement \citep{and10}.


\subsection{Scaling relations}

In this Section we investigate other relations between $D_{\rm A}^2\,\Yv$ and X-ray quantities.  Note that since $\Mv$ is derived from $Y_{\rm X,500}$, its dependence on $D_{\rm A}^2\,\Yv$ is directly linked to the $D_{\rm A}^2\,\Yv - Y_{\rm X,500}$ relation with the exception of differing $E(z)$ dependencies.  Moreover, $M_{\rm g,500}$ and $\TX$ are not independent; they are related via Eqn.~\ref{eqn:Yx}. It is still useful to investigate these relations, though, both for completeness and for comparison to recent results from ground-based studies.

Relations between $D_{\rm A}^2\, \Yv$ and gas mass $M_{\rm g,500}$ and the X-ray temperature $\TX$ are shown in the upper panels of Fig.~\ref{fig:scaling}.  The bias-corrected relations yield slopes that are consistent with self-similar (5/3 and 5/2, respectively) to high accuracy.  Scatter about the $D_{\rm A}^2\, \Yv - M_{\rm g,500}$ relation is small, at $\sigma_{\rm log,i} = 0.09\pm0.01$, while that about the $D_{\rm A}^2\, \Yv - \TX$ relation is among the largest of the relations, at $\sigma_{\rm log,i} = 0.14\pm0.02$.  Once again, cool core systems are fully consistent with the other clusters, and there is no particular evidence that cool core systems show less scatter than the sample as a whole.  

The bottom left-hand panel of Fig.~\ref{fig:scaling} shows the relation between $D_{\rm A}^2\, \Yv$ and mass.  Here again, the slope of the bias-corrected relation is fully consistent with self-similar (5/3).  The scatter is small ($\sigma_{\rm log,i} = 0.10\pm0.01$),  although it is a lower limit since the scatter between $Y_{\rm X,500}$ and total mass is not taken into account in deriving $\Mv$ (it is in fact the same as that about the $D_{\rm A}^2\, \Yv - Y_{\rm X,500}$ relation except for the different $E(z)$ scaling).  Both slope and normalisation are in excellent agreement with X-ray predictions from \rexcess, as expected from the good agreement in the $D_{\rm A}^2\, \Yv - Y_{\rm X,500}$ relation.  The slight offset in normalisation of the relation found by {\it SPT} \citep{and10} can be explained by the different calibration of the $M_{500} - Y_{\rm X,500}$ relation used in their study; it is not a significant offset given their larger normalistion uncertainties. 

The ease of detecting clusters through their X-ray emission makes the X-ray luminosity an important quantity, and its calibration with the SZ signal is imperative for maximising the synergy between the \planck\ all-sky survey and  previous all-sky X-ray surveys such as the RASS and the upcoming {\it eROSITA} survey.  The slope of the $D_{\rm A}^2\, \Yv - L_{\rm X,500}$ relation for the present sample, $1.12\pm0.08$, is in excellent agreement with the slope predicted from X-ray observations alone ($1.07 \pm 0.08$, \citealt{arn10} from \rexcess), and the normalisation is also consistent within the uncertainties.  The slight offset in the best fitting normalisation for the present sample relative to the \rexcess\ prediction can be attributed to the relative lack of strong cooling core clusters in the present sample compared to \rexcess\ (see Fig.~\ref{fig:neprofs}).  The scatter, $\sigma_{\rm log,i} = 0.14\pm0.02$, is largest about this relation due to the influence of cool cores, which are segregated from the other systems and all lie to the high-luminosity side.  Indeed, as Fig.~\ref{fig:scaling} shows and Table~\ref{tab:screl} quantifies, while the vast majority of the dispersion about the $D_{\rm A}^2\, \Yv - L_{\rm X,500}$ relation is due to cool cores, these systems do not contribute significantly to the dispersion about the $D_{\rm A}^2\, \Yv - Y_{\rm X,500}$ relation. Thus while the X-ray luminosity is very sensitive to the presence of cool cores, $D_{\rm A}^2\, \Yv$ appears to be less so. 

The slope of our best fitting $D_{\rm A}^2\, \Yv -L_{\rm X,500}$ relation is also fully consistent within 1$\sigma$ with that derived by \citet{planck2011-5.2a}, which is based on a bin-averaging analysis at the position of known X-ray clusters in the MCXC \citep{pif10}. As X-ray selection is more sensitive to the presence of cool cores (due to the density squared dependence of the X-ray luminosity), one might expect the \citet{planck2011-5.2a} best-fitting $D_{\rm A}^2\, \Yv - L_{\rm X,500}$ relation to be shifted to slightly higher luminosities (i.e., a slightly lower normalisation), as is seen. However, since the MCXC selection function is both complex and unknown, it is impossible to correct their relation for Malmquist bias effects. Thus some part of the normalisation difference between the two studies arises from correction for selection effects.  However the good agreement between our results and those from \citet{planck2011-5.2a}'s bin-averaged analysis argues that sample selection does not have a strong effect on the scaling relations derived from the latter analysis.

Finally, the results we have derived show that X-ray and SZ measurements give a fully coherent view of cluster structure out to moderately large scales.  In particular, they indicate that effects of clumping in the X-ray gas are not significant, at least in the mass and radial range we have probed in the present study.  Furthermore, the excellent agreement between the observed $D_{\rm A}^2\,\Yv - Y_{\rm X,500}$ relation and the X-ray predictions argue that the SZ and X-ray calibrations we have used are fundamentally sound.  


\section{Conclusions}

We have presented SZ and X-ray data from a sample of 62 local ($z<0.5$) galaxy clusters detected at high S/N in the \planck\ survey and observed by \xmm.  The objects range over approximately a decade in mass ($M_{500} \sim 2-20\times 10^{14}\,$~M$_\odot$), and, while the sample is neither  representative nor complete, it constitutes the largest, highest-quality SZ--X-ray data set currently available.  This study has been undertaken in the framework of a series of papers on cluster astrophysics based on the first \planck\ all-sky data set \citep{planck2011-5.1a,planck2011-5.1b,planck2011-5.2a,planck2011-5.2c}.  

SZ and X-ray quantities have been extracted within $R_{500}$ and we have presented a detailed study of the resulting SZ scaling relations.  Moreover, we have investigated how selection effects influence the results of the scaling relation fits.  Their influence is subtle, but the slopes and normalisations of the scaling relations are generally in good agreement with X-ray predictions and other results after accounting for the selection effects.  For the fundamental $D_{\rm A}^2\, \Yv - Y_{\rm X,500}$ relation, we measure a remarkably small logarithmic intrinsic scatter of only $(10\pm1)$ per cent, consistent with the idea that both quantities are low-scatter mass proxies.

The results are fully consistent with the predictions from X-ray observations \citep{arn10} and with recent measurements from a smaller sample spanning a wider redshift range observed with {\it SPT\/} \citep{and10}.  The results are also in excellent agreement with the statistical analysis undertaken at the positions of known X-ray clusters \citep{planck2011-5.2a}.  This excellent agreement between observed SZ quantities and X-ray-based predictions underlines the robustness and consistency of our overall view of ICM properties.  It is difficult to reconcile with the claim, based on a recent {\it WMAP} 7-year  analysis, that X-ray data over-predict the SZ signal \citep{kom11}.  

The results presented here, derived from only 62 systems, provide a maximally-robust local reference for evolution studies or for the use of SZ clusters for  cosmology. Overall, the agreement between the present results, ground-based results and  X-ray predictions augurs well for our understanding of cluster astrophysics and for the use of clusters for cosmology.  Future work will make use of the individual pressure profile shape as derived from X-rays to further improve the SZ flux extraction.  Comparison of X-ray and SZ pressure profiles will also be undertaken, as will comparison of measurements with independent mass estimation methods.  We will also extend our analysis to the full \planck\ catalogue, observing higher-redshift systems, to constrain evolution, and lower-mass objects, to better probe cluster astrophysics.  


\begin{acknowledgements}

The Planck Collaboration acknowledges the support of: ESA; CNES and CNRS/INSU-IN2P3-INP (France); ASI, CNR, and INAF (Italy); NASA and DoE (USA); STFC and UKSA (UK); CSIC, MICINN and JA (Spain); Tekes, AoF and CSC (Finland); DLR and MPG (Germany); CSA (Canada); DTU Space (Denmark); SER/SSO (Switzerland); RCN (Norway); SFI (Ireland); FCT/MCTES (Portugal); and DEISA (EU). The present work is partly based on observations obtained with \xmm, an ESA science mission with instruments and contributions directly funded by ESA Member States and the USA (NASA). This research has made use of the following databases: SIMBAD, operated at CDS, Strasbourg, France; the NED database, which is operated by the Jet Propulsion Laboratory, California Institute of Technology, under contract with the National Aeronautics and Space Administration; BAX, which is operated by the Laboratoire d'Astrophysique de Tarbes-Toulouse (LATT), under contract with the Centre National d'Etudes Spatiales (CNES). A description of the Planck Collaboration and a list of its members, including the technical or scientific activities in which they have been involved, can be found at \url{http://www.rssd.esa.int/Planck}.

\end{acknowledgements}

\bibliographystyle{aa}
\bibliography{16458.bib,Planck_bib.bib}

\begin{thebibliography}{87}
\expandafter\ifx\csname natexlab\endcsname\relax\def\natexlab#1{#1}\fi

\bibitem[{{Afshordi} {et~al.}(2007){Afshordi}, {Lin}, {Nagai}, \&
  {Sanderson}}]{afs07}
{Afshordi}, N., {Lin}, Y., {Nagai}, D., \& {Sanderson}, A.~J.~R. 2007, \mnras,
  378, 293

\bibitem[{{Aghanim} {et~al.}(2009){Aghanim}, {da Silva}, \& {Nunes}}]{agh09}
{Aghanim}, N., {da Silva}, A.~C., \& {Nunes}, N.~J. 2009, \aap, 496, 637

\bibitem[{{Aghanim} {et~al.}(2005){Aghanim}, {Hansen}, \& {Lagache}}]{agh05}
{Aghanim}, N., {Hansen}, S.~H., \& {Lagache}, G. 2005, \aap, 439, 901

\bibitem[{{Akritas} \& {Bershady}(1996)}]{akr96}
{Akritas}, M.~G. \& {Bershady}, M.~A. 1996, \apj, 470, 706

\bibitem[{Andersson {et~al.}(2010)Andersson, Benson, Ade, Aird, Armstrong,
  Bautz, Bleem, Brodwin, Carlstrom, Chang, Crawford, Crites, de~Haan, Desai,
  Dobbs, Dudley, Foley, Forman, Garmire, George, Gladders, Halverson, High,
  Holder, Holzapfel, Hrubes, Jones, Joy, Keisler, Knox, Lee, Leitch, Lueker,
  Marrone, Mcmahon, Mehl, Meyer, Mohr, Montroy, Murray, Padin, Plagge, Pryke,
  Reichardt, Rest, Ruel, Ruhl, Schaffer, Shaw, Shirokoff, Song, Spieler,
  Stalder, Staniszewski, Stark, Stubbs, Vanderlinde, Vieira, Vikhlinin,
  Williamson, Yang, \& Zahn}]{and10}
Andersson, K., Benson, B., Ade, P., {et~al.} 2010, {\tt arXiv:1006.3068}

\bibitem[{Arnaud {et~al.}(2005)Arnaud, Pointecouteau, \& Pratt}]{arn05}
Arnaud, M., Pointecouteau, E., \& Pratt, G.~W. 2005, \aap, 441, 893

\bibitem[{Arnaud {et~al.}(2007)Arnaud, Pointecouteau, \& Pratt}]{arn07}
Arnaud, M., Pointecouteau, E., \& Pratt, G.~W. 2007, \aap, 474, L37

\bibitem[{{Arnaud} {et~al.}(2010){Arnaud}, {Pratt}, {Piffaretti},
  {B{\"o}hringer}, {Croston}, \& {Pointecouteau}}]{arn10}
{Arnaud}, M., {Pratt}, G.~W., {Piffaretti}, R., {et~al.} 2010, \aap, 517, A92

\bibitem[{{Bersanelli} {et~al.}(2010){Bersanelli}, {Mandolesi}, {Butler},
  {Mennella}, {Villa}, {Aja}, {Artal}, {Artina}, {Baccigalupi}, {Balasini},
  {Baldan}, {Banday}, {Bastia}, {Battaglia}, {Bernardino}, {Blackhurst},
  {Boschini}, {Burigana}, {Cafagna}, {Cappellini}, {Cavaliere}, {Colombo},
  {Crone}, {Cuttaia}, {D'Arcangelo}, {Danese}, {Davies}, {Davis}, {de Angelis},
  {de Gasperis}, {de La Fuente}, {de Rosa}, {de Zotti}, {Falvella}, {Ferrari},
  {Ferretti}, {Figini}, {Fogliani}, {Franceschet}, {Franceschi}, {Gaier},
  {Garavaglia}, {Gomez}, {Gorski}, {Gregorio}, {Guzzi}, {Herreros},
  {Hildebrandt}, {Hoyland}, {Hughes}, {Janssen}, {Jukkala}, {Kettle},
  {Kilpi{\"a}}, {Laaninen}, {Lapolla}, {Lawrence}, {Lawson}, {Leahy},
  {Leonardi}, {Leutenegger}, {Levin}, {Lilje}, {Lowe}, {Lubin}, {Maino},
  {Malaspina}, {Maris}, {Marti-Canales}, {Martinez-Gonzalez}, {Mediavilla},
  {Meinhold}, {Miccolis}, {Morgante}, {Natoli}, {Nesti}, {Pagan}, {Paine},
  {Partridge}, {Pascual}, {Pasian}, {Pearson}, {Pecora}, {Perrotta},
  {Platania}, {Pospieszalski}, {Poutanen}, {Prina}, {Rebolo}, {Roddis},
  {Rubi{\~n}o-Martin}, {Salmon}, {Sandri}, {Seiffert}, {Silvestri},
  {Simonetto}, {Sjoman}, {Smoot}, {Sozzi}, {Stringhetti}, {Taddei}, {Tauber},
  {Terenzi}, {Tomasi}, {Tuovinen}, {Valenziano}, {Varis}, {Vittorio}, {Wade},
  {Wilkinson}, {Winder}, {Zacchei}, \& {Zonca}}]{Bersanelli2010}
{Bersanelli}, M., {Mandolesi}, N., {Butler}, R.~C., {et~al.} 2010, \aap, 520,
  A4+

\bibitem[{Bertschinger(1985)}]{ber85}
Bertschinger, E. 1985, \apjs, 58, 39

\bibitem[{{Bielby} \& {Shanks}(2007)}]{bie07}
{Bielby}, R.~M. \& {Shanks}, T. 2007, \mnras, 382, 1196

\bibitem[{{Bock} {et~al.}(1999){Bock}, {Large}, \& {Sadler}}]{boc99}
{Bock}, D., {Large}, M.~I., \& {Sadler}, E.~M. 1999, \aj, 117, 1578

\bibitem[{{B{\"o}hringer} {et~al.}(2007){B{\"o}hringer}, {Schuecker}, {Pratt},
  {Arnaud}, {Ponman}, {Croston}, {Borgani}, {Bower}, {Briel}, {Collins},
  {Donahue}, {Forman}, {Finoguenov}, {Geller}, {Guzzo}, {Henry}, {Kneissl},
  {Mohr}, {Matsushita}, {Mullis}, {Ohashi}, {Pedersen}, {Pierini}, {Quintana},
  {Raychaudhury}, {Reiprich}, {Romer}, {Rosati}, {Sabirli}, {Temple}, {Viana},
  {Vikhlinin}, {Voit}, \& {Zhang}}]{boe07}
{B{\"o}hringer}, H., {Schuecker}, P., {Pratt}, G.~W., {et~al.} 2007, \aap, 469,
  363

\bibitem[{{Bonamente} {et~al.}(2008){Bonamente}, {Joy}, {LaRoque}, {Carlstrom},
  {Nagai}, \& {Marrone}}]{bon08}
{Bonamente}, M., {Joy}, M., {LaRoque}, S.~J., {et~al.} 2008, \apj, 675, 106

\bibitem[{{Bourdin} \& {Mazzotta}(2008)}]{bou08}
{Bourdin}, H. \& {Mazzotta}, P. 2008, \aap, 479, 307

\bibitem[{{Burns} {et~al.}(2008){Burns}, {Hallman}, {Gantner}, {Motl}, \&
  {Norman}}]{bur08}
{Burns}, J.~O., {Hallman}, E.~J., {Gantner}, B., {Motl}, P.~M., \& {Norman},
  M.~L. 2008, \apj, 675, 1125

\bibitem[{{Carlstrom} {et~al.}(2011){Carlstrom}, {Ade}, {Aird}, {Benson},
  {Bleem}, {Busetti}, {Chang}, {Chauvin}, {Cho}, {Crawford}, {Crites}, {Dobbs},
  {Halverson}, {Heimsath}, {Holzapfel}, {Hrubes}, {Joy}, {Keisler}, {Lanting},
  {Lee}, {Leitch}, {Leong}, {Lu}, {Lueker}, {Luong-van}, {McMahon}, {Mehl},
  {Meyer}, {Mohr}, {Montroy}, {Padin}, {Plagge}, {Pryke}, {Ruhl}, {Schaffer},
  {Schwan}, {Shirokoff}, {Spieler}, {Staniszewski}, {Stark}, {Tucker},
  {Vanderlinde}, {Vieira}, \& {Williamson}}]{car11}
{Carlstrom}, J.~E., {Ade}, P.~A.~R., {Aird}, K.~A., {et~al.} 2011, \pasp, 123,
  568

\bibitem[{{Condon} {et~al.}(1998){Condon}, {Cotton}, {Greisen}, {Yin},
  {Perley}, {Taylor}, \& {Broderick}}]{con98}
{Condon}, J.~J., {Cotton}, W.~D., {Greisen}, E.~W., {et~al.} 1998, \aj, 115,
  1693

\bibitem[{Croston {et~al.}(2008)Croston, Pratt, B{\"o}hringer, Arnaud,
  Pointecouteau, Ponman, Sanderson, Temple, Bower, \& Donahue}]{cro08}
Croston, J.~H., Pratt, G.~W., B{\"o}hringer, H., {et~al.} 2008, \aap, 487, 431

\bibitem[{da~Silva {et~al.}(2004)da~Silva, Kay, Liddle, \& Thomas}]{das04}
da~Silva, A.~C., Kay, S.~T., Liddle, A.~R., \& Thomas, P.~A. 2004, \mnras, 348,
  1401

\bibitem[{{Douspis} {et~al.}(2006){Douspis}, {Aghanim}, \& {Langer}}]{dou06}
{Douspis}, M., {Aghanim}, N., \& {Langer}, M. 2006, \aap, 456, 819

\bibitem[{{Evrard} {et~al.}(1996){Evrard}, {Metzler}, \& {Navarro}}]{evr96}
{Evrard}, A.~E., {Metzler}, C.~A., \& {Navarro}, J.~F. 1996, \apj, 469, 494

\bibitem[{{Fabian} {et~al.}(1994){Fabian}, {Crawford}, {Edge}, \&
  {Mushotzky}}]{fab94}
{Fabian}, A.~C., {Crawford}, C.~S., {Edge}, A.~C., \& {Mushotzky}, R.~F. 1994,
  \mnras, 267, 779

\bibitem[{{Jenkins} {et~al.}(2001){Jenkins}, {Frenk}, {White}, {Colberg},
  {Cole}, {Evrard}, {Couchman}, \& {Yoshida}}]{jen01}
{Jenkins}, A., {Frenk}, C.~S., {White}, S.~D.~M., {et~al.} 2001, \mnras, 321,
  372

\bibitem[{{Jones} \& {Forman}(1984)}]{jon84}
{Jones}, C. \& {Forman}, W. 1984, \apj, 276, 38

\bibitem[{{Kaiser}(1986)}]{kai86}
{Kaiser}, N. 1986, \mnras, 222, 323

\bibitem[{Komatsu {et~al.}(1999)Komatsu, Kitayama, Suto, Hattori, Kawabe,
  Matsuo, Schindler, \& Yoshikawa}]{kom99}
Komatsu, E., Kitayama, T., Suto, Y., {et~al.} 1999, \apjl, 516, L1

\bibitem[{Komatsu {et~al.}(2001)Komatsu, Matsuo, Kitayama, Hattori, Kawabe,
  Kohno, Kuno, Schindler, Suto, \& Yoshikawa}]{kom01}
Komatsu, E., Matsuo, H., Kitayama, T., {et~al.} 2001, \pasj, 53, 57

\bibitem[{{Komatsu} {et~al.}(2011){Komatsu}, {Smith}, {Dunkley}, {Bennett},
  {Gold}, {Hinshaw}, {Jarosik}, {Larson}, {Nolta}, {Page}, {Spergel},
  {Halpern}, {Hill}, {Kogut}, {Limon}, {Meyer}, {Odegard}, {Tucker}, {Weiland},
  {Wollack}, \& {Wright}}]{kom11}
{Komatsu}, E., {Smith}, K.~M., {Dunkley}, J., {et~al.} 2011, \apjs, 192, 18

\bibitem[{{Kosowsky}(2003)}]{kos03}
{Kosowsky}, A. 2003, \nar, 47, 939

\bibitem[{{Kravtsov} {et~al.}(2006){Kravtsov}, {Vikhlinin}, \& {Nagai}}]{kra06}
{Kravtsov}, A.~V., {Vikhlinin}, A., \& {Nagai}, D. 2006, \apj, 650, 128

\bibitem[{{Lamarre} {et~al.}(2010){Lamarre}, {Puget}, {Ade}, {Bouchet},
  {Guyot}, {Lange}, {Pajot}, {Arondel}, {Benabed}, {Beney}, {Beno{\^i}t},
  {Bernard}, {Bhatia}, {Blanc}, {Bock}, {Br{\'e}elle}, {Bradshaw}, {Camus},
  {Catalano}, {Charra}, {Charra}, {Church}, {Couchot}, {Coulais}, {Crill},
  {Crook}, {Dassas}, {de Bernardis}, {Delabrouille}, {de Marcillac}, {Delouis},
  {D{\'e}sert}, {Dumesnil}, {Dupac}, {Efstathiou}, {Eng}, {Evesque},
  {Fourmond}, {Ganga}, {Giard}, {Gispert}, {Guglielmi}, {Haissinski},
  {Henrot-Versill{\'e}}, {Hivon}, {Holmes}, {Jones}, {Koch}, {Lagard{\`e}re},
  {Lami}, {Land{\'e}}, {Leriche}, {Leroy}, {Longval},
  {Mac{\'{\i}}as-P{\'e}rez}, {Maciaszek}, {Maffei}, {Mansoux}, {Marty}, {Masi},
  {Mercier}, {Miville-Desch{\^e}nes}, {Moneti}, {Montier}, {Murphy},
  {Narbonne}, {Nexon}, {Paine}, {Pahn}, {Perdereau}, {Piacentini}, {Piat},
  {Plaszczynski}, {Pointecouteau}, {Pons}, {Ponthieu}, {Prunet}, {Rambaud},
  {Recouvreur}, {Renault}, {Ristorcelli}, {Rosset}, {Santos}, {Savini},
  {Serra}, {Stassi}, {Sudiwala}, {Sygnet}, {Tauber}, {Torre}, {Tristram},
  {Vibert}, {Woodcraft}, {Yurchenko}, \& {Yvon}}]{Lamarre2010}
{Lamarre}, J., {Puget}, J., {Ade}, P.~A.~R., {et~al.} 2010, \aap, 520, A9+

\bibitem[{{Leahy} {et~al.}(2010){Leahy}, {Bersanelli}, {D'Arcangelo}, {Ganga},
  {Leach}, {Moss}, {Keih{\"a}nen}, {Keskitalo}, {Kurki-Suonio}, {Poutanen},
  {Sandri}, {Scott}, {Tauber}, {Valenziano}, {Villa}, {Wilkinson}, {Zonca},
  {Baccigalupi}, {Borrill}, {Butler}, {Cuttaia}, {Davis}, {Frailis},
  {Francheschi}, {Galeotta}, {Gregorio}, {Leonardi}, {Mandolesi}, {Maris},
  {Meinhold}, {Mendes}, {Mennella}, {Morgante}, {Prezeau}, {Rocha},
  {Stringhetti}, {Terenzi}, \& {Tomasi}}]{Leahy2010}
{Leahy}, J.~P., {Bersanelli}, M., {D'Arcangelo}, O., {et~al.} 2010, \aap, 520,
  A8+

\bibitem[{{Leccardi} \& {Molendi}(2008)}]{lec08}
{Leccardi}, A. \& {Molendi}, S. 2008, \aap, 486, 359

\bibitem[{{Lieu} {et~al.}(2006){Lieu}, {Mittaz}, \& {Zhang}}]{lie06}
{Lieu}, R., {Mittaz}, J.~P.~D., \& {Zhang}, S. 2006, \apj, 648, 176

\bibitem[{{Mandolesi} {et~al.}(2010){Mandolesi}, {Bersanelli}, {Butler},
  {Artal}, {Baccigalupi}, {Balbi}, {Banday}, {Barreiro}, {Bartelmann},
  {Bennett}, {Bhandari}, {Bonaldi}, {Borrill}, {Bremer}, {Burigana}, {Bowman},
  {Cabella}, {Cantalupo}, {Cappellini}, {Courvoisier}, {Crone}, {Cuttaia},
  {Danese}, {D'Arcangelo}, {Davies}, {Davis}, {de Angelis}, {de Gasperis}, {de
  Rosa}, {de Troia}, {de Zotti}, {Dick}, {Dickinson}, {Diego}, {Donzelli},
  {D{\"o}rl}, {Dupac}, {En{\ss}lin}, {Eriksen}, {Falvella}, {Finelli},
  {Frailis}, {Franceschi}, {Gaier}, {Galeotta}, {Gasparo}, {Giardino}, {Gomez},
  {Gonzalez-Nuevo}, {G{\'o}rski}, {Gregorio}, {Gruppuso}, {Hansen}, {Hell},
  {Herranz}, {Herreros}, {Hildebrandt}, {Hovest}, {Hoyland}, {Huffenberger},
  {Janssen}, {Jaffe}, {Keih{\"a}nen}, {Keskitalo}, {Kisner}, {Kurki-Suonio},
  {L{\"a}hteenm{\"a}ki}, {Lawrence}, {Leach}, {Leahy}, {Leonardi}, {Levin},
  {Lilje}, {L{\'o}pez-Caniego}, {Lowe}, {Lubin}, {Maino}, {Malaspina}, {Maris},
  {Marti-Canales}, {Martinez-Gonzalez}, {Massardi}, {Matarrese}, {Matthai},
  {Meinhold}, {Melchiorri}, {Mendes}, {Mennella}, {Morgante}, {Morigi},
  {Morisset}, {Moss}, {Nash}, {Natoli}, {Nesti}, {Paine}, {Partridge},
  {Pasian}, {Passvogel}, {Pearson}, {P{\'e}rez-Cuevas}, {Perrotta}, {Polenta},
  {Popa}, {Poutanen}, {Prezeau}, {Prina}, {Rachen}, {Rebolo}, {Reinecke},
  {Ricciardi}, {Riller}, {Rocha}, {Roddis}, {Rohlfs}, {Rubi{\~n}o-Martin},
  {Salerno}, {Sandri}, {Scott}, {Seiffert}, {Silk}, {Simonetto}, {Smoot},
  {Sozzi}, {Sternberg}, {Stivoli}, {Stringhetti}, {Tauber}, {Terenzi},
  {Tomasi}, {Tuovinen}, {T{\"u}rler}, {Valenziano}, {Varis}, {Vielva}, {Villa},
  {Vittorio}, {Wade}, {White}, {White}, {Wilkinson}, {Zacchei}, \&
  {Zonca}}]{Mandolesi2010}
{Mandolesi}, N., {Bersanelli}, M., {Butler}, R.~C., {et~al.} 2010, \aap, 520,
  A3+

\bibitem[{{Mantz} {et~al.}(2010){Mantz}, {Allen}, {Ebeling}, {Rapetti}, \&
  {Drlica-Wagner}}]{man10}
{Mantz}, A., {Allen}, S.~W., {Ebeling}, H., {Rapetti}, D., \& {Drlica-Wagner},
  A. 2010, \mnras, 406, 1773

\bibitem[{{Marrone} {et~al.}(2009){Marrone}, {Smith}, {Richard}, {Joy},
  {Bonamente}, {Hasler}, {Hamilton-Morris}, {Kneib}, {Culverhouse},
  {Carlstrom}, {Greer}, {Hawkins}, {Hennessy}, {Lamb}, {Leitch}, {Loh},
  {Miller}, {Mroczkowski}, {Muchovej}, {Pryke}, {Sharp}, \& {Woody}}]{mar09}
{Marrone}, D.~P., {Smith}, G.~P., {Richard}, J., {et~al.} 2009, \apjl, 701,
  L114

\bibitem[{{Maughan}(2007)}]{mau07}
{Maughan}, B.~J. 2007, \apj, 668, 772

\bibitem[{{Melin} {et~al.}(2006){Melin}, {Bartlett}, \& {Delabrouille}}]{mel06}
{Melin}, J., {Bartlett}, J.~G., \& {Delabrouille}, J. 2006, \aap, 459, 341

\bibitem[{{Melin} {et~al.}(2011){Melin}, {Bartlett}, {Delabrouille}, {Arnaud},
  {Piffaretti}, \& {Pratt}}]{mel11}
{Melin}, J., {Bartlett}, J.~G., {Delabrouille}, J., {et~al.} 2011, \aap, 525,
  A139+

\bibitem[{{Mennella et al.}(2011)}]{planck2011-1.4}
{Mennella et al.} 2011, {Planck early results 03: First assessment of the Low
  Frequency Instrument in-flight performance} ({Submitted to \aap,
  [arXiv:astro-ph/1101.2038]})

\bibitem[{{Motl} {et~al.}(2005){Motl}, {Hallman}, {Burns}, \& {Norman}}]{mot05}
{Motl}, P.~M., {Hallman}, E.~J., {Burns}, J.~O., \& {Norman}, M.~L. 2005,
  \apjl, 623, L63

\bibitem[{{Nagai}(2006)}]{nag06}
{Nagai}, D. 2006, \apj, 650, 538

\bibitem[{{Nagai} {et~al.}(2007){Nagai}, {Kravtsov}, \& {Vikhlinin}}]{nag07}
{Nagai}, D., {Kravtsov}, A.~V., \& {Vikhlinin}, A. 2007, \apj, 668, 1

\bibitem[{{Piffaretti} {et~al.}(2010){Piffaretti}, {Arnaud}, {Pratt},
  {Pointecouteau}, \& {Melin}}]{pif10}
{Piffaretti}, R., {Arnaud}, M., {Pratt}, G.~W., {Pointecouteau}, E., \&
  {Melin}, J. 2010, {\tt arXiv:1007.1916}

\bibitem[{{Planck Collaboration}(2011{\natexlab{a}})}]{planck2011-1.1}
{Planck Collaboration}. 2011{\natexlab{a}}, {Planck early results 01: The
  Planck mission} ({Submitted to \aap, [arXiv:astro-ph/1101.2022]})

\bibitem[{{Planck Collaboration}(2011{\natexlab{b}})}]{planck2011-1.3}
{Planck Collaboration}. 2011{\natexlab{b}}, {Planck early results 02: The
  thermal performance of Planck} ({Submitted to \aap,
  [arXiv:astro-ph/1101.2023]})

\bibitem[{{Planck Collaboration}(2011{\natexlab{c}})}]{planck2011-1.10}
{Planck Collaboration}. 2011{\natexlab{c}}, {Planck early results 07: The Early
  Release Compact Source Catalogue} ({Submitted to \aap,
  [arXiv:astro-ph/1101.2041]})

\bibitem[{{Planck Collaboration}(2011{\natexlab{d}})}]{planck2011-5.1a}
{Planck Collaboration}. 2011{\natexlab{d}}, {Planck early results 08: The
  all-sky early Sunyaev-Zeldovich cluster sample} ({Submitted to \aap,
  [arXiv:astro-ph/1101.2024]})

\bibitem[{{Planck Collaboration}(2011{\natexlab{e}})}]{planck2011-5.1b}
{Planck Collaboration}. 2011{\natexlab{e}}, {Planck early results 09:
  XMM-Newton follow-up for validation of Planck cluster candidates} ({Submitted
  to \aap, [arXiv:astro-ph/1101.2025]})

\bibitem[{{Planck Collaboration}(2011{\natexlab{f}})}]{planck2011-5.2a}
{Planck Collaboration}. 2011{\natexlab{f}}, {Planck early results 10:
  Statistical analysis of Sunyaev-Zeldovich scaling relations for X-ray galaxy
  clusters} ({Submitted to \aap, [arXiv:astro-ph/1101.2043]})

\bibitem[{{Planck Collaboration}(2011{\natexlab{g}})}]{planck2011-5.2b}
{Planck Collaboration}. 2011{\natexlab{g}}, {Planck early results 11:
  Calibration of the local galaxy cluster Sunyaev-Zeldovich scaling relations}
  ({Submitted to \aap, [arXiv:astro-ph/1101.2026]})

\bibitem[{{Planck Collaboration}(2011{\natexlab{h}})}]{planck2011-5.2c}
{Planck Collaboration}. 2011{\natexlab{h}}, {Planck early results 12: Cluster
  Sunyaev-Zeldovich optical Scaling relations} ({Submitted to \aap,
  [arXiv:astro-ph/1101.2027]})

\bibitem[{{Planck Collaboration}(2011{\natexlab{i}})}]{planck2011-6.1}
{Planck Collaboration}. 2011{\natexlab{i}}, {Planck early results 13:
  Statistical properties of extragalactic radio sources in the Planck Early
  Release Compact Source Catalogue} ({Submitted to \aap,
  [arXiv:astro-ph/1101.2044]})

\bibitem[{{Planck Collaboration}(2011{\natexlab{j}})}]{planck2011-6.2}
{Planck Collaboration}. 2011{\natexlab{j}}, {Planck early results 14: Early
  Release Compact Source Catalogue validation and extreme radio sources}
  ({Submitted to \aap, [arXiv:astro-ph/1101.1721]})

\bibitem[{{Planck Collaboration}(2011{\natexlab{k}})}]{planck2011-6.3a}
{Planck Collaboration}. 2011{\natexlab{k}}, {Planck early results 15: Spectral
  energy distributions and radio continuum spectra of northern extragalactic
  radio sources} ({Submitted to \aap, [arXiv:astro-ph/1101.2047]})

\bibitem[{{Planck Collaboration}(2011{\natexlab{l}})}]{planck2011-6.4a}
{Planck Collaboration}. 2011{\natexlab{l}}, {Planck early results 16: The
  Planck view of nearby galaxies} ({Submitted to \aap,
  [arXiv:astro-ph/1101.2045]})

\bibitem[{{Planck Collaboration}(2011{\natexlab{m}})}]{planck2011-6.4b}
{Planck Collaboration}. 2011{\natexlab{m}}, {Planck early results 17: Origin of
  the submillimetre excess dust emission in the Magellanic Clouds} ({Submitted
  to \aap, [arXiv:astro-ph/1101.2046]})

\bibitem[{{Planck Collaboration}(2011{\natexlab{n}})}]{planck2011-6.6}
{Planck Collaboration}. 2011{\natexlab{n}}, {Planck early results 18: The power
  spectrum of cosmic infrared background anisotropies} ({Submitted to \aap,
  [arXiv:astro-ph/1101.2028]})

\bibitem[{{Planck Collaboration}(2011{\natexlab{o}})}]{planck2011-7.0}
{Planck Collaboration}. 2011{\natexlab{o}}, {Planck early results 19: All-sky
  temperature and dust optical depth from Planck and IRAS --- constraints on
  the ``dark gas" in our Galaxy} ({Submitted to \aap,
  [arXiv:astro-ph/1101.2029]})

\bibitem[{{Planck Collaboration}(2011{\natexlab{p}})}]{planck2011-7.2}
{Planck Collaboration}. 2011{\natexlab{p}}, {Planck early results 20: New light
  on anomalous microwave emission from spinning dust grains} ({Submitted to
  \aap, [arXiv:astro-ph/1101.2031]})

\bibitem[{{Planck Collaboration}(2011{\natexlab{q}})}]{planck2011-7.3}
{Planck Collaboration}. 2011{\natexlab{q}}, {Planck early results 21:
  Properties of the interstellar medium in the Galactic plane} ({Submitted to
  \aap, [arXiv:astro-ph/1101.2032]})

\bibitem[{{Planck Collaboration}(2011{\natexlab{r}})}]{planck2011-7.7a}
{Planck Collaboration}. 2011{\natexlab{r}}, {Planck early results 22: The
  submillimetre properties of a sample of Galactic cold clumps} ({Submitted to
  \aap, [arXiv:astro-ph/1101.2034]})

\bibitem[{{Planck Collaboration}(2011{\natexlab{s}})}]{planck2011-7.7b}
{Planck Collaboration}. 2011{\natexlab{s}}, {Planck early results 23: The
  Galactic cold core population revealed by the first all-sky survey}
  ({Submitted to \aap, [arXiv:astro-ph/1101.2035]})

\bibitem[{{Planck Collaboration}(2011{\natexlab{t}})}]{planck2011-7.12}
{Planck Collaboration}. 2011{\natexlab{t}}, {Planck early results 24: Dust in
  the diffuse interstellar medium and the Galactic halo} ({Submitted to \aap,
  [arXiv:astro-ph/1101.2036]})

\bibitem[{{Planck Collaboration}(2011{\natexlab{u}})}]{planck2011-7.13}
{Planck Collaboration}. 2011{\natexlab{u}}, {Planck early results 25: Thermal
  dust in nearby molecular clouds} ({Submitted to \aap,
  [arXiv:astro-ph/1101.2037]})

\bibitem[{{Planck Collaboration}(2011{\natexlab{v}})}]{planck2011-1.10sup}
{Planck Collaboration}. 2011{\natexlab{v}}, {The Explanatory Supplement to the
  Planck Early Release Compact Source Catalogue} ({ESA})

\bibitem[{{Planck HFI Core Team}(2011{\natexlab{a}})}]{planck2011-1.5}
{Planck HFI Core Team}. 2011{\natexlab{a}}, {Planck early results 04: First
  assessment of the High Frequency Instrument in-flight performance}
  ({Submitted to \aap, [arXiv:astro-ph/1101.2039]})

\bibitem[{{Planck HFI Core Team}(2011{\natexlab{b}})}]{planck2011-1.7}
{Planck HFI Core Team}. 2011{\natexlab{b}}, {Planck early results 06: The High
  Frequency Instrument data processing} ({Submitted to \aap,
  [arXiv:astro-ph/1101.2048]})

\bibitem[{Pointecouteau {et~al.}(1999)Pointecouteau, Giard, Benoit, D{\'e}sert,
  Aghanim, Coron, Lamarre, \& Delabrouille}]{poi99}
Pointecouteau, E., Giard, M., Benoit, A., {et~al.} 1999, \apj, 519, L115

\bibitem[{Pointecouteau {et~al.}(2001)Pointecouteau, Giard, Benoit, D{\'e}sert,
  Bernard, Coron, \& Lamarre}]{poi01}
Pointecouteau, E., Giard, M., Benoit, A., {et~al.} 2001, \apj, 552, 42

\bibitem[{{Pratt} {et~al.}(2010){Pratt}, {Arnaud}, {Piffaretti},
  {B{\"o}hringer}, {Ponman}, {Croston}, {Voit}, {Borgani}, \& {Bower}}]{pra10}
{Pratt}, G.~W., {Arnaud}, M., {Piffaretti}, R., {et~al.} 2010, \aap, 511, A85+

\bibitem[{Pratt {et~al.}(2007)Pratt, B{\"o}hringer, Croston, Arnaud, Borgani,
  Finoguenov, \& Temple}]{pra07}
Pratt, G.~W., B{\"o}hringer, H., Croston, J.~H., {et~al.} 2007, \aap, 461, 71

\bibitem[{{Pratt} {et~al.}(2009){Pratt}, {Croston}, {Arnaud}, \&
  {B{\"o}hringer}}]{pra09}
{Pratt}, G.~W., {Croston}, J.~H., {Arnaud}, M., \& {B{\"o}hringer}, H. 2009,
  \aap, 498, 361

\bibitem[{{Rosset} {et~al.}(2010){Rosset}, {Tristram}, {Ponthieu}, {Ade},
  {Aumont}, {Catalano}, {Conversi}, {Couchot}, {Crill}, {D{\'e}sert}, {Ganga},
  {Giard}, {Giraud-H{\'e}raud}, {Ha{\"i}ssinski}, {Henrot-Versill{\'e}},
  {Holmes}, {Jones}, {Lamarre}, {Lange}, {Leroy}, {Mac{\'{\i}}as-P{\'e}rez},
  {Maffei}, {de Marcillac}, {Miville-Desch{\^e}nes}, {Montier}, {Noviello},
  {Pajot}, {Perdereau}, {Piacentini}, {Piat}, {Plaszczynski}, {Pointecouteau},
  {Puget}, {Ristorcelli}, {Savini}, {Sudiwala}, {Veneziani}, \&
  {Yvon}}]{Rosset2010}
{Rosset}, C., {Tristram}, M., {Ponthieu}, N., {et~al.} 2010, \aap, 520, A13+

\bibitem[{{Sehgal} {et~al.}(2011){Sehgal}, {Trac}, {Acquaviva}, {Ade},
  {Aguirre}, {Amiri}, {Appel}, {Barrientos}, {Battistelli}, {Bond}, {Brown},
  {Burger}, {Chervenak}, {Das}, {Devlin}, {Dicker}, {Bertrand Doriese},
  {Dunkley}, {D{\"u}nner}, {Essinger-Hileman}, {Fisher}, {Fowler}, {Hajian},
  {Halpern}, {Hasselfield}, {Hern{\'a}ndez-Monteagudo}, {Hilton}, {Hilton},
  {Hincks}, {Hlozek}, {Holtz}, {Huffenberger}, {Hughes}, {Hughes}, {Infante},
  {Irwin}, {Jones}, {Baptiste Juin}, {Klein}, {Kosowsky}, {Lau}, {Limon},
  {Lin}, {Lupton}, {Marriage}, {Marsden}, {Martocci}, {Mauskopf}, {Menanteau},
  {Moodley}, {Moseley}, {Netterfield}, {Niemack}, {Nolta}, {Page}, {Parker},
  {Partridge}, {Reid}, {Sherwin}, {Sievers}, {Spergel}, {Staggs}, {Swetz},
  {Switzer}, {Thornton}, {Tucker}, {Warne}, {Wollack}, \& {Zhao}}]{seh11}
{Sehgal}, N., {Trac}, H., {Acquaviva}, V., {et~al.} 2011, \apj, 732, 44

\bibitem[{{Staniszewski} {et~al.}(2009){Staniszewski}, {Ade}, {Aird}, {Benson},
  {Bleem}, {Carlstrom}, {Chang}, {Cho}, {Crawford}, {Crites}, {de Haan},
  {Dobbs}, {Halverson}, {Holder}, {Holzapfel}, {Hrubes}, {Joy}, {Keisler},
  {Lanting}, {Lee}, {Leitch}, {Loehr}, {Lueker}, {McMahon}, {Mehl}, {Meyer},
  {Mohr}, {Montroy}, {Ngeow}, {Padin}, {Plagge}, {Pryke}, {Reichardt}, {Ruhl},
  {Schaffer}, {Shaw}, {Shirokoff}, {Spieler}, {Stalder}, {Stark},
  {Vanderlinde}, {Vieira}, {Zahn}, \& {Zenteno}}]{sta10}
{Staniszewski}, Z., {Ade}, P.~A.~R., {Aird}, K.~A., {et~al.} 2009, \apj, 701,
  32

\bibitem[{Sunyaev \& Zeldovich(1972)}]{sun72}
Sunyaev, R.~A. \& Zeldovich, Y.~B. 1972, Comments on Astrophysics and Space
  Physics, 4, 173

\bibitem[{{Tauber} {et~al.}(2010){Tauber}, {Mandolesi}, {Puget}, {Banos},
  {Bersanelli}, {Bouchet}, {Butler}, {Charra}, {Crone}, {Dodsworth}, \&
  et~al.}]{tauber2010a}
{Tauber}, J.~A., {Mandolesi}, N., {Puget}, J., {et~al.} 2010, \aap, 520, A1+

\bibitem[{{Vanderlinde} {et~al.}(2010){Vanderlinde}, {Crawford}, {de Haan},
  {Dudley}, {Shaw}, {Ade}, {Aird}, {Benson}, {Bleem}, {Brodwin}, {Carlstrom},
  {Chang}, {Crites}, {Desai}, {Dobbs}, {Foley}, {George}, {Gladders}, {Hall},
  {Halverson}, {High}, {Holder}, {Holzapfel}, {Hrubes}, {Joy}, {Keisler},
  {Knox}, {Lee}, {Leitch}, {Loehr}, {Lueker}, {Marrone}, {McMahon}, {Mehl},
  {Meyer}, {Mohr}, {Montroy}, {Ngeow}, {Padin}, {Plagge}, {Pryke}, {Reichardt},
  {Rest}, {Ruel}, {Ruhl}, {Schaffer}, {Shirokoff}, {Song}, {Spieler},
  {Stalder}, {Staniszewski}, {Stark}, {Stubbs}, {van Engelen}, {Vieira},
  {Williamson}, {Yang}, {Zahn}, \& {Zenteno}}]{van10}
{Vanderlinde}, K., {Crawford}, T.~M., {de Haan}, T., {et~al.} 2010, \apj, 722,
  1180

\bibitem[{{Vikhlinin} {et~al.}(2009){Vikhlinin}, {Burenin}, {Ebeling},
  {Forman}, {Hornstrup}, {Jones}, {Kravtsov}, {Murray}, {Nagai}, {Quintana}, \&
  {Voevodkin}}]{vik09}
{Vikhlinin}, A., {Burenin}, R.~A., {Ebeling}, H., {et~al.} 2009, \apj, 692,
  1033

\bibitem[{{Vikhlinin} {et~al.}(2006){Vikhlinin}, {Kravtsov}, {Forman}, {Jones},
  {Markevitch}, {Murray}, \& {Van Speybroeck}}]{vik06}
{Vikhlinin}, A., {Kravtsov}, A., {Forman}, W., {et~al.} 2006, \apj, 640, 691

\bibitem[{Voit(2005)}]{voi05}
Voit, G.~M. 2005, Reviews of Modern Physics, 77, 207

\bibitem[{{White} {et~al.}(2002){White}, {Hernquist}, \& {Springel}}]{whi02}
{White}, M., {Hernquist}, L., \& {Springel}, V. 2002, \apj, 579, 16

\bibitem[{{Wik} {et~al.}(2008){Wik}, {Sarazin}, {Ricker}, \& {Randall}}]{wik08}
{Wik}, D.~R., {Sarazin}, C.~L., {Ricker}, P.~M., \& {Randall}, S.~W. 2008,
  \apj, 680, 17

\bibitem[{{Zacchei et al.}(2011)}]{planck2011-1.6}
{Zacchei et al.} 2011, {Planck early results 05: The Low Frequency Instrument
  data processing} ({Submitted to \aap, [arXiv:astro-ph/1101.2040]})

\end{thebibliography}


\appendix

\section{Optimised SZ extraction and comparison with X-ray predictions}
\label{ap:opt}

\begin{figure*}[!ht]
\begin{centering}
\includegraphics[width=0.33\textwidth]{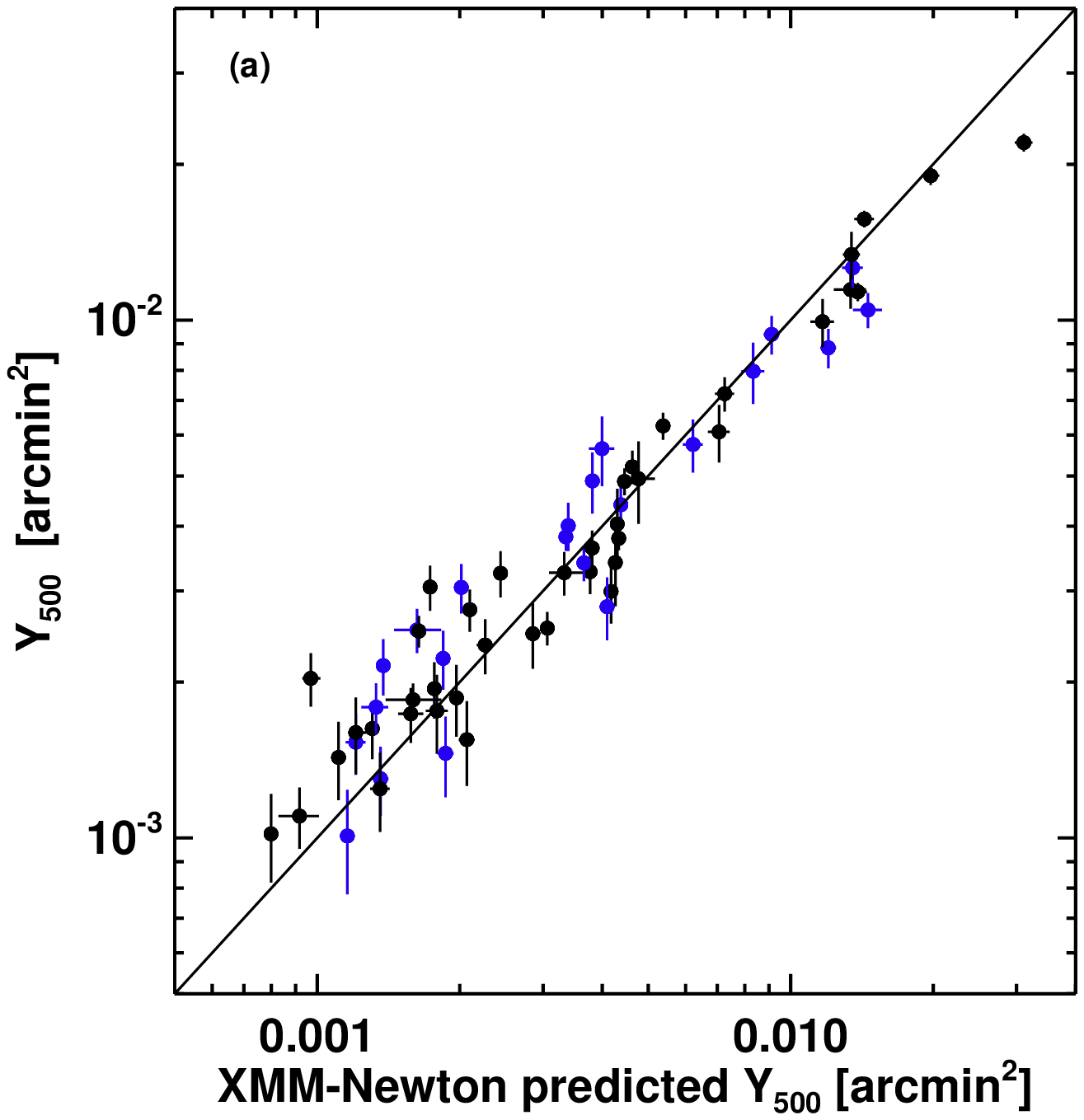}
\hfill
\includegraphics[width=0.33\textwidth]{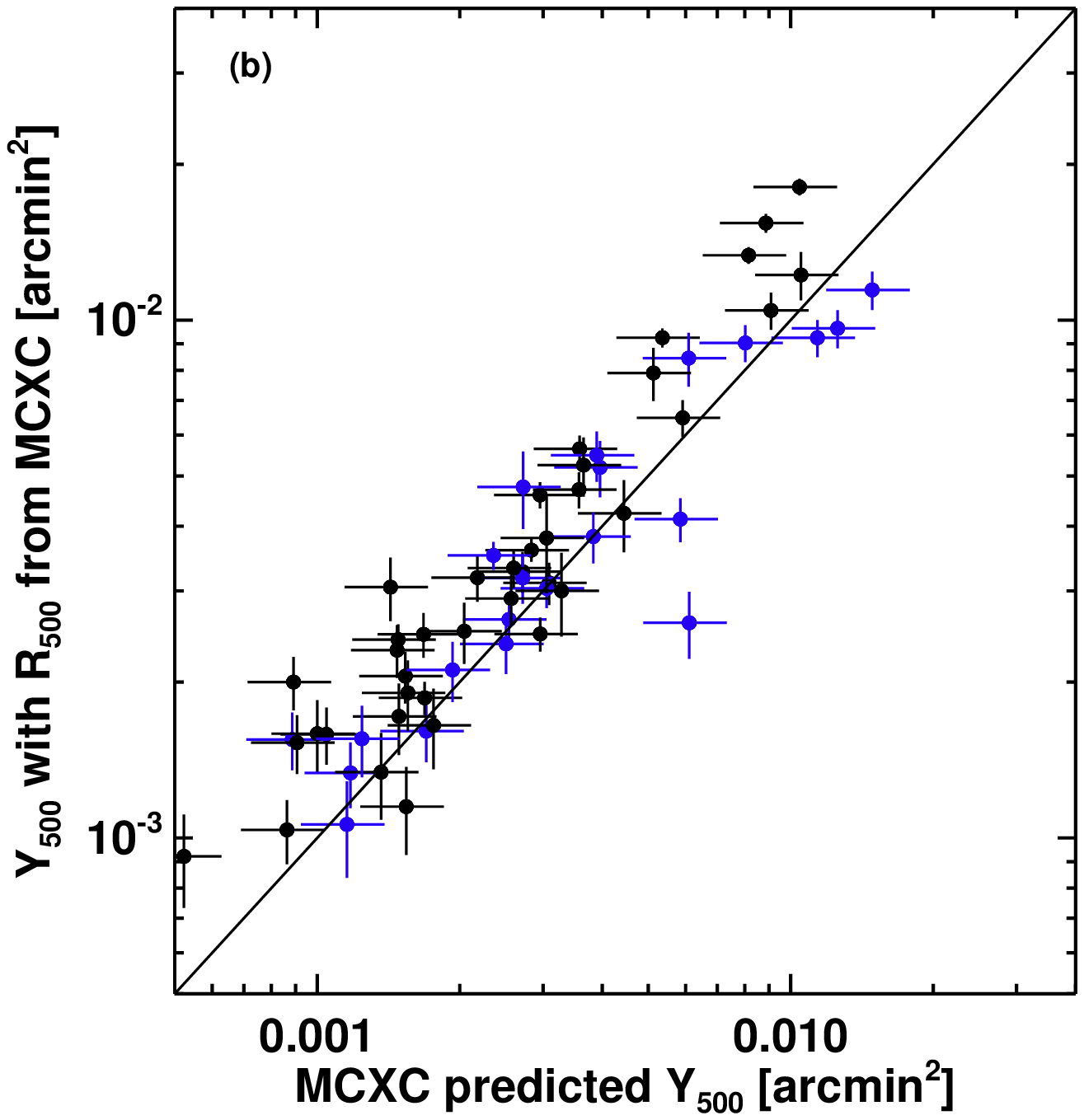}
\hfill
\includegraphics[width=0.33\textwidth]{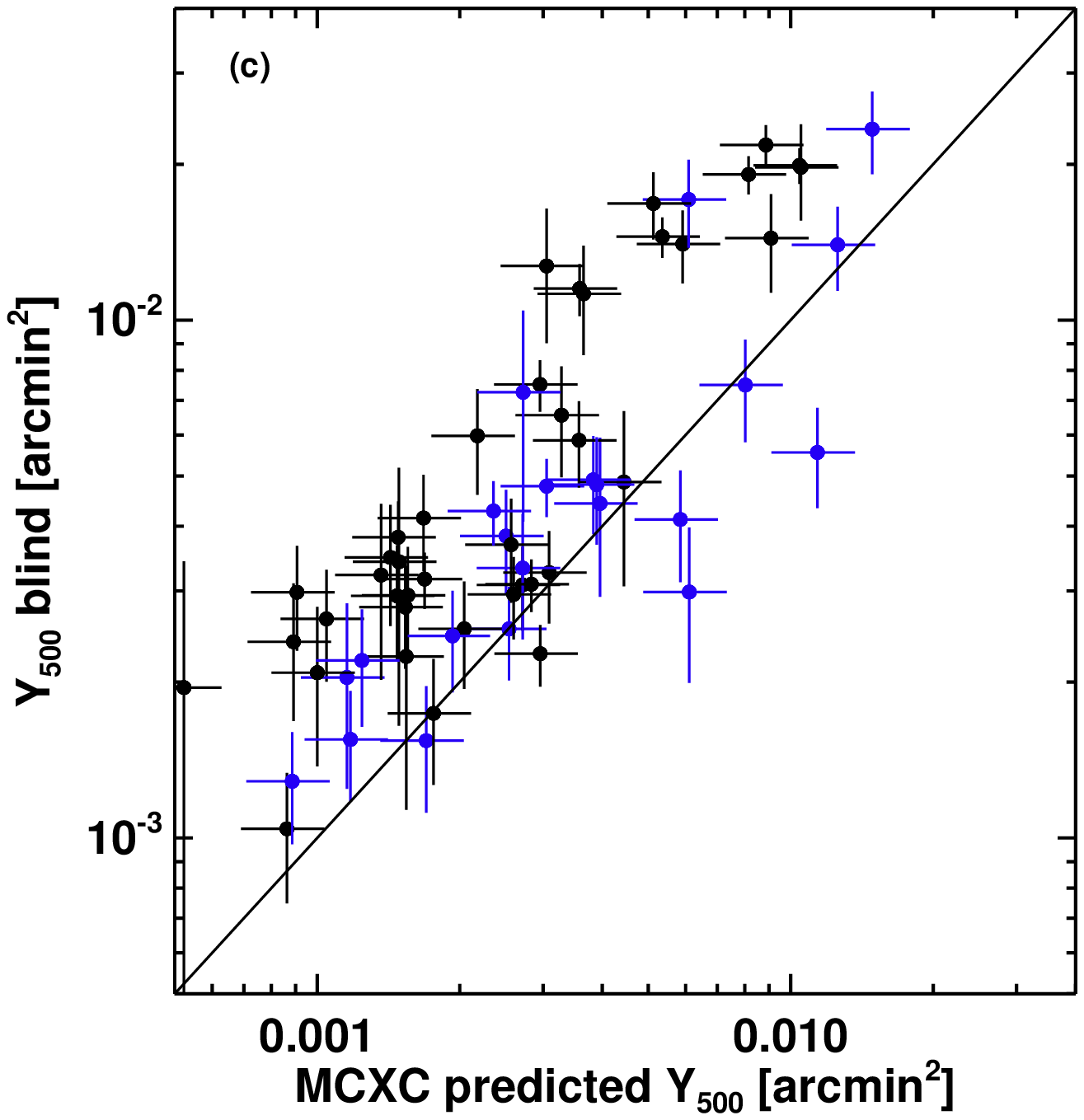}
\end{centering}
\caption{\footnotesize Comparison of the \planck\ measured SZ flux, $\Yv$ and the predictions from X-ray measurements, $(\sigma_T/m_{\rm e} c^2)/(\mu_{\rm e} m_{\rm p}) D_{\rm A}^{-2} \YX$.  Cool core systems are marked as blue stars, other systems as black dots.  {(\it Left)} $\YX$ from \xmm\ (see top-left panel of Fig.~\ref{fig:scaling}), and $\Yv$ computed at the position and $\theta_{500}$ derived from \xmm\ measurements.  {(\it Middle)} $\YX$ and $\Yv$ respectively from $L_{{\rm X, 500}}$, and  position and $\theta_{500}$ as given in the MCXC \citep{pif10}.  {(\it Right)} $\YX$ from $L_{{\rm X,500}}$ as given in the MCXC, while $\Yv$ are blind \planck\ measurements.}\label{fig:ysznew}
\end{figure*}

As discussed in the main text, with the present cluster sample we have optimised the SZ photometry by using the higher-quality estimate of the X-ray size $\theta_{500}$, derived from $R_{500}$ measured using the $M_{500}- \YX$ relation as detailed in Sect.~\ref{sec:xquan}.  

In Fig.~\ref{fig:ysznew} we examine the change in $\Yv$ when derived using different characteristic sizes $\theta_{500}$ to extract the SZ signal.  We also compare the SZ signal predicted using X-ray observations (from the $\Yv /\YX$ relation of \citealt{arn10}) to the observed SZ signal.  In all cases, the ICM pressure is assumed to follow the baseline universal profile of \citet{arn10}.  As extensively described in \citet{planck2011-5.1a}, the SZ flux $\Yv$ is computed by integrating along the line-of-sight and normalising the universal pressure profile.  Each profile is truncated at $5 \times \Rv$, effectively giving a measure of the flux within a cylinder of aperture radius $5 \times \Rv$, and then converted to the value in a sphere of radius $\Rv$ for direct comparison with the X-ray prediction.  

The left hand panel (a) of Fig.~\ref{fig:ysznew} shows the relation between $\Yv$ and that predicted from \xmm\ observations as used in the present paper, illustrating the tight agreement between the two quantities.  We recall that here, the $R_{500}$ within which the SZ signal is extracted is derived from the measured $Y_{\rm X,500}$ using the $M_{500} - Y_{\rm X,500}$ relation given in Eqn.~\ref{eqn:Yx}.

The middle panel (b) of Fig.~\ref{fig:ysznew} assumes that only the X-ray position and luminosity of the cluster are known.  In this case the mass is derived from the $M_{500}-L_{\rm X,500}$ relation as described in the MCXC of \citet{pif10}, thus yielding the characteristic size used to extract the SZ signal, $\theta_{500}$.  This mass is also used to predict $Y_{\rm X,500}$ via the $M_{500}-Y_{\rm X,500}$ relation in Eqn.~\ref{eqn:Yx}.  Consistently, the expected SZ signal is extracted from a region of size $\theta_{500}$ centred on the X-ray position given in MCXC \citep[as in][]{planck2011-5.2a}.  

In the right hand panel (c) of Fig.~\ref{fig:ysznew} the position and size of the cluster are unknown, thus $\Yv$ is devired blindly together with the SZ flux.  However, the predicted SZ flux is derived as above in panel (b).
The agreement between measured and predicted values clearly degrades dramatically from panels (a) to (c). Comparing panels (b) and (a),
there is a systematic shift to lower predicted $Y_{\rm X,500}$ values, with a segregation now appearing between cool cores and the other systems.  This can be explained by the fact that using the luminosity as a simple mass proxy leads to an underestimate of the mass for morphologically disturbed systems in view of their position with respect to the mean $L_{X, 500} - M_{500}$ relation \citep{pra09}.  The inverse effect is seen for the cool cores.  In addition, there is a smaller impact on the measured $\Yv$ via the effect of the assumed $\theta_{500}$.  However the effect is smaller:
the average ratio of \xmm~ and MCXC characteristic sizes $\theta_{500,{\rm MCXC}}/\theta_{500,{\rm XMM}}$ is $0.95\pm 0.06$, corresponding to a change in area of $\sim 10$ per cent, which translates into a similar variation in SZ flux.  This shows that the X-ray luminosity in the MCXC is a sufficiently good mass proxy for a reliable size estimate.  

Finally, panel (c) of Fig.~\ref{fig:ysznew} illustrates the size-flux degeneracy in blind \planck\ measurements.  When $\Yv$ is measured blindly, the size is on average overestimated \citep[see also][]{planck2011-5.1a}, and so the disagreement with predictions is even more apparent.

We see that as a result of the size-flux degeneracy, an accurate estimate of the characteristic size is mandatory in order to derive an accurate measure of $\Yv$.  A similar conclusion was reached in \citet[][see their Fig.~11]{planck2011-5.1a}, where the effect was demonstrated using the full sample of 158 clusters known in X-rays (i.e., those included in the MCXC).  These authors found that, in addition to a reduction in intrinsic scatter (from 43 to 34 per cent), knowledge of the cluster size dramatically reduced the offset of the measured $\Yv$ to that predicted from X-rays (from 84 to 14 per cent -- compare panels (c) and (b) of Fig.~\ref{fig:ysznew} above).

However as noted in \citet{planck2011-5.1a}, there is still a small but
  systematic discrepancy.  This is mostly due to the use of $L_{\rm X,500}$, a quantity which shows considerable scatter with
  mass, as a mass proxy.  The superior constraints provided by the 
  \xmm\ observations on the cluster size and on $\YX$ suppress most
  of this remaining systematic effect (compare panels (b) and
  (a)).   A smaller contribution is liked to effects due to the
nature of the sample selection.  This illustrates that a fully coherent approach is needed when undertaking a proper comparison between SZ and X-ray predictions.

\raggedright
\end{document}